\pdfoutput=1
\documentclass[11pt]{article}
\usepackage[utf8]{inputenc}
\usepackage{cite}
\usepackage{hyperref}
\hypersetup{colorlinks=true,linkcolor=bluecyan,citecolor=red3,urlcolor=green3,pdfencoding=auto,linktocpage}
\usepackage{amsmath,amssymb,amsbsy,amstext,amsthm,simplewick,amsfonts}
\usepackage{mathrsfs}
\usepackage{graphicx}
\usepackage{wrapfig}
\usepackage{upgreek}
\usepackage{bm} 
\usepackage{framed}
\usepackage{bbm}
\usepackage{textcomp}
\usepackage{adjustbox}
\usepackage{makecell}
\usepackage{tcolorbox}
\usepackage{empheq}
\usepackage[normalem]{ulem}
\usepackage{enumitem}
\usepackage{braket} 
\usepackage{array}
\usepackage{dsfont}
\usepackage{physics}
\usepackage{ulem}
\usepackage{xcolor}
\usepackage{caption}
\usepackage{subcaption}
\usepackage{tocloft}
\usepackage{tikz}

\usepackage[framemethod=default]{mdframed}

\newmdenv[backgroundcolor=gray!15,%
skipabove=5pt,%
skipbelow=5pt,%
leftmargin=2pt,%
rightmargin=2pt,%
innertopmargin=-6pt,%
innerbottommargin=5pt,%
innerleftmargin=5pt,%
innerrightmargin=5pt,%
splittopskip=0pt,%
splitbottomskip=0pt,%
linewidth=0pt,%
nobreak=true]%
{keyeqn}

\newmdenv[backgroundcolor=gray!15,%
skipabove=5pt,%
skipbelow=5pt,%
leftmargin=2pt,%
rightmargin=2pt,%
innertopmargin=-2pt,%
innerbottommargin=5pt,%
innerleftmargin=5pt,%
innerrightmargin=5pt,%
splittopskip=0pt,%
splitbottomskip=0pt,%
linewidth=0pt,%
nobreak=true]%
{keythrm}

\graphicspath{{Fig/}}

\definecolor{lightgreen}{cmyk}{0.2, 0, 0.2, 0.2}
\definecolor{lightgray}{cmyk}{0.1,0.2,0,0.1}
\definecolor{lightgray2}{cmyk}{0.1,0.1,0,0.1}
\definecolor{greyish2}{rgb}{.96,.96,.96}
\definecolor{bluecyan}{RGB}{0, 100, 200}
\definecolor{blue3}{RGB}{31,119,180}
\definecolor{red3}{RGB}{214,39,40}
\definecolor{orange3}{RGB}{255,127,14}
\definecolor{green3}{RGB}{44,160,44}
\definecolor{red2}{RGB}{255,0,0}
\definecolor{green2}{RGB}{0,170,0}
\definecolor{blue2}{RGB}{0,128,255}
\definecolor{magenta2}{RGB}{191,64,191}
\definecolor{purple2}{RGB}{112,48,160}
\definecolor{orange2}{RGB}{255,192,0}
\definecolor{blue2}{RGB}{117,223,230}
\definecolor{red4}{RGB}{186,60,71}

\newcommand{\email}[1]{\footnote{\href{mailto:#1}{\nolinkurl{#1}}}}

\renewcommand{\eqref}[1]{(\ref{#1})}

\def \i{\mathrm{i}}
\def \a{\alpha}
\def \b{\beta}
\def \c{\gamma}
\def \saa{\mathsf{a}}
\def \sb{\mathsf{b}}
\def \sc{\mathsf{c}}
\newcommand{\m}[1]{\mathsf{#1}}

\numberwithin{equation}{section}

\allowdisplaybreaks[1]
\begin{document}

	\begin{titlepage}
\begin{flushright}
CTPU-PTC-24-10
\end{flushright}
		\setcounter{page}{1} \baselineskip=15.5pt 
		\thispagestyle{empty}

		\begin{center}
            {\fontsize{15.8}{16} \bf Cosmological Correlators with Double Massive Exchanges:}\\[14pt] 
            {\fontsize{13.3}{13.2} \bf Bootstrap Equation and Phenomenology}
            \\[14pt] 
		\end{center}
		\vskip 15pt
		\begin{center}
			\noindent
			{\fontsize{12}{18} \selectfont 
            Shuntaro Aoki 
   \footnote[1]{\href{mailto:shuntaro1230@gmail.com}{shuntaro1230@gmail.com}}$^{,a}$, 
            Lucas Pinol 
   \footnote[2]{\href{mailto:lucas.pinol@phys.ens.fr}{lucas.pinol@phys.ens.fr}}$^{,b}$, 
            Fumiya Sano 
   \footnote[3]{\href{mailto:sanof.cosmo@gmail.com}{sanof.cosmo@gmail.com}}$^{,c,d}$, 
			\\ Masahide Yamaguchi 
   \footnote[4]{\href{mailto:gucci@ibs.re.kr}{gucci@ibs.re.kr}}$^{,d,c}$
            and Yuhang Zhu 
    \footnote[5]{\href{mailto:yhzhu@ibs.re.kr}{yhzhu@ibs.re.kr}}$^{,d}$}
		\end{center}
		
		\begin{center}
			\vskip 8pt
			$a$ \textit{Particle Theory and Cosmology Group, Center for Theoretical Physics of the Universe,\\
				Institute for Basic Science, Daejeon, 34126, Korea} \\
			$b$ \textit{Laboratoire de Physique de l’École Normale Supérieure, ENS, CNRS, Université PSL,\\ Sorbonne Université, Université Paris Cité, F-75005, Paris, France} \\
            $c$ \textit{Department of Physics, Tokyo Institute of Technology, Tokyo, 152-8551, Japan} \\
			$d$ \textit{Cosmology, Gravity and Astroparticle Physics Group, Center for Theoretical Physics of the Universe,	Institute for Basic Science, Daejeon, 34126, Korea}
		\end{center}
		
\noindent\rule{\textwidth}{0.4pt}
 \begin{center}
     \noindent \textbf{Abstract}
 \end{center}
Using the recently developed cosmological bootstrap method, we compute the exact analytical solution for the seed integral appearing in cosmological correlators with \textit{double} massive scalar exchanges.
The result is explicit, valid in any kinematic configuration, and free from spurious divergences.
It is applicable to any number of fields' species with any masses.
With an appropriate choice of variables, the results contain only single-layer summations. We also propose simple approximate formulas valid in different limits, enabling direct and instantaneous evaluation.
Supported by exact numerical results using \textsf{CosmoFlow}, we explore the phenomenology of double massive exchange diagrams.
Contrary to single-exchange diagrams with ubiquitous Lorentz-covariant interactions, the size of the cubic coupling constant can be large while respecting perturbativity bounds.
Because of this property, the primordial bispectrum from double-exchange diagrams can be as large as, coincidentally, current observational constraints. In addition to being sizable on equilateral configurations, we show that the primordial bispectrum exhibits a large cosmological collider signal in the squeezed limit, making the double massive exchanges interesting channels for the detection of massive primordial fields.
We propose to decisively disentangle double-exchange channels from single-exchange ones with cosmological observations by exploiting the phase information of the cosmological collider signal, the inflationary flavor oscillations from multiple fields' species exchanges and the double soft limit in the primordial trispectrum.
	
		
	\end{titlepage} 
	
	
	\newpage
	\setcounter{page}{2}
	{
		\tableofcontents
	}
	
	\newpage

	
\section{Introduction}

\paragraph{Cosmological correlators.}

Key predictions of inflationary cosmology are encoded as statistical information in the \textit{cosmological correlators}, namely, the $n$-point correlation functions of ubiquitous massless fields: the curvature perturbation and the gravitons. 
The final values of these correlation functions at the end of inflation depend on the dynamics and interactions of quantum fluctuations in the bulk of the inflationary spacetime.
Because these correlators provide the initial conditions for the subsequent evolution of the universe, observing the correlation functions in the Cosmic Microwave Background (CMB) and Large Scale Structures (LSS) offers the intriguing possibility of gaining insights into high-energy processes inaccessible through terrestrial collider experiments.
The correlators involving the exchange of massive and possibly spinning particles are of particular interest. 
The inherently high-energy scale of inflation creates an environment favorable for producing massive fields during this remarkable epoch. 
Through their interactions with the curvature perturbation, these fields generate distinct patterns in the correlation functions, encoding crucial information about their masses, spins, mixing angles, and interactions.
This concept was initially introduced in pioneering works~\cite{Chen:2009zp,Baumann:2011nk,Noumi:2012vr}, and was
subsequently elaborated and termed as the \textit{Cosmological Collider} (CC)~\cite{Arkani-Hamed:2015bza}. Numerous recent studies have utilized the CC signal to investigate models of high-energy physics~\cite{Chen:2009we,Sefusatti:2012ye,Norena:2012yi,Chen:2012ge,Pi:2012gf,Cespedes:2013rda,Gong:2013sma,Emami:2013lma,Kehagias:2015jha,Liu:2015tza,Dimastrogiovanni:2015pla,Schmidt:2015xka,Chen:2015lza,Delacretaz:2015edn,Bonga:2015urq,Chen:2016nrs,Flauger:2016idt,Lee:2016vti,Delacretaz:2016nhw,Meerburg:2016zdz,Chen:2016uwp,Chen:2016hrz,Kehagias:2017cym,An:2017hlx,Tong:2017iat,MoradinezhadDizgah:2017szk,Iyer:2017qzw,An:2017rwo,Kumar:2017ecc,Franciolini:2017ktv,RiquelmeM:2017qhp,Tong:2018tqf,Chen:2018sce,Saito:2018omt,Cabass:2018roz,MoradinezhadDizgah:2018ssw,Wang:2018tbf,Chen:2018xck,Bartolo:2018hjc,Dimastrogiovanni:2018uqy,Bordin:2018pca,Chen:2018cgg,Achucarro:2018ngj,Chua:2018dqh,Kumar:2018jxz,Goon:2018fyu,Wu:2018lmx,Anninos:2019nib,Li:2019ves,McAneny:2019epy,Kim:2019wjo,Alexander:2019vtb,Lu:2019tjj,Hook:2019zxa,Hook:2019vcn,Kumar:2019ebj,Liu:2019fag,Wang:2019gbi,Welling:2019bib,Wang:2019gok,Wang:2020uic,Li:2020xwr,Wang:2020ioa,Fan:2020xgh,Kogai:2020vzz,Bodas:2020yho, Aoki:2020zbj,Maru:2021ezc,Kim:2021pbr,Lu:2021gso,Sou:2021juh, Lu:2021wxu,Pinol:2021aun,Cui:2021iie,Tong:2022cdz,Reece:2022soh,Chen:2022vzh,Qin:2022lva,Cabass:2022rhr,Cabass:2022oap,Niu:2022quw,Niu:2022fki,Tran:2022euk,Aoki:2023tjm,Tong:2023krn,Jazayeri:2023xcj,Yin:2023jlv,Chakraborty:2023qbp,Chen:2023txq,Jazayeri:2023kji,Chakraborty:2023eoq,Ema:2023dxm,McCulloch:2024hiz,Craig:2024qgy,Wu:2024wti} along with developing both analytical and numerical techniques aiming to deepen the understanding of underlying fundamental properties of these correlators~\cite{Maldacena:2011nz,Assassi:2012zq, Baumann:2017jvh, Arkani-Hamed:2018bjr, Sleight:2019mgd, Sleight:2019hfp,Arkani-Hamed:2018kmz, Baumann:2019oyu, Baumann:2020dch, Pajer:2020wnj,  Goodhew:2020hob, Meltzer:2021zin, Goodhew:2021oqg, Jazayeri:2021fvk,  Baumann:2021fxj, Melville:2021lst,   DiPietro:2021sjt, Gomez:2021qfd, Premkumar:2021mlz, Bonifacio:2021azc,  Hogervorst:2021uvp,  Sleight:2020obc,Sleight:2021plv,Sleight:2021iix, Cabass:2021fnw, Tong:2021wai, Wang:2021qez, Gomez:2021ujt, Baumann:2022jpr, Heckelbacher:2022hbq,Pajer:2020wxk, Pimentel:2022fsc,  Jazayeri:2022kjy,Bonifacio:2022vwa,Lee:2022fgr, Qin:2022fbv, Xianyu:2022jwk, Wang:2022eop, Qin:2023ejc, Qin:2023bjk, Qin:2023nhv, Xianyu:2023ytd, Green:2023ids, DuasoPueyo:2023viy,De:2023xue,Loparco:2023rug,Lee:2023jby,Aoki:2023dsl,Arkani-Hamed:2023kig,Arkani-Hamed:2023bsv,Stefanyszyn:2023qov,Donath:2024utn,Fan:2024iek,Grimm:2024mbw,Du:2024hol,Melville:2024ove}, see also the recent first searches for CC signals in real cosmological data~\cite{Cabass:2024wob,Sohn:2024xzd}.
\paragraph{Analytical and numerical methods.}
Conventional methods for calculating inflationary correlation functions, based on the so-called in-in (or Schwinger-Keldysh (SK)) formalism, typically involve multiple layers of nested integrals in the time domain (see~\cite{Chen:2010xka,Wang:2013zva,Chen:2017ryl} for comprehensive reviews).
Remarkably, in the recent years we have seen significant developments in various analytical and numerical methods. 
For instance, newly developed wavefunction approaches have made the analytical properties behind correlators more transparent~\cite{Anninos:2014lwa,Arkani-Hamed:2017fdk,Hillman:2019wgh,Goodhew:2020hob,Cespedes:2020xqq,Baumann:2021fxj,Goodhew:2021oqg,Jazayeri:2021fvk,Hillman:2021bnk,Cabass:2021fnw,Meltzer:2021zin,Bonifacio:2022vwa,Salcedo:2022aal,Stefanyszyn:2023qov,Albayrak:2023hie,Agui-Salcedo:2023wlq,Cespedes:2023aal}. 
Indeed, evaluating directly the nested time integrals---involving massive propagators expressed as complicated special functions---can be extremely challenging. 
Given that the inflationary background is quasi-de Sitter (dS), by restricting the calculation of correlators to their values at the end of inflation only, we can bring a drastic simplification.
This so-called ``boundary perspective'' was introduced in a program known as the \textit{cosmological bootstrap}~\cite{Arkani-Hamed:2018kmz,Baumann:2019oyu,Baumann:2020dch}, see\cite{Baumann:2022jpr} for a review.
Initially relying on the exact dS symmetries, this method has led to the full analytical form of tree-level correlators built off single-exchange channels only.
Later, it has been extended to symmetry-breaking cases, such as those with non-unit sound speed~\cite{Pajer:2020wxk, Pimentel:2022fsc, Jazayeri:2022kjy},
chemical potential~\cite{Qin:2022fbv}, with IR effects~\cite{Wang:2022eop}, and time-dependent mass~\cite{Aoki:2023dsl}. Additionally, fundamental properties including unitarity, locality, causality, and analyticity were invoked to comprehend the underlying structure of these cosmological correlators~\cite{Goodhew:2020hob,Melville:2021lst,Goodhew:2021oqg,DiPietro:2021sjt,Baumann:2021fxj,Jazayeri:2021fvk,Meltzer:2021zin}. 
The Mellin transformation also proved to be powerful for understanding inflationary correlators~\cite{Sleight:2021iix,Sleight:2021plv,Sleight:2020obc}. 
The so-called partially Mellin-Barnes (MB) method was introduced over the last two years~\cite{Qin:2022lva,Qin:2022fbv,Qin:2023ejc,Qin:2023bjk,Qin:2023nhv,Xianyu:2023ytd} and makes it easier to calculate bulk time integrals, and subsequently many exact results for inflationary correlators were obtained.
In a different direction, a systematic program called the ``Cosmological Flow'' was proposed last year~\cite{Werth:2023pfl,Pinol:2023oux}.
It enables us to trace the time evolution of the correlators in the bulk of the inflationary evolution and is valid even in the strong quadratic mixing regime that remains so far inaccessible with traditional methods.
The numerical implementation of this method is called \textsf{CosmoFlow} and a user-friendly guide with non-trivial examples is provided in~\cite{Werth:2024aui}.

\paragraph{Beyond single-exchange diagrams.}

After these efforts, the analytical structure of correlators associated with a single massive (spinning) field exchange is now well understood.
Even in theories involving several different species of massive fluctuations with non-trivial interactions, the single-exchange channel was computed, and a new phenomenon, dubbed inflationary flavor oscillations, was discovered~\cite{Pinol:2021aun}. 
Nevertheless, \textit{More is different.}
Moving beyond the single-exchange tree-level diagrams is highly challenging due to the significant increase in complexity. For example, our understanding of loop diagrams is still far from satisfactory. 
Recently, some progress has been made towards understanding the structure of loop diagrams involving massive fields.  
For example, using the techniques of spectral decomposition in dS to obtain the one-loop results with the pair of massive scalars~\cite{Xianyu:2022jwk}, and using the MB representation to extract the non-analytical signals within loop diagrams~\cite{Qin:2023bjk,Qin:2023nhv}. 
Another important but complicated case arises from including several massive propagators and more interactions in the tree-level diagrams. 
Those diagrams are crucial because of their rich phenomenology, whereas their analytical understanding is still limited. 
From the perspective of the bulk time integral, incorporating more massive fields and more interaction vertices results in exceedingly complicated integrals. 
The integrand comprises products of several special functions, and the number of layers of nested time integrals has increased.
Thus, it seems hopeless to perform the time integration directly. 
To address the problem, several possible methods have been proposed, each with their own unique advantages.
First, an approximate method, called the bulk-version cutting rule, has proven to be useful. 
This method treats one massive propagator at a time while mimicking others using some local operators in an effective way, which enables us to extract the leading-order CC signals \cite{Tong:2021wai}.
However, sub-leading contributions, as well as the background signal that is non-negligible in kinematic configurations relevant for comparison with cosmological observations, are missed.
Regarding exact calculations, the Cosmological Flow and its numerical implementation \textsf{CosmoFlow} are useful tools as they can give the exact result for any massive spinless exchange channel, including CC and background signals, and also including the regime of strong quadratic mixing~\cite{Werth:2023pfl,Pinol:2023oux,Werth:2024aui}. 
As far as exact analytical methods are concerned, only one very recent work has successfully generalized the MB method to include more than just single-exchange correlators~\cite{Xianyu:2023ytd}, although the exact results are formulated with several layers of series summation.
As for the bootstrap equations, to the best of our knowledge, no result has been published so far that is associated to diagrams beyond the single exchange.

\paragraph{Content of this work.}

In this work, we present the first success in applying the cosmological bootstrap equations for deriving the exact solutions of correlators that involve \textit{double} massive fields exchanges.
Moreover, we perform for the first time a detailed comparison of the bootstrap result with the Cosmological Flow method---completely independent since not relying on a perturbative scheme in terms of the quadratic mixing---by utilizing \textsf{CosmoFlow}.
To be more specific, we consider the double-exchange four- and three-point functions depicted in Figure~\ref{double_diagram}. 
In full generality, we allow those two exchanged fields to have different masses, and the obtained results are applicable to theories with various types of interactions, as long as they do not lead to secular divergences. 
Although the concrete calculation concerns correlators of massless scalar fluctuations, introduced as external fields, the extension to graviton correlators is straightforward. 

\subparagraph{Bootstrap equations.}
The first part of our work focuses on deriving the bootstrap equations satisfied by the \textit{seed integral} associated with these diagrams, and obtaining analytical solutions for the resulting differential equations. More precisely, we can apply a differential operator to each massive propagator individually, resulting in two second-order differential equations with a source term being the single-exchange channel in the absence of the corresponding propagator.
This whole procedure is summarized in the schematic diagram in Figure \ref{eq_diagram}, and is fully explained in Section~\ref{sec: BS_equation}.
Here, we stand on the shoulders of giants, benefiting from the well-established result of the single-exchange diagram, particularly the exact and closed-form expression of the corresponding seed integral~\cite{Qin:2023ejc} which does not require any series summation. 

\subparagraph{Solutions to the bootstrap equations.}
We show in Section~\ref{sec:BS_solution} that we can solve these complicated bootstrap equations. 
The homogeneous equations can be solved using a specific type of two-variable hypergeometric function known as the Appell function $F_4$. We have observed that a particular choice of variables is quite useful, namely $r_i\equiv 2 k_i/(k_1+k_2+k_3+k_4)$, where $k_i=|\mathbf{k}_i|$. 
By employing these variables and choosing appropriate ansatz, we find particular solutions and also the integration constants, ultimately deriving the exact general solutions for these bootstrap equations. 
Notably, these solutions involve only \textit{one} series summation which makes the analysis of their analytical structure more transparent.
As is now well known, the single-exchange channel exhibits two distinct contributions characterized by different behaviors: one contribution involves imaginary powers of momentum ratios and is known as the celebrated CC signal, while the other contributions display analytic dependence on the momentum ratios and are typically regarded as the featureless background.  
As could be expected, the double-massive fields exchanges channels consist of three distinct terms: one arises from signals generated by both massive fields, another one from the mixing between the background and the signal originating from only one massive field and the remaining one represents purely the background. Each term has different mass dependence as will be emphasized in the main text. 
This iterative procedure, employing simpler diagrams as the source to bootstrap equations for seed integrals involving more massive fields exchanges, can be generalized to more complicated diagrams.  
In particular, we demonstrate that the results of the double-exchange diagram can be used to calculate triple-exchange correlators.

\subparagraph{Explicit cancellations of spurious divergences.}
Unfortunately, the primary results of the four-point correlation function are not applicable to all kinematic regions, as also observed in~\cite{Xianyu:2023ytd}. 
The situation is exacerbated when considering the three-point function, typically derived by taking the soft limit of one of the external lines (e.g., $\mathbf{k}_4\rightarrow0$ in Figure~\ref{double_diagram}).
The primary result obtained from taking this limit does not converge for kinematic regions that satisfy the triangle inequality. 
In Section~\ref{Sec_Continuation}, we will perform some transformations of the special functions in it, and carefully examine the divergent behaviour under various limits (e.g. folded limit). 
We observed that, by regrouping the divergences of homogeneous and particular solutions together, the whole expression achieves convergent. In brief, by using the continuation of certain special functions and appropriately organizing and treating all contributions as a whole, we are able to obtain a final result that is explicitly convergent and valid for all physical kinematic regions, rendering it perfectly usable for concrete purposes.

\subparagraph{Large primordial bispectrum and cosmological collider signal, inflationary flavor oscillations and the primordial trispectrum.}
The second part of our work, in Section~\ref{Sec_pheno}, precisely digs into the rich phenomenology associated with double massive exchange (DE) diagrams. We recall how to relate correlators of the massless scalar fluctuation $\varphi$ to cosmological correlation functions of the primordial curvature fluctuation $\zeta$. We also propose two natural embeddings for the theory under study, with the massive scalar fluctuations $\sigma_\alpha$ identified as the massive eigenstates of the isocurvature fluctuations during inflation.
We explain that, in the large class of general non-linear sigma models of inflation, one should generically expect cubic interactions involving one massless field and two massive ones, leading to DE diagrams.
Moreover, from generic effective field theory perspectives, we also expect such massive fluctuations to be coupled via an interaction of this form.
Importantly, the size of this cubic interaction---set by the curvature of the target space in non-linear sigma models---is not dictated by the quadratic mixing alone, contrary to the ubiquitous Lorentz-covariant one leading to single-exchange (denoted as SE in the following) diagrams.
Therefore, we show that the resulting primordial bispectrum can be large, first on equilateral configurations for which we display a simple fitting formula, and also in squeezed configurations for which we derive the CC signal in a closed and explicit form for the first time.
Importantly, the DE CC signal \textit{is not} suppressed by additional factors of the Boltzmann factor ($e^{-\pi \mu}$ where $\mu=\sqrt{m^2/H^2-9/4}$ is the mass parameter of the double exchanged massive field) compared to the SE CC signal, and we show that its relative polynomial suppression ($\mu^{-2}$) can easily be compensated by a larger cubic coupling constant for masses $m \gtrsim H$ relevant for cosmological observations.
We explain that the bootstrap result is exact and valid for any kinematic configuration, showcasing its usefulness to predict the full shape of the bispectrum and making contact with observations.
For each of these steps, we precisely compare our analytical predictions with exact numerical calculations using \textsf{CosmoFlow} and not relying on a perturbative scheme for the quadratic mixing.
Impressively, the two independent methods yield almost identical results for any kinematic configuration, and any values of the mass of the double exchanged field.
Moreover, supported by the exact numerical result, we propose a first naive extrapolation of the bootstrap result to the strong quadratic mixing regime, highlighting the weaknesses of this procedure.
We also propose ways to disentangle DE channels from SE ones from cosmological observations, beyond the simple observation that the DE signal may be observable while respecting perturbativity bounds, contrary to the SE one.
First, we showcase the utility of the CC phase information to tell those channels apart, since the relation between frequency and phase are well different for masses close to the Hubble scale.
Second, we explain that in the more generic situation where there exist different species of massive fields, each with their own masses, one can use the inflationary flavor oscillations to tell apart the two channels.
Third, by moving to the primordial trispectrum we show a unique feature of DE diagrams that cannot be mimicked by a SE one, with CC signal oscillations transitioning from a frequency $\mu$ to $2 \mu$ in the double soft limit.

\paragraph{Organization of this work.}
This paper is organized as follows. In Section \ref{sec: BS_equation}, we define the seed integral related to the double-exchange inflation correlators and then derive the bootstrap equations satisfied by these seed integrals. In Section \ref{sec:BS_solution}, we derive the exact solution of the bootstrap equations step by step, and the final exact expression is summarized in Section \ref{Summary_sol}. The results are extended to all physical regions in Section \ref{Sec_Continuation}, where we also discuss their behaviors under various limits. In Section \ref{Sec_pheno}, we delve into the intriguing phenomenology underlying the double-exchange correlators. 
Finally, we conclude and outline future directions in Section \ref{sec:conclusion}.

\paragraph{Conventions and notations.}  Throughout this paper, we adopt  the $(-,+,+,+)$ metric sign convention. The background spacetime is fixed as: $\mathrm{d}s^2=a^2(\tau)(-\mathrm{d}\tau^2+\mathrm{d}\mathbf{x}^2)~$, with the scale factor $ a(\tau)=-{1}/{H\tau}$ and $\tau\in(-\infty,0)$. Here $H$ represents  the Hubble parameter that is fixed as the constant.  A prime on quantum average values $\langle\cdots\rangle'$ indicates  the momentum $\delta$-function and the factor $(2\pi)^3$ is omitted. We frequently use the shorthand notations, for example $k_{1234}\equiv \sum_{i=1}^{4} k_i$ and $p_{123}\equiv p_{1}+p_2+p_3$, with similar shorthand subscripts following the same convention. Some frequently used variables are defined as 
\begin{align}
    u\equiv\frac{k_1}{k_{24}},\qquad v\equiv\frac{k_3}{k_{24}},\qquad r_i\equiv\frac{2 k_i}{k_{1234}},\qquad \tilde{r}_{i}\equiv\frac{2k_i}{k_{123}}~.\nonumber
\end{align}
The majority of the special functions utilized in this work are referenced in the mathematical functions handbook \cite{NIST:DLMF}. Additionally, in Appendix~\ref{sec: formula}, we provide the definitions and useful formulae related to these special functions. Other variables and functions will be defined in the main text as they are introduced.
\label{sec:intro}
\section{Seed Integral and Bootstrap Equations}\label{sec: BS_equation}

\begin{figure}[ht]
\centering
\includegraphics[width=0.75\textwidth]{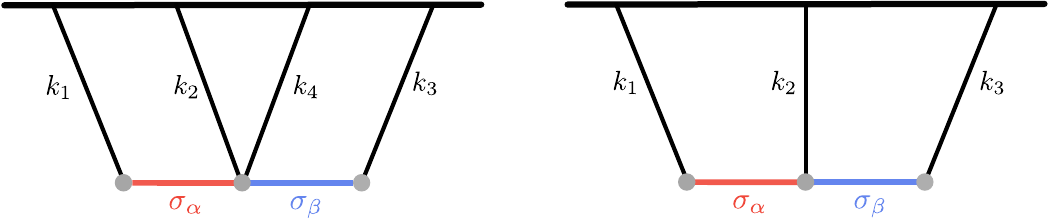}\\
	\caption{Four-point and three-point diagrams with double massive exchange. For generality, here we allow internal massive scalar fields $\sigma_\a$ and $\sigma_\b$ to have different masses. }\label{double_diagram}
\end{figure}

This part serves as an introduction to the basic ingredients needed for later discussion. Experts who are familiar with these descriptions can directly skip to Section~\ref{def_seed}.\\
As mentioned in the introduction, in this work, we will mainly focus on the four-point and three-point inflation correlators shown in Figure \ref{double_diagram}. 

The external lines here always represent the massless inflaton fluctuation $\varphi$ 
(or, equivalently, the curvature perturbation $\zeta$), and the internal lines are associated with massive scalar fields $\sigma_{\alpha}\ (\alpha=1,2,\cdots)$. Distinguished by different colors for clarity, we allow the exchange of massive scalar fields with distinct masses $m_\alpha$, where $\alpha$ (or $\beta$) serves as the label for different species.
By taking the free Hamiltonian to be quadratic and diagonal, each of the fields can be quantized separately as
\begin{align}\Phi^A(\tau,\mathbf{x})&= \int \frac{\mathrm{d}^3 \mathbf{k}}{(2 \pi)^3}\left(\Phi^A_k(\tau) a^A_{\mathbf{k}}+\Phi_k^{A*}(\tau) a_{-\mathbf{k}}^{A \dagger }\right)e^{\i\mathbf{k} \cdot \mathbf{x}} \quad \text{(no sum on $A$)} \,, \label{Q_phi}
\end{align}
where $\Phi^A$ can denote either the inflaton perturbation $\varphi$ or the massive fields $\sigma_\alpha$ and $a^{A\dagger}_{\mathbf{k}}(a_{\mathbf{k}}^{A})$ are the creation (annihilation) operators satisfying the usual commutation relations $[a^A_{\mathbf{k}},a^{\dagger B}_{\mathbf{k}^\prime}] = (2\pi)^3 \delta^{AB} \delta^{(3)}({\mathbf{k}}-{\mathbf{k}}^\prime) $.
Mode functions $\Phi^A_k(\tau)=\{\varphi_k, \sigma^\alpha_k\} $ are specified below,
\begin{align}
    &{\varphi}_k=\frac{H}{\sqrt{2 k^3}}(1+\i k \tau) e^{-\i k \tau}~, \label{sol_u}\\
    &\sigma^\alpha_k=-\i\,e^{-\frac{\pi}{2} \mu_{\alpha}+\i \frac{\pi}{4}} \frac{\sqrt{\pi}}{2} H(-\tau)^{3 / 2} H_{\i \mu_{\alpha}}^{(1)}(-k \tau)~,\label{eom_v}
\end{align}
where $\mu_{\alpha}=\sqrt{m_{\alpha}^2/H^2-9/4}$ and $H_\nu^{(1)}(\cdot)$ is the Hankel function of the first kind of order $\nu$. 
In all the following sections, we show our results for the case with massive fields in the principal series ($ m_{\alpha}>3 H/2$)  which is particularly relevant to cosmological collider physics.
The results related to imaginary $\mu_\a$ 
($ m_{\alpha}<3/2 H $) can then be simply obtained by the substitution $\mu_\alpha \rightarrow - i \nu_\alpha$ with $\nu_\alpha=\sqrt{9/4-m_\alpha^2/H^2}$, and we will use that for a specific example later on in Sec.~\ref{Sec_pheno}.
To formulate the seed integral, we adopt the Schwinger--Keldysh (SK) diagrammatic conventions outlined in \cite{Chen:2017ryl}, where readers can find more details. The bulk-to-boundary propagators $K$ associated with the inflaton fluctuation $\varphi$ are
\begin{align}
    K_{\sf{a}}(k,\tau)=\frac{H^2}{2k^3}(1-\i{\sf{a}}k\tau)e^{\i{\sf{a}} k\tau}~,\label{Bulk-boundary}
\end{align}
where ${\sf{a}}=\pm 1$ is the SK vertex index, and the bulk-to-bulk propagators $D_{\m{ab}}^{\alpha}$  associated with massive fields~$\sigma_{\alpha}$ are explicitly given by 
\begin{align}
&D^{\alpha}_{-+}\left(k ; \tau_1, \tau_2\right)=\sigma_k^{\alpha}\left(\tau_1\right) \sigma_k^{\alpha*}\left(\tau_2\right)= \frac{H^2\pi e^{-\pi \mu_{\alpha}}}{4}\left(\tau_1 \tau_2\right)^{3 / 2} H_{\i \mu_{\alpha}}^{(1)}\left(-k \tau_1\right) H_{-\i \mu_{\alpha}}^{(2)}\left(-k \tau_2\right),\label{D_mp}\\
&D_{+-}^{\alpha}\left(k ; \tau_1, \tau_2\right)=\left(D^{\alpha}_{-+}\left(k ; \tau_1, \tau_2\right)\right)^*,\\
&D^{\alpha}_{ \pm \pm}\left(k ; \tau_1, \tau_2\right)=D^{\alpha}_{\mp\pm}\left(k ; \tau_1, \tau_2\right) \theta\left(\tau_1-\tau_2\right)+D^{\alpha}_{\pm\mp}\left(k ; \tau_1, \tau_2\right) \theta\left(\tau_2-\tau_1\right).\label{D_ppmm}
\end{align}
Here $\tau_1$ and $\tau_2$ denote  the conformal time at the vertex where bulk-to-bulk propagators connect.
Note that propagators $D_{\mp\pm}^{\alpha}$ are factorised in time and satisfy the homogeneous differential equation, whereas the time ordering in the $D_{\pm\pm}$ introduced an additional $\delta$-function source to this equation, explicitly 
\begin{align}
&\left[\tau_i^2 \partial_{\tau_i}^2-2 \tau_i \partial_{\tau_i}+k^2 \tau_i^2+\mu_{\alpha}^2+\frac{9}{4}\right] D^{\alpha}_{ \pm \mp}\left(k ; \tau_1, \tau_2\right)=0~,\label{green1}\\
&\left[\tau_i^2 \partial_{\tau_i}^2-2 \tau_i \partial_{\tau_i}+k^2 \tau_i^2+\mu_{\alpha}^2+\frac{9}{4}\right] D^{\alpha}_{ \pm \pm}\left(k ; \tau_1, \tau_2\right)=\mp \mathrm{i} H^2 \tau_1^2 \tau_2^2 \delta\left(\tau_1-\tau_2\right).\label{green2}
\end{align}
where $i=1,2$.
These differential equations will play an important role in deriving the bootstrap equations later and one can easily check that they are indeed satisfied by (\ref{eom_v}).

\subsection{Definition of the seed integral and typical examples}\label{def_seed}
Based on the quantities defined in the previous sections, we now introduce a seed integral $\mathcal{I}$ for four-point inflaton correlators involving {\it{double}} massive field exchange: 
\begin{keyeqn}
\begin{align}
\nonumber \mathcal{I}_{{\saa\sb{\sf{c}},\a\b}}^{p_1p_2p_3}=&\  H^{-4}k_{24}^{9+p_{123}}(-\i \saa\sb{\sf{c}})  \int^0_{-\infty}  \mathrm{d} \tau_1\mathrm{d} \tau_2\mathrm{d} \tau_3(-\tau_1)^{p_1}(-\tau_2)^{p_2}(-\tau_3)^{p_3}\\
&\times e^{\mathrm{i}\mathsf{a} k_{1} \tau_1+\i\mathsf{b} k_{24} \tau_2+\i\mathsf{c} k_{3} \tau_3} D^{\a}_{\saa\sb}\left(k_1 ; \tau_1, \tau_2\right)D^{\b}_{\sb\mathsf{c}}\left(k_3 ; \tau_2, \tau_3\right),\label{I_abc}
\end{align}
\end{keyeqn}
where $p_i\ (i=1,2,3)$ are constant numbers specifying types of $\varphi$-$\sigma$ vertices and $k_j\ (j=1,2,3,4)$ are the four external momenta of  inflatons $\varphi_{k_j}$ and the prefactor $H^{-4}$ is introduced to render the seed integral dimensionless. We use the abbreviation  $p_{123}\equiv p_1+p_2+p_3$, 
$k_{24}\equiv k_2+k_4$, and assume $p_i>-5/2$ for the time-integrals to converge at late times $\tau_i \rightarrow 0$.\footnote{ As long as $\Re[p_i]>-5/2$, it can be analytically continued to imaginary $p_i$ in principle~\cite{Qin:2023ejc}.} We also remind $\saa,\sb,\mathsf{c}=\pm$ correspond to the SK indices and $\a,\b=1,2,\cdots$ label the massive fields. The propagators for massive fields $D_{\saa\sb}^\a$ are defined in \eqref{D_mp}--\eqref{D_ppmm}. The seed integral~\eqref{I_abc} can be diagrammatically shown on the left of Figure \ref{double_diagram}, and other channels can be obtained by simply taking the permutation of the momentum variables.

To get an analytical expression of the seed integral~\eqref{I_abc} is one of the main purposes of this paper. We derive the differential equations (bootstrap equations) for Eq.~\eqref{I_abc}  in the next part Section \ref{BS_double} and solve them in Section~\ref{sec:BS_solution}.
Furthermore, by taking the limit $k_4\rightarrow 0$ in \eqref{I_abc}, one can relate it to the expression for the three-point function (as depicted in the schematic diagram of the right of Figure \ref{double_diagram}). 
Taking this limit is a non-trivial task due to cancellations of apparent divergences, and therefore this will be discussed separately in Section~\ref{Sec_Continuation}.

\subsubsection*{Example}
Once the analytical expressions for the seed integral~\eqref{I_abc} and its various limits are obtained, we then have all the necessary information  to compute two-, three-, and four-point correlators with double exchanges for various $\varphi$-$\sigma $ interactions. 
Among them, the three-point correlators with interactions corresponding to $(p_1,p_2,p_3)=(-2,-2,-2)$ are of particular phenomenological interest since they generically arise from UV motivated multi-field inflation models with curved field space metric~\cite{Garcia-Saenz:2019njm,Pinol:2020kvw}.
As we will show in greater details in Section~\ref{Sec_pheno}, this interaction can also lead to a bispectrum sufficiently large to be constrained by current and upcoming cosmological observations while remaining under theoretical control, thus making it an interesting channel for discovery of new physics at high energies.

Let us look more at this example (three-point correlators with $(p_1,p_2,p_3)=(-2,-2,-2)$) and how it is expressed by the seed integral~\eqref{I_abc}. In this case, the interactions between inflaton's fluctuations~$\varphi$ and massive scalars~$\sigma_\a$ are given by~\cite{Pinol:2020kvw, Pinol:2021aun}

\begin{align}
&S_{\rm{int},2}=\sum_{\alpha}\int \mathrm{d}\tau \mathrm{d}^3\mathbf{x}\ \rho_{\alpha}\,a^{3}\sigma_{\alpha} \varphi^{\prime},\label{int_2}\\
&S_{\rm{int},3}=\sum_{\alpha,\beta}\int \mathrm{d}\tau \mathrm{d}^3\mathbf{x} \ \lambda_{\a\b}\,a^{3}\sigma_{\alpha}\sigma_{\beta} \varphi^{\prime},\label{int_3}
\end{align}
where $\rho_{\alpha}$ and $\lambda_{\a\b}$ are coupling constants with mass dimensions $1$ and $0$ respectively, and the prime denotes the derivative with respect to conformal time $\tau$. Note that $a=(-H\tau)^{-1}$ in de Sitter spacetime.
Then, by the SK diagrammatic rule~\cite{Chen:2017ryl}, one can write down the three-point correlator of double-exchange with interactions~\eqref{int_2} and~\eqref{int_3} as
\begin{align}
\langle\varphi_{\bf{k}_1} \varphi_{\bf{k}_2}\varphi_{\bf{k}_3}\rangle '
\nonumber =&-2\,\i \sum_{\alpha,\beta}\frac{\rho_\a \rho_\b \lambda_{\a\b}}{H^9}\int^0_{-\infty}  \mathrm{d} \tau_1\mathrm{d} \tau_2\mathrm{d} \tau_3( -\tau_1)^{-3}( -\tau_2)^{-3}( -\tau_3)^{-3} \\
&\times \sum_{\saa, \sb,\sc=\pm} (\saa\sb\sc) \,K_{\saa}^{\prime}(k_1;\tau_1)K_{\sb}^{\prime}(k_2;\tau_2) K_{\sc}^{\prime}(k_3;\tau_3)D^{\a}_{\saa\sb}\left(k_1 ; \tau_1, \tau_2\right)D^{\b}_{\sb\sc}\left(k_3 ; \tau_2, \tau_3\right)+{\rm{2\ perms}} ,
\end{align} where the prime on the left-hand side means that a momentum conservation factor $(2\pi)^3\delta^{(3)}({\bf{k}}_1 +{\bf{k}}_2+{\bf{k}}_3)$ is extracted, and ``$2 \text { perms}$'' represents the permutations $(k_2 \leftrightarrow k_1)$ and $(k_2 \leftrightarrow k_3)$. The propagators $K_{\saa}$ and $D_{\saa\sb}^{\alpha}$ are given in \eqref{Bulk-boundary} and \eqref{D_mp}--\eqref{D_ppmm},
respectively. With $K^{\prime}_{\saa}(k,\tau)=H^2\tau/{2 k}\,e^{\i \saa k \tau}$, this can be further simplified
as
\begin{align}
\langle\varphi_{\bf{k}_1} \varphi_{\bf{k}_2}\varphi_{\bf{k}_3}\rangle '
\nonumber =\,& 2\,\i \sum_{\alpha,\beta}\frac{\rho_\a \rho_\b \lambda_{\a\b}}{H^3}\cdot \frac{1}{8 k_1 k_2 k_3} \int^0_{-\infty}  \mathrm{d} \tau_1\mathrm{d} \tau_2\mathrm{d} \tau_3( -\tau_1)^{-2}( -\tau_2)^{-2}( -\tau_3)^{-2}\\
&\times \sum_{\saa, \sb,\sc=\pm} (\saa\sb\sc) e^{\mathrm{i}\saa k_{1} \tau_1+\i\sb k_{2} \tau_2+\i\sc k_{3} \tau_3} D^{\a}_{\saa\sb}\left(k_1 ; \tau_1, \tau_2\right)D^{\b}_{\sb\sc}\left(k_3 ; \tau_2, \tau_3\right)+{\rm{2\ perms}}.    
\end{align}
Now, it is clear that this three-point correlator can be directly related to the seed integral~\eqref{I_abc} with 
\begin{align}
\label{Bispectrum1}
\langle\varphi_{\bf{k}_1} \varphi_{\bf{k}_2}\varphi_{\bf{k}_3}\rangle '=-\sum_{\alpha,\beta}\rho_\a \rho_\b \lambda_{\a\b}\cdot \frac{H}{4 k_1 k_2^4 k_3} \sum_{\saa, \sb,\sc=\pm}\lim _{k_4 \rightarrow 0}\mathcal{I}_{\saa\sb\sc,\a\b}^{-2-2-2}+{\rm{2\ perms}}.
\end{align}
\\
Another example that will be discussed in this work is the trispectrum with four-point interaction like
\begin{align}
&S_{\rm{int},4}=\sum_{\alpha,\beta}\int \mathrm{d}\tau \mathrm{d}^3\mathbf{x} \ \tilde{\lambda}_{\a\b}\,a^{2}\sigma_{\alpha}\sigma_{\beta} (\varphi^{\prime})^2~,\label{int_4}
\end{align}
where $\tilde{\lambda}_{\a\b}$ is a coupling constant with mass dimension $-2$.
Following the same procedure as before, the trispectrum involving two massive propagators can be expressed as
\begin{align}
\langle\varphi_{\bf{k}_1} \varphi_{\bf{k}_2}\varphi_{\bf{k}_3}\varphi_{\bf{k}_4}\rangle '
\nonumber =\,& -4\,\i \sum_{\alpha,\beta}\rho_\a \rho_\b \tilde{\lambda}_{\a\b}\cdot \frac{1}{16 k_1 k_2 k_3 k_4} \int^0_{-\infty}  \mathrm{d} \tau_1\mathrm{d} \tau_2\mathrm{d} \tau_3( -\tau_1)^{-2}(-\tau_2)^0( -\tau_3)^{-2}\\
&\times \sum_{\saa, \sb,\sc=\pm} (\saa\sb\sc) e^{\mathrm{i}\saa k_{1} \tau_1+\i\sb k_{24} \tau_2+\i\sc k_{3} \tau_3} D^{\a}_{\saa\sb}\left(k_1 ; \tau_1, \tau_2\right)D^{\b}_{\sb\sc}\left(k_3 ; \tau_2, \tau_3\right)+{\rm{5\ perms}}  \,,
\end{align}
where the total of 6=4!/4 permutations correspond to the different ways to form unordered pairs out of the four $k_i$.
Then, the trispectrum can be connected to the seed integral (\ref{I_abc}) through a simple relation, such as
\begin{align}
\label{Trispectrum1}
\langle\varphi_{\bf{k}_1} \varphi_{\bf{k}_2}\varphi_{\bf{k}_3}\varphi_{\bf{k}_4}\rangle '=\sum_{\alpha,\beta}\rho_\a \rho_\b \tilde{\lambda}_{\a\b}\cdot \frac{H^4}{4 k_1 k_2 k_3 k_4 k_{24}^5} \sum_{\saa, \sb,\sc=\pm}\mathcal{I}_{\saa\sb\sc,\a\b}^{-2~0~-2}+{\rm{5\ perms}}.
\end{align}

\subsection{Derivation of bootstrap equations with double-exchange}\label{BS_double}
Now, our task is to compute the seed integral~\eqref{I_abc}. However, the nested time integral of the  special function makes it difficult to perform a direct integration. Instead of doing so,  here we derive differential equations that the seed integral satisfies and subsequently solve them with appropriate boundary conditions. To facilitate the later derivation, let us first change the
integration variables from $\tau_i$ to $z_i$ by 
\begin{align}
-k_1\tau_1=z_1, \quad -k_{24}\tau_2=z_2, \quad  -k_3\tau_3=z_3, \label{z_i}  
\end{align}
and hence $z_i \in(0,+\infty)$.
Then, the propagator~$D^{\a}_{\m{ab}}$ inside the seed integral~\eqref{I_abc} can be rewritten as
\begin{align}
D_{\saa\sb}^{\a}\left(k_1; \tau_1, \tau_2\right)&=D_{\saa\sb}^{\a}\left(k_1;-\frac{z_1}{k_1},-\frac{z_2}{k_{24}}\right)\equiv k_1^{-3}\widehat{D}^{\a}_{\saa\sb}\left(z_1, u z_2\right),
\end{align}
where we have defined momentum ratios as
\begin{align}
    u\equiv \frac{k_1}{k_{24}}~,\qquad\qquad v\equiv \frac{k_3}{k_{24}}~,
\end{align}
and ``hat''  propagators $\widehat{D}_{\saa\sb}$ by
\begin{align}
&\widehat{D}^{\a}_{-+}\left(z_1, u z_2\right)= \frac{H^2 \pi e^{-\pi \mu_\a}}{4} z_1^{3 / 2}\left(u z_2\right)^{3 / 2} H_{\i\mu_\a}^{(1)}\left(z_1\right) H_{-\i \mu_\a}^{(2)}\left(u z_2\right), \label{hat_mp}\\
&\widehat{D}^{\a}_{+-}\left(z_1, u z_2\right)=\left(\widehat{D}^{\a}_{-+}\left(z_1, u z_2\right)\right)^*,\\
&\widehat{D}^{\a}_{\pm\pm}\left(z_1, u z_2\right)=\theta\left(u z_2-z_1\right)\widehat{D}^{\a}_{\mp\pm}\left(z_1, u z_2\right)+\theta\left(z_1-u z_2\right)\widehat{D}^{\a}_{\pm\mp}\left(z_1, u z_2\right).\label{hat_pp}
\end{align}
In the same way, for 
$D^{\b}_{\sb\sc}$ \eqref{I_abc}, we have 
\begin{align}
D_{\sb\sc}^{\b}\left(k_3; \tau_2, \tau_3\right)=D_{\sb\sc}^{\b}\left(k_3;-\frac{z_2}{k_{24}},-\frac{z_3}{k_3}\right)= k_3^{-3}\widehat{D}^{\b}_{\sb\sc}\left(vz_2, z_3\right).
\end{align}
With the redefinition above, the seed integral~\eqref{I_abc} is written as  
\begin{align}
\nonumber \mathcal{I}_{\saa\sb\sc,\a\b}^{p_1p_2p_3}&= \frac{H^{-4}(-\i \saa\sb\sc)}{u^{4+p_1}v^{4+p_3}} \int_0^{\infty}  \mathrm{d} z_1\mathrm{d} z_2\mathrm{d} z_3\  z_1^{p_1}z_2^{p_2}z_3^{p_3}e^{-\mathrm{i}\mathrm{a} z_{1} -\i\mathrm{b} z_2-\i\mathrm{c} z_3} \widehat{D}^{\a}_{\saa\sb}\left(z_1, uz_2\right)\widehat{D}^{\b}_{\sb\sc}\left(vz_2, z_3\right),\\
&\equiv \frac{1}{u^{4+p_1}v^{4+p_3}} \widehat{\mathcal{I}}_{\saa\sb\sc,\a\b}^{p_1p_2p_3}(u,v)~,
\end{align}
where we have introduced the hatted seed integral $\widehat{\mathcal{I}}$ in the second line. Note that the seed integral depends on the two combinations of momentum, $u=k_1/k_{24}$ and $v=k_3/k_{24}$. The important observation here is that $\widehat{D}^{\a}_{\saa\sb} \,(\widehat{D}^{\b}_{\sb\sc})$ depends on a specific combination~$uz_2~(vz_2)$, which allows us to derive differential equations for the hat propagators with respect to $u\,(v)$ instead of $\tau_2$ or $z_2$. For example, from Eqs.~\eqref{green1} and~\eqref{green2} with change of variables~\eqref{z_i}, we find that $\widehat{D}^{\a}_{\saa\sb}\left(z_1, u z_2\right)$ satisfy 
\begin{align}
&\left[z_2^2 \partial_{z_2}^2-2 z_2 \partial_{z_2}+(uz_2)^2+\mu_\a^2+\frac{9}{4}\right] \widehat{D}^{\a}_{\pm\mp}\left(z_1, u z_2\right)=0,\label{de_hat_u1}\\
&\left[z_2^2 \partial_{z_2}^2-2 z_2 \partial_{z_2}+(uz_2)^2+\mu_\a^2+\frac{9}{4}\right] \widehat{D}^{\a}_{\pm\pm}\left(z_1, u z_2\right)=\mp \i H^2 z_1^2 (uz_2)^2 \delta\left(z_1-u z_2\right)~.\label{de_hat_u2}
\end{align}
Then, noting $z_2\partial_{z_2}f(uz_2)=u\partial_u f(uz_2)$, one can rewrite the equations above to those with respect to $u$ that is
\begin{align}
&\left[u^2 \partial_u^2-2 u \partial_u+(uz_2)^2+\mu_\a^2+\frac{9}{4}\right] \widehat{D}^{\a}_{\pm\mp}\left(z_1, uz_2\right)=0, \label{de_hat_u3}  \\
&\left[u^2 \partial_u^2-2 u \partial_u+(uz_2)^2+\mu_\a^2+\frac{9}{4}\right] \widehat{D}^{\a}_{\pm\pm}\left(z_1, uz_2\right)=\mp \i H^2z_1^2\left(u z_2\right)^2 \delta\left(z_1-uz_2\right)~.\label{de_hat_u4} 
\end{align}
In the same way, for $\widehat{D}^{\b}_{\sb\sc}\left(vz_2,  z_3\right)$, we obtain
\begin{align}
&\left[v^2 \partial_v^2-2 v \partial_v+(vz_2)^2+\mu_\b^2+\frac{9}{4}\right] \widehat{D}^{\b}_{\pm\mp}\left(vz_2, z_3\right)=0,  \\
&\left[v^2 \partial_v^2-2 v \partial_v+(vz_2)^2+\mu_\b^2+\frac{9}{4}\right]\widehat{D}^{\b}_{\pm\pm}\left(vz_2, z_3\right)=\mp \i H^2\left(vz_2\right)^2z_3^2 \delta\left(vz_2-z_3\right)~.
\end{align}

The seed integrals consist of massive propagators. By utilizing the differential equations mentioned above and employing integration by parts, we can derive the differential equations that the seed integral satisfies.
Let us begin with the simplest case, $\widehat{\mathcal{I}}_{\pm\mp\pm}$ which involves only the non-time-ordered propagators~$\widehat{D}_{\pm\mp}$, where the bulk time integral at each vertex is completely factorised. By applying some derivative operators, it can be easily shown
\begin{align}
\nonumber &\left[u^2 \partial_u^2-2 u \partial_u+\mu_\a^2+\frac{9}{4}\right]  \widehat{\mathcal{I}}_{\pm\mp\pm,\a\b}^{p_1p_2p_3}\\
\nonumber &= \pm\i\, H^{-4}\int_0^{\infty} \mathrm{d} z_1 \mathrm{d} z_2\mathrm{d} z_3 \,z_1^{p_1}z_2^{p_2}z_3^{p_3}e^{\mp\i  z_1\pm\i z_2\mp\i  z_3} \left[u^2 \partial_u^2-2 u \partial_u+\mu_\a^2+\frac{9}{4}\right] \widehat{D}^{\a}_{\pm\mp}\left( z_1, uz_2\right)\widehat{D}^{\b}_{\mp\pm}\left( vz_2, z_3\right)\\
&=\mp\i\,u^2H^{-4}\int_0^{\infty} \mathrm{d} z_1 \mathrm{d} z_2\mathrm{d} z_3\,z_1^{p_1}z_2^{p_2+2}z_3^{p_3}e^{\mp\i  z_1\pm\i z_2\mp\i  z_3} \widehat{D}^{\a}_{\pm\mp}\left( z_1, uz_2\right)\widehat{D}^{\b}_{\mp\pm}\left( vz_2, z_3\right),\label{mid}
\end{align}
where we used homogeneous equation~\eqref{de_hat_u3} from the second to the third line. The equation~\eqref{mid} is proportional to the original $\widehat{\mathcal{I}}^{p_1,p_2+2,p_3}$ where the power index $p_2$  is increased by a factor of two, and one can relate it to $\widehat{\mathcal{I}}^{p_1p_2p_3}$ by the following formula
\begin{align}
\int_0^{\infty} \mathrm{d} z\,z^{p+2} e^{-\i \saa z}  f(u z) g(v z)=-\left(u \partial_u+v \partial_v+p+2\right)\left(u \partial_u+v \partial_v+p+1\right) \int_0^{\infty} \mathrm{d} z\,z^pe^{-\i \saa z}  f(u z) g(v z)~,   \label{int_formula_b}
\end{align}
for well-behaved functions $f(uz)$ and $g(vz)$, and with $\saa=\pm 1$.\footnote{The proof is as follows. First, we note
\begin{align}
\nonumber 0&=\int_0^{\infty} \mathrm{d} z\ \partial_z\left[z^{p+1} e^{-\i \saa z} f(u z) g(v z)\right]\\
&=\int_0^{\infty}  \mathrm{d} z\ z^p e^{-\i \saa z}\left[p+1-\i \saa z+u \partial_u+v \partial_v\right] f(u z) g(v z),
\end{align}
by $z\partial_{z}f(uz)=u\partial_u f(uz)$ and so on. Therefore, we have
\begin{align}
\int_0^{\infty} \mathrm{d} z\ z^{p+1} e^{-\i \saa z}  f(u z) g(v z)=-\i \saa\left(u \partial_u+v\partial_v+p+1\right) \int_0^{\infty}\mathrm{d} z\  z^p e^{-\i \saa z} f(u z) g(v z).
\end{align}
Placing $p\rightarrow p+1$ in the above equation, we arrive at
\begin{align}
\nonumber &\int_0^{\infty} \mathrm{d} z\ z^{p+2} e^{-\i \saa z}  f(u z) g(v z) \\
\nonumber &=-\i \saa\left(u \partial_u+v\partial_v+p+2\right) \int_0^{\infty}\mathrm{d}z\  z^{p+1} e^{-\i \saa z} f(u z) g(v z)\\
&= -\left(u \partial_u+v \partial_v+p+2\right)\left(u \partial_u+v \partial_v+p+1\right) \int_0^{\infty} \mathrm{d} z\ z^pe^{-\i \saa z}  f(u z) g(v z).  
\end{align}}
It turns out that~\eqref{mid} can be written as
\begin{align}
\left[u^2 \partial_u^2-2 u \partial_u+\mu_\a^2+\frac{9}{4}\right]  \widehat{\mathcal{I}}_{\pm\mp\pm,\a\b}^{p_1p_2p_3}=u^2\left(u \partial_u+v \partial_v+p_2+2\right)\left(u \partial_u+v \partial_v+p_2+1\right)\widehat{\mathcal{I}}_{\pm\mp\pm,\a\b}^{p_1p_2p_3}~.    
\end{align}
By moving the right-hand side to the left, we arrive at the homogeneous differential equations satisfied by ${\mathcal{I}}_{\pm\mp\pm}$, which are free from source terms,
\begin{align}
\mathcal{D}_{u}^{\a}\,\widehat{\mathcal{I}}_{\pm \mp \pm,\a\b}^{p_1p_2p_3}(u,v)=0~,  \label{BS_1}
\end{align}
where the differential operator $\mathcal{D}$ is defined as 
\begin{keyeqn}
\begin{align}
\nonumber \mathcal{D}_{u}^{\a}&\equiv   u^2\partial_u^2-2u\partial_u+\mu_\a^2+\frac{9}{4}-u^2\left(u \partial_u+v \partial_v+p_2+2\right)\left(u \partial_u+v \partial_v+p_2+1\right)\\
\nonumber &=u^2\left(1-u^2\right) \partial_u^2-u^2 v^2 \partial_v^2-2 u\Big[1+u^2\left( p_2+2\right)\Big]\partial_u-2u^2 v\left( p_2+2\right) \partial_v\\
&~~~-2 u^3 v \partial_u \partial_v+\left(\mu_\a^2+\frac{9}{4}\right)-u^2(p_2+2)(p_2+1)~,\label{def_Du}
\end{align}
\end{keyeqn}
giving the bootstrap equation for $\widehat{\mathcal{I}}_{\pm \mp \pm}$ with respect to $u$.
In the same way, one can derive the one with respect to $v$, 
\begin{align}
\mathcal{D}_{v}^{\b}\,\widehat{\mathcal{I}}_{\pm \mp \pm,\a\b}^{p_1p_2p_3}(u,v)=0~,  \label{BS_2}
\end{align}
where the operator $\mathcal{D}_{v}^{\b}$ is obtained by replacing $u\leftrightarrow v$ and $\a\leftrightarrow\b$ in \eqref{def_Du}. 
In summary, we find that the seed integral $\widehat{\mathcal{I}}_{\pm \mp \pm}$ simultaneously satisfies the homogeneous partial differential equations~\eqref{BS_1} and~\eqref{BS_2}.
\\
The similar procedure is applicable to the other seed integrals. However, a distinction arises when the derivative operators are applied to the time-ordered propagators $\widehat{D}_{\pm\pm}$, where the $\delta$-function on the right-hand side of Eq.~\eqref{green2} introduces additional source terms. For instance, in the case of $\widehat{\mathcal{I}}_{\pm\pm\mp}$, we obtain:
\begin{align}
\nonumber &\left[u^2 \partial_u^2-2 u \partial_u+\mu_\a^2+\frac{9}{4}\right]  \widehat{\mathcal{I}}_{\pm\pm\mp,\a\b}^{p_1p_2p_3}\\
\nonumber &=u^2\left(u \partial_u+v \partial_v+p_2+2\right)\left(u \partial_u+v \partial_v+p_2+1\right)\widehat{\mathcal{I}}_{\pm\pm\mp,\a\b}^{p_1p_2p_3}\\
&~~~+H^{-2}u^2\int_0^{\infty} \mathrm{d} z_1 \mathrm{d}z_2\mathrm{d} z_3\  z_1^{p_1+2}z_2^{p_2+2}z_3^{p_3}e^{\mp\i  z_1\mp\i z_2\pm\i  z_3} \delta\left(z_1-u z_2\right)\widehat{D}^{\b}_{\pm\mp}\left( vz_2, z_3\right),
\end{align}
where the second line appears due to the  $\delta$-function term in~\eqref{de_hat_u4}. Performing the $z_1$-integral and changing a variable by $\tilde{z}_2=(u+1)z_2$, this source can be simplified as
\begin{align}
\frac{H^{-2} u^{p_1+4}}{(u+1)^{p_{12}+5}} \int_0^{\infty} \mathrm{d}\tilde{z}_2 \mathrm{d} z_3 \ \tilde{z}_2^{p_{12}+4} z_3^{p_3} e^{\mp \i \tilde{z}_2 \pm \i z_3} \widehat{D}_{ \pm \mp}^{\b}\left(\frac{v}{u+1} \tilde{z}_2, z_3\right).\label{mid_2}
\end{align}
Remarkably, this source term now contains only one massive propagator and two time integrals, which is actually associated to the seed integral for \textit{single} massive exchange with one linear mixing vertex $\mathcal{I}_{\saa\sb,\a}^{p_1p_2}$ and its exact closed analytical expression has already been investigated in details in the previous work \cite{Qin:2023ejc} (also see Appendix~\ref{diagram_a} for a summary of the results). To be more explicit, the seed integral with single-exchange that we will frequently utilize in this work is
\begin{keyeqn}
\begin{align}
\mathcal{I}_{\saa\sb,\a}^{p_1p_2}&= H^{-2}k_3^{5+p_{12}} (-\saa\sb)\int^0_{-\infty}  \mathrm{d} \tau_1\mathrm{d} \tau_2 (-\tau_1)^{p_1}(-\tau_2)^{p_2}e^{\mathrm{i}\saa k_{124} \tau_1+\i\sb k_{3} \tau_2} D^\a_{\saa\sb}\left(k_3 ; \tau_1, \tau_2\right), \label{I_dS_a}
\end{align}
\end{keyeqn}
where the notation used here is the same as that we defined in the double-exchange seed integral.\footnote{Due to a non-linearly realized symmetry, whenever there exists a quadratic interaction of the form~\eqref{int_2}, it must come also with the following cubic interaction, denoted as the ``Lorentz-covariant'' combination
\begin{equation}
    S_{\text{int}}=-\sum_{\alpha}\int \mathrm{d}\tau \mathrm{d}^3\mathbf{x}\, \sqrt{-g} \frac{\rho_\a}{2f_\pi^2}g^{\mu\nu}\partial_\mu\varphi\partial_\nu\varphi\sigma_\a~,\label{int_LI}
\end{equation}
where $f_\pi$ is an energy scale denoting the normalisation of the primordial curvature fluctuation, see Sec.~\ref{Sec_pheno} for more details.
With the insertion of one massive propagator and a quadratic vertex, this cubic interaction gives rise to a single-exchange bispectrum diagram related to the seed integral~\eqref{I_dS_a} via
\begin{align}
\nonumber \langle\varphi_{\bf{k}_1} \varphi_{\bf{k}_2}\varphi_{\bf{k}_3}\rangle ' 
= &-\sum_\a\frac{\rho_\a^2H^3}{f_\pi^2}  \cdot \frac{1}{16\left(k_1 k_2 k_3\right)^2} \cdot\sum_{\mathsf{a}, \mathsf{b}=\pm}\lim _{k_4 \rightarrow 0}\left[\frac{k_{12}^2-k_3^2}{k_3^2}\mathcal{I}_{\m{a b}}^{0-2}+\i \m{a} \frac{k_{12}\left(k_3^2-k_1^2-k_2^2\right)  }{k_1 k_2k_3}\mathcal{I}_{\m{a b}}^{-1-2}+\frac{k_3^2-k_1^2-k_2^2}{k_1 k_2} \mathcal{I}_{\m{a b}}^{-2-2}\right]  \nonumber \\
    &+{\rm{2\,perms.}}
\end{align}
}

It turns out that the seed integral $\mathcal{I}_{\saa\sb,\a}^{p_1p_2}(R)$ depends on a specific momentum ratio $R={2k_3}/{k_{1234}}$.
By using the single-exchange seed integral $\mathcal{I}_{\m{ab},\a}^{p_1p_2}(R)$, Eq.~\eqref{mid_2} can be expressed as (see Eq.~\eqref{I_dS_a_2} for detailed derivation)
\begin{align}
\frac{u^{p_1+4}}{v^{p_{12}+5}}\,\mathcal{I}_{ \pm \mp,\b}^{p_{12}+4, p_3}\left(R=\frac{2 v}{1+u+v}\right).
\end{align}
Therefore, we obtain the bootstrap equation for $\widehat{\mathcal{I}}_{\pm\pm\mp}$ with respect to $u$, 
\begin{align}
\mathcal{D}_{u}^{\a}\,\widehat{\mathcal{I}}_{\pm\pm\mp,\a\b}^{p_1p_2p_3}(u,v)= \frac{u^{p_1+4}}{v^{p_{12}+5}}\,\mathcal{I}_{ \pm \mp,\b}^{p_{12}+4, p_3}\left(R=\frac{2 v}{1+u+v}\right),  
\end{align}
which is an inhomogeneous differential equation.  In addition, $\widehat{\mathcal{I}}_{\pm\pm\mp}$ should also satisfy another differential equation when the operator $\mathcal{D}^{\b}_{v}$ is applied. One difference is that $\mathcal{D}^{\b}_{v}$ will act on the non-time ordering propagator $D^\beta_{\pm\mp}$ which has no source term contribution. Consequently, the differential equation related to~$\mathcal{D}^{\b}_{v}$ should be homogeneous. Following the same procedure, we can derive all bootstrap equations, which are summarized below:
\begin{tcolorbox}[toggle enlargement=none, colback=white]
The summary of bootstrap equations of different seed integrals.
\begin{itemize}
\item The fully factorised seed integral $\widehat{\mathcal{I}}_{\pm \mp \pm}$:
\begin{align}
&\mathcal{D}_{u}^{\a}\,\widehat{\mathcal{I}}_{\pm \mp \pm,\a\b}^{p_1p_2p_3}(u,v)=0,\label{BS_b_dS_1}\\ &\mathcal{D}_{v}^{\b}\,\widehat{\mathcal{I}}_{\pm \mp \pm,\a\b}^{p_1p_2p_3}(u,v)=0.\label{BS_b_dS_2}    
\end{align}

\item The partially factorised partially nested seed integral $\widehat{\mathcal{I}}_{\pm\pm\mp}$:
\begin{align}
&\mathcal{D}_{u}^{\a}\,\widehat{\mathcal{I}}_{\pm\pm\mp,\a\b}^{p_1p_2p_3}(u,v)= \frac{u^{p_1+4}}{v^{p_{12}+5}}\,\mathcal{I}_{ \pm \mp,\b}^{p_{12}+4, p_3}\left(R=\frac{2 v}{1+u+v}\right), \label{BS_b_dS_ppm1}\\
&\mathcal{D}_{v}^{\b}\,\widehat{\mathcal{I}}_{\pm \pm \mp,\a\b}^{p_1p_2p_3}(u,v)=0.\label{BS_b_dS_ppm2}    
\end{align}
\item The fully nested seed integral $\widehat{\mathcal{I}}_{\pm\pm\pm}$:
\begin{align}
&\mathcal{D}_{u}^{\a}\,\widehat{\mathcal{I}}_{\pm\pm\pm,\a\b}^{p_1p_2p_3}(u,v)= \frac{u^{p_1+4}}{v^{p_{12}+5}}\,\mathcal{I}_{ \pm \pm,\b}^{p_{12}+4, p_3}\left(R=\frac{2 v}{1+u+v}\right), \label{BS_b_dS_ppp1}\\
&\mathcal{D}_{v}^{\b}\,\widehat{\mathcal{I}}_{\pm \pm \pm,\a\b}^{p_1p_2p_3}(u,v)= \frac{v^{p_3+4}}{u^{p_{23}+5}}\,\mathcal{I}_{ \pm \pm,\a}^{p_{23}+4, p_1}\left(R=\frac{2 u}{1+u+v}\right).\label{BS_b_dS_ppp2}    
\end{align}
\end{itemize}
The differential operator $\mathcal{D}_u^\a$ is defined in \eqref{def_Du}. The explicit expressions of $\mathcal{I}_{\m{ab},\a}^{p_1p_2}(R)$ can be found in ~\eqref{result_Ipm} and~\eqref{result_Ipp}.
\end{tcolorbox}
Here ``factorised" or ``nested" is associated with the time integral. For the other partially factorised partially nested 
seed integral $\widehat{\mathcal{I}}_{\mp\pm\pm,\a\b}(u,v)$, the expression can be easily obtained through replacing $u\leftrightarrow v$, $\alpha\leftrightarrow \beta$ and $p_1\leftrightarrow p_3$ in the result of $\widehat{\mathcal{I}}_{\pm\pm\mp,\a\b}$, ~\eqref{BS_b_dS_ppm1} and \eqref{BS_b_dS_ppm2}.
\begin{figure}[h]
\centering
\includegraphics[width=0.7\textwidth]{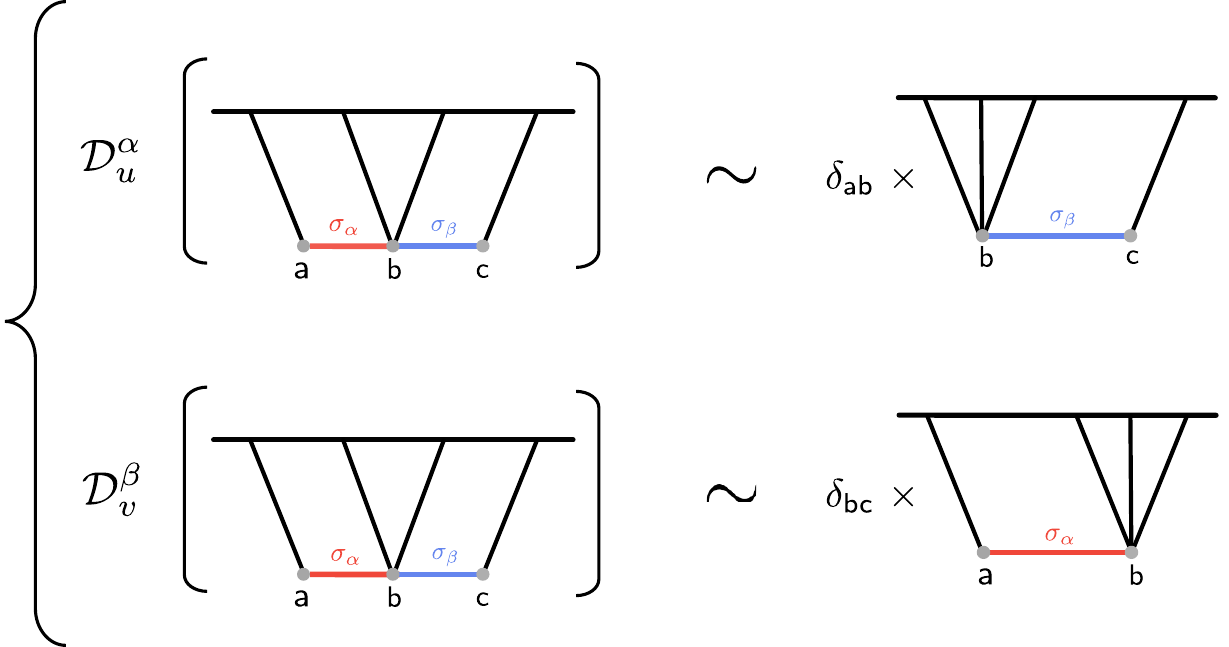}\\
	\caption{The schematic diagram illustrates the double-exchange bootstrap equations. When acting on it with the differential operator $\mathcal{D}^\alpha_u$ ($\mathcal{D}^\beta_v$) associated with the massive fields $\sigma_\alpha$ ($\sigma_\beta$), the double-exchange diagram reduces to a single-exchange diagram without the corresponding massive propagator.}
    \label{eq_diagram}
\end{figure}

In Figure \ref{eq_diagram}, we show the schematic diagram of these differential equations. Once applying the differential operators $\mathcal{D}_u$ ($\mathcal{D}_v$), the double-exchange diagram is reduced to the single-exchange one for which exact \textit{closed} results are already known from the previous work \cite{Qin:2023ejc}. To be more specific, when the two SK indices of one massive propagator are opposite, applying the corresponding differential operator causes the right-hand side of Figure \ref{eq_diagram} to vanish, resulting in homogeneous equations. Conversely, if the two indices are the same, an additional $\delta$-term introduces a source term, which can be expressed using the seed integral of the single-exchange diagram.
\\
Alternatively, as mentioned in the Appendix of \cite{Pimentel:2022fsc}, one can apply another differential operator to both sides of Figure \ref{eq_diagram} to further reduce another massive propagator. In doing so, the source becomes the simple contact one, but the trade-off is that the differential equations become of order four, and we noticed that it is then difficult to solve them. Instead, in this work, we stand on the shoulders of giants, leveraging the known results of the single-exchange diagram to derive those of the double-exchange one.
By following a similar procedure, we can apply analogous differential operators for more complicated diagrams. In the double-exchange trispectrum depicted in Figure \ref{double_diagram} which we calculate in this paper,  the interactions have two quadratic mixings along with one quartic coupling of the schematic form $\varphi^2\sigma^2$. Additionally, our method could be used to calculate another type of double-exchange four-point correlator, that generated by two cubic interactions, one $\varphi^2\sigma$ and one $\varphi\sigma^2$, and one quadratic mixing. Another important example is the seed integral for triple-exchange diagrams, with a four-point correlator generated by a quartic interaction  $\varphi\sigma^3$ and three quadratic mixings.
In Appendix~\ref{triple}, we present the first steps of this calculation, leaving the detailed work for a future publication.

Before moving to the solutions of those bootstrap equations, let us first address the limitations related to the procedure employed here.
Indeed, although our analytical results will represent the exact solutions to the bootstrap equations we just derived, those equations were found using working hypotheses.
The most crucial assumption is that the free Hamiltonian is diagonal, i.e. there is no linear mode mixing and, rather, quadratic mixings are treated as vertices of the perturbation theory.
If we relax this assumption, we would need to consider a coupled system of quantum operators $\hat{\Phi}_{\bf k}^A(\tau)=\sum_a \Phi^A_{a,\bf k}(\tau) \hat{a}^a_{\bf k} + \text{h.c.} $ with independent (creation and) annihilation ($ \hat{a}^{\dagger a}_{\bf k}$ and) $ \hat{a}^a_{\bf k}$ operators and defining a matrix of mode functions $\Phi^A_{a,\bf k}(\tau)$ and related propagators.
Our assumption forces us to consider the quadratic mixings of the form $\Omega_{AB} \Phi^{A\prime} \Phi^B$, with $\Omega_{AB}$ an anti-symmetrix matrix, as quadratic vertices, therefore assuming their effect is negligible compared to the kinetic and mass terms (we can without loss of generality assume that the mass matrix is diagonal in the basis of these $\Phi^A$~\cite{Pinol:2021aun}).
This assumption imposes upper bounds on the quadratic mixings that we consider explicitly in this work, $\rho_\alpha \lesssim H$.
By consistency, it also justifies our neglect of the other quadratic mixings between the massive fields $\sigma_\alpha$ identified in~\cite{Pinol:2020kvw}, as those would result in higher-order corrections suppressed by additional powers of small coupling constants.
Having delineated the regime of validity of the analytical results will be important when we will dig into the phenomenology of double-exchange diagrams and compare predictions of the theory to observational constrains.
It will also be important to compare our analytical results to an independent method not relying on this perturbative scheme for the quadratic mixings, as provided by the Cosmological Flow framework~\cite{Werth:2023pfl,Pinol:2023oux} and its numerical implementation \textsf{CosmoFlow}~\cite{Werth:2024aui}.
These endeavours are undertaken in Section~\ref{Sec_pheno} in great detail.
We finish this section by mentioning another hypothesis leaving room for future works: we assumed that the inflaton fluctuations $\varphi$ have unit speed of sound, while be it in general single-field inflation or in the framework of the effective field theory of inflationary fluctuations, $c_s^2 \neq 1$ is ubiquitous.
In particular, it was recently understood that a small speed of sound could result in interesting phenomenology in the soft limits of single-exchange diagrams, denoted as ``low-speed collider''~\cite{Jazayeri:2022kjy,Pimentel:2022fsc, Jazayeri:2023xcj}, and it would be interesting to extend our results to this intriguing scenario.

\section{Analytical Computations for Bootstrap Equations}\label{sec:BS_solution}
In this section, we will solve all the bootstrap equations derived in Section~\ref{sec: BS_equation}. Readers who are not interested in technical details can directly skip to Section~\ref{Summary_sol} for the summary of final results.

Let us start from $\widehat{\mathcal{I}}_{\pm\mp\pm}$, which satisfy the homogeneous partial differential equations~\eqref{BS_b_dS_1} and~\eqref{BS_b_dS_2}, or explicitly, 
\begin{align}
\nonumber &\bigg\{u^2\left(1-u^2\right) \partial_u^2-u^2 v^2 \partial_v^2-2 u\Big[1+u^2\left( p_2+2\right)\Big]\partial_u-2u^2 v\left( p_2+2\right) \partial_v\\
&~~~-2 u^3 v \partial_u \partial_v+\left(\mu_\a^2+\frac{9}{4}\right)-u^2(p_2+2)(p_2+1)\bigg\}\,\widehat{\mathcal{I}}_{\pm\mp\pm,\a\b}^{p_1p_2p_3}=0~,\\
\nonumber &\bigg\{v^2\left(1-v^2\right) \partial_v^2-v^2 u^2 \partial_u^2-2 v\Big[1+v^2\left( p_2+2\right)\Big] \partial_v-2v^2 u\left(p_2+2\right) \partial_u\\
&~~~-2 v^3 u \partial_v \partial_u+\left(\mu_\b^2+\frac{9}{4}\right)-v^2(p_2+2)(p_2+1)\bigg\}\,\widehat{\mathcal{I}}_{\pm\mp\pm,\a\b}^{p_1p_2p_3}=0~.
\end{align}
We find that this system of partial differential equations can be solved analytically by
\begin{align}
\widehat{\mathcal{I}}_{\pm\mp\pm,\a\b}^{p_1p_2p_3}=\sum_{\m{a,b}=\pm}c_{\pm\mp\pm,\m{ab}}\,u^{\frac{3}{2}-\i\saa \mu_\a} v^{\frac{3}{2}-\i\m{b} \mu_\b} \ \mathcal{F}_4\left[\begin{array}{c|c}
\frac{4+p_2-\i\left(\saa\mu_\a+\m{b}\mu_\b\right)}{2}, \frac{5+p_2-\i\left(\saa\mu_\a+\m{b}\mu_\b\right)}{2} \\
1-\i\saa \mu_\a, 1-\i \m{b}\mu_\b
\end{array}\  u^2, v^2\right],\label{Homo_sol}
\end{align}
where $c_{\pm\mp\pm,\m{ab}}$ with $\m{a,b}=\pm$ are four undetermined coefficients, which will be fixed later by imposing proper boundary conditions. $\mathcal{F}_4[\cdots]$ is the dressed Appell series defined in~\eqref{def_f4}. 
Note that $\mathcal{F}_4[\cdots]$ with this definition is only convergent for $u+v<1$, and we will discuss in detail the continuation to all physical kinematic regions in the next section.

The other seed integrals satisfy inhomogeneous equations, and therefore, the general solutions can be expressed as the sum of homogeneous solutions, as shown in~\eqref{Homo_sol}, and particular solutions, which will be discussed in the following subsections.

\subsection{Particular solutions} 
Here we aim to find particular solutions for the inhomogeneous bootstrap equations represented by~\eqref{BS_b_dS_ppm1}--\eqref{BS_b_dS_ppp2}. These particular solutions are denoted as $\widehat{\mathcal{P}}_{\pm\pm\mp,\a\b}^{p_1p_2p_3}$ and $\widehat{\mathcal{P}}_{\pm\pm\pm,\a\b}^{p_1p_2p_3}$, respectively.
\paragraph{The particular solution $\widehat{\mathcal{P}}_{\pm \pm \mp}~.$}\,
\\
This solution contributes to the partially factorised  partially nested seed integral and must satisfy a system of specific equations
\begin{align}
&\mathcal{D}_{u}^{\a}\,\widehat{\mathcal{P}}_{\pm\pm\mp,\a\b}^{p_1p_2p_3}(u,v)=\frac{u^{p_1+4}}{v^{p_{12}+5}}\,\mathcal{I}_{ \pm \mp,\b}^{p_{12}+4, p_3}\left(R=\frac{2 v}{1+u+v}\right)~, \label{ppm1}\\
&\mathcal{D}_{v}^{\b}\,\widehat{\mathcal{P}}_{\pm \pm \mp,\a\b}^{p_1p_2p_3}(u,v)=0~.\label{ppm2}    
\end{align}
To find solutions, we begin by rescaling
\begin{align}
\widehat{\mathcal{P}}_{\pm\pm\mp} =\frac{u^{p_1+4}}{v^{p_{12}+5}}\widetilde{\mathcal{P}}_{\pm\pm\mp},\label{rescale_P}
\end{align}
and change variables by $r_i=2k_i/k_{1234}$, then
\begin{align}
\frac{2 u}{1+u+v}=\frac{2k_1}{k_{1234}}\equiv r_1,\quad \frac{2 v}{1+u+v}=\frac{2k_3}{k_{1234}}\equiv r_3,\label{defUV}
\end{align}
such that equations~\eqref{ppm1} and~\eqref{ppm2} become
\begin{align}
\nonumber &\Bigg[r_1^2(1-r_1) \partial_{r_1}^2 -r_1^2 r_3 \partial_{r_1}\partial_{r_3} -\big(\left(4+p_1\right)r_1-2\left(3+p_1\right) \big) r_1\partial_{r_1} \\
&-\left(3+p_1\right)r_1 r_3 \partial_{r_3} +\mu_\a^2+\left( p_1+\frac{5}{2}\right)^2\Bigg]\widetilde{\mathcal{P}}_{\pm\pm\mp,\a\b}^{p_1p_2p_3} (r_1,r_3)=\mathcal{I}_{\pm\mp,\beta}^{p_{12}+4, p_3}\left(r_3\right),\label{ppm_UV}\\
\nonumber &\Bigg[r^2_3(1-r_3) \partial_{r_3}^2 -r_1{r^2_3} \partial_{r_1} \partial_{r_3} +\big(\left(5+p_{12}\right)r_3-2\left(6+p_{12}\right) \big) r_3\partial_{r_3} \\
&+\left(6+p_{12}\right)r_1 r_3 \partial_{r_1} +\mu_\b^2+\left( p_{12}+\frac{13}{2}\right)^2\Bigg]\widetilde{\mathcal{P}}_{\pm\pm\mp,\a\b}^{p_1p_2p_3}(r_1,r_3)=0~.\label{ppm_UV2}
\end{align}
The differential operator appears asymmetric at the moment due to the unbalanced rescaling of $\widetilde{\mathcal{P}}$ as defined in \eqref{rescale_P}. Referring to \eqref{result_Ipm}, the right-hand side of \eqref{ppm_UV} originates from the factorised component of the single-exchange seed integral and is explicitly provided by
\begin{align}
\nonumber \mathcal{I}_{ \pm \mp,\b}^{p_{12}+4, p_3}\left(r_3\right)=&\ \sum_{\saa=\pm}\frac{e^{\mp\i\frac{\pi}{2} (p_{12}-p_3)}}{2^{10+p_{123}-2\i  \saa\mu_\b} \pi^{\frac{1}{2}}} \Gamma\left[\begin{array}{c}
\frac{5}{2}+p_3+\i \mu_\b, \frac{5}{2}+p_3-\i \mu_\b, \frac{13}{2}+p_{12}-\i \saa\mu_\b, \i \saa \mu_\b \\
3+p_3
\end{array}\right] \\
&~~~~~~\times{r_3}^{\frac{13}{2}+p_{12}-\i   \saa\mu_\b} { }_2 \mathrm{F}_1\left[\begin{array}{c|c}
\frac{1}{2}-\i  \saa\mu_\b, \frac{13}{2}+p_{12}-\i \saa \mu_\b \\
1-2\i \saa\mu_\b
\end{array}\ r_3\right].\label{result_Ipm2}
\end{align}
The source term exhibits non-analytic behavior in $r_3$ (like a $r_3^{\i\saa\mu_\b}$ term). Motivated by this observation and the series expansion of the hypergeometric
function~\eqref{def_hyper}, we propose the following ansatz for $\widetilde{\mathcal{P}}$,
\begin{align}
\widetilde{\mathcal{P}}_{\pm\pm\mp,\a\b}^{p_1p_2p_3}(r_1,r_3)=\frac{ e^{\mp\i\frac{\pi}{2} (p_{12}-p_3)}}{2^{10+p_{123}} \pi^{\frac{1}{2}}} \Gamma\left[\begin{array}{c}
\frac{5}{2}+p_3+\i \mu_\b, \frac{5}{2}+p_3-\i \mu_\b\\
3+p_3
\end{array}\right]\sum_{\saa=\pm}\sum_{m, n=0}^{\infty} \mathcal{A}_{m, n}^{\pm\pm\mp(\saa)} r_1^m r_3^{n+p_{12}+\frac{13}{2}+\i \saa\mu_\b}~. \label{ansatz_ppm}
\end{align}
We usually denote the oscillatory terms as CC signals, which can help identify the mass of exchanging particles based on the frequency of oscillation patterns. 
On the other hand, the featureless analytic terms are commonly termed  as the background, upon which the signals are situated, and its effect can often be mimicked by inflaton self-interactions. Here we note that this ansatz is designed to be analytical along the $r_1$ direction but exhibit non-analytical oscillatory behavior along the $r_3$ direction as stated above.  This characteristic suggests that it represents the contribution that arises from the mixing between the background and the signal.
The coefficients $\mathcal{A}_{m, n}^{\pm\pm\mp(\saa)}$ are determined by the differential equations~\eqref{ppm_UV} and~\eqref{ppm_UV2}. Upon substituting the ansatz, we obtain the following recurrence relations,
\begin{align}
&\mathcal{A}_{m+1, n}^{\pm\pm\mp(\saa)}=\frac{\left(3+p_1+m\right)\left(m+n+p_{12}+13/2+ \i\saa \mu_\b\right)}{(m+p_1+7/2)^2+\mu_\a^2}\mathcal{A}_{m, n}^{\pm\pm\mp(\saa)},\label{A_m+1}\\
&\mathcal{A}_{0, n}^{\pm\pm\mp(\saa)}= \frac{\i \pi^{1 / 2} \operatorname{csch}(\saa\pi \mu_\b)}{\mu_\a^2+(p_1+5/2)^2} \Gamma\left[\begin{array}{cc}
\frac{1}{2}+n + \i\saa \mu_\b, & \frac{13}{2}+p_{12}+n+ \i\saa \mu_\b\\
1+n, & 1+n +2 \i\saa \mu_\b
\end{array}\right],\label{A_0n}\\
&\mathcal{A}_{m, n+1}^{\pm\pm\mp(\saa)}=\frac{(1/2+ n+\i\saa \mu_\b)\left( m+ n+ p_{12}+13/2+ \i\saa \mu_\b\right)}{(n+1)(n+1+2 \i\saa \mu_\b)}\mathcal{A}_{m, n}^{\pm\pm\mp(\saa)},\label{A_n+1}
\end{align}
where the first two are obtained from~\eqref{ppm_UV} while the last one is from~\eqref{ppm_UV2}.
These recurrence relations can be solved\footnote{The solution is completely fixed by \eqref{A_m+1} and~\eqref{A_0n}. One can explicitly check that the remaining relation~\eqref{A_n+1} is automatically satisfied by the solution~\eqref{sol_A_ppm}.} 
\begin{align}
\nonumber \mathcal{A}_{m, n}^{\pm\pm\mp(\saa)}=&\ \frac{\i \pi^{\frac{1}{2}} \operatorname{csch}(\saa\pi \mu_\b)}{\mu_\a^2+(m+p_1+5/2)^2}\frac{(3+p_1)_m}{\left(\frac{5}{2}+p_1-\i \mu_\a\right)_m\left(\frac{5}{2}+p_1+\i \mu_\a\right)_m}\\
&\times\Gamma\left[\begin{array}{cc}
\frac{1}{2}+n+\i\saa \mu_\b, & \frac{13}{2}+m+n+p_{12}+\i\saa \mu_\b \\
1+n, & 1+n+2 \i\saa \mu_\b
\end{array}\right], \label{sol_A_ppm}
\end{align}
where $(a)_m\equiv\Gamma(a+m)/\Gamma(a)$ is the Pochhammer symbol. Upon inserting it back into~\eqref{ansatz_ppm}, we derive the particular solution. It is noteworthy that one of the infinite summations, for instance, the $n$-summation, can be explicitly performed. After performing the summation over $n$, we obtain 

\begin{keyeqn}
\begin{align}
\nonumber &\widehat{\mathcal{P}}_{\pm\pm\mp,\a\b}^{p_1p_2p_3}=\sum_{\saa=\pm} \frac{u^{p_1+4}}{v^{p_{12}+5}}\cdot\frac{\i\ e^{\mp\i\frac{\pi}{2} (p_{12}-p_3)}}{2^{10+p_{123}} } r_3^{p_{12}+\frac{13}{2}} \times\Gamma\left[\begin{array}{c}
\frac{5}{2}+p_1+\i \mu_\a, \frac{5}{2}+p_1-\i \mu_\a,\frac{5}{2}+p_3+\i \mu_\b, \frac{5}{2}+p_3-\i \mu_\b\\
3+p_1, 3+p_3
\end{array}\right]\\
\times&\sum_{m=0}^{\infty} \operatorname{csch}(\pi \saa\mu_\b)r_1^m r_3^{\i\saa\mu_\b}\Gamma\left[\begin{array}{cc}
3+m+p_1\\
\frac{7}{2}+m+p_1-\i\mu_\a, \frac{7}{2}+m+p_1+\i\mu_\a
\end{array}\right]  { }_2 \mathcal{F}_1\left[\begin{array}{c|c}
\frac{1}{2}+\i\saa\mu_\b,\frac{13}{2}+m+p_{12}+\i \saa \mu_\b \\
1+2\i\saa\mu_\b
\end{array}\ r_3 \right]. \label{sol_P_ppm}   
\end{align}
\end{keyeqn}

\paragraph{The particular solution $\widehat{\mathcal{P}}_{\pm \pm \pm}.$}\,
\\
Lastly, we examine the particular solution for the fully nested seed integral $\widehat{\mathcal{P}}_{\pm\pm\pm,\a\b}^{p_1p_2p_3}$, which satisfies the system of differential equations,
\begin{align}
&\mathcal{D}_{u}^{\a}\widehat{\mathcal{P}}_{\pm\pm\pm,\a\b}^{p_1p_2p_3}(u,v)=\frac{u^{p_1+4}}{v^{p_{12}+5}} \mathcal{I}_{ \pm \pm,\b}^{p_{12}+4, p_3}\left(R=\frac{2 v}{1+u+v}\right)~, \\
&\mathcal{D}_{v}^{\b} \widehat{\mathcal{P}}_{\pm \pm \pm,\a\b}^{p_1p_2p_3}(u,v)= \frac{v^{p_3+4}}{u^{p_{23}+5}} \mathcal{I}_{ \pm \pm,\a}^{p_{23}+4, p_1}\left(R=\frac{2 u}{1+u+v}\right)~.  
\end{align}
Similarly to the previous discussion, it is advantageous to perform rescaling in the following way
\begin{align}
\widehat{\mathcal{P}}_{\pm\pm\pm} =\frac{u^{p_1+4}}{v^{p_{12}+5}}\widetilde{\mathcal{P}}_{\pm\pm\pm}~,
\end{align}
and changing variables by \eqref{defUV}. Then, 
the bootstrap equations are transformed into
\begin{align}
\nonumber &\Bigg[r_1^2(1-r_1) \partial_{r_1}^2 -r_1^2 {r_3} \partial_{r_1} \partial_{r_3} -\big(\left(4+p_1\right) r_1-2\left(3+p_1\right)\big) r_1\partial_{r_1} \\
&-\left(3+p_1\right)r_1 {r_3} \partial_{r_3} +\mu_\a^2+\left( p_1+\frac{5}{2}\right)^2 \Bigg]\widetilde{\mathcal{P}}_{\pm\pm\pm,\a\b}^{p_1p_2p_3}(r_1,{r_3})= \mathcal{I}_{\pm\pm,\b}^{p_{12}+4, p_3}\left({r_3}\right),\label{ppp_1}\\
\nonumber &\Bigg[{r_3}^2(1-{r_3}) \partial_{r_3}^2 -r_1 {r_3}^2 \partial_{r_1} \partial_{r_3} +\Big(\left(5+p_{12}\right) {r_3}-2\left(6+p_{12}\right)\Big) {r_3}\partial_{r_3} \\
&+\left(6+p_{12}\right)r_1 {r_3} \partial_{r_1} +\mu_\b^2+\left( p_{12}+\frac{13}{2}\right)^2\Bigg]\widetilde{\mathcal{P}}_{\pm\pm\pm,\a\b}^{p_1p_2p_3}(r_1,{r_3})= \left(\frac{{r_3}}{r_1}\right)^{p_{123}+9}\mathcal{I}_{\pm\pm,\a}^{p_{23}+4, p_1}\left(r_1\right),\label{ppp_2}
\end{align}
where the source term originates from the nested part of the single-exchange diagram, and its exact expression can be found in \eqref{result_Ipp}, 
\begin{align}
\nonumber &\mathcal{I}_{ \pm\pm,\b}^{p_{12}+4, p_3}({r_3})\\
\nonumber&=  \frac{ \pm \i\ e^{\mp \i\frac{\pi}{2} p_{123}}}{2^{10+p_{123}} \pi^{\frac{1}{2}}} \Gamma\left[\begin{array}{c}
\frac{5}{2}+p_3+\i \mu_\b, \frac{5}{2}+p_3-\i \mu_\b \\
3+p_3
\end{array}\right]\times \Bigg\{\mp 2\,\i\,  \pi^{\frac{1}{2}} {r_3}^{p_{123}+9}{ }_3 \mathcal{F}_2\left[\left.\begin{array}{c}
1, p_{123}+9,3+p_3 \\
\frac{7}{2}+p_3-\i \mu_\b,  \frac{7}{2}+p_3+\i \mu_\b
\end{array} \right\rvert\, {r_3}\right]\\
&+\sum_{\saa=\pm}e^{\mp \pi \saa\mu_\b} 2^{2 \i \saa\mu_\b} \Gamma\left[\frac{13}{2}+p_{12}-\i \saa\mu_\b, \i \saa\mu_\b\right]{r_3}^{\frac{13}{2}+p_{12}-\i \saa\mu_\b} {}_2{\mathrm{F}}_1\left[\left.\begin{array}{c}
\frac{1}{2}-\i \saa\mu_\b, \frac{13}{2}+p_{12}-\i \saa\mu_\b \\
1-2 \i \saa\mu_\b
\end{array} \right\rvert\, {r_3}\right]\Bigg\}~,\label{RHSppp}
\end{align}
where ${_p\mathcal{F}_q}$ is the dressed hypergeometric function defined in \eqref{dhyper}, and $\mathcal{I}_{ \pm\pm,\a}^{p_{23}+4, p_1}(r_1)$ is given by the same form as \eqref{RHSppp} with replacement $r_1\leftrightarrow {r_3} $, $p_1\leftrightarrow p_3 $, and $\alpha\leftrightarrow \beta$. 
Inspired by the structure of the source term and the series expansions~\eqref{def_hyper}--\eqref{dhyper}, we choose the following ansatz for the particular solutions:
\begin{align}
\nonumber &\widetilde{\mathcal{P}}_{\pm\pm\pm,\a\b}^{p_1p_2p_3}(r_1,{r_3})\\&=\frac{ \pm\i\,e^{\mp\i\frac{\pi}{2} p_{123}}\,r_3^{p_{123}}}{2^{10+p_{123}} \pi^{\frac{1}{2}}}\sum_{m, n=0}^{\infty}\Biggl[\mathcal{C}_{m,n}^{\pm\pm\pm}r_1^m{r_3}^{n+9}+\sum_{\saa=\pm}\left\{ \mathcal{A}_{m, n}^{\pm\pm\pm(\saa)} r_1^m {r_3}^{n-p_3+\frac{13}{2}+\i \saa\mu_\b}+\mathcal{B}_{m, n}^{\pm\pm\pm(\saa)} r_1^{n-p_1-\frac{5}{2}+\i\saa\mu_\a} {r_3}^{m+9}\right\}\Biggr], \label{ansatz_ppp}
\end{align}
which comprises three terms, each exhibiting distinct analytic structures about $r_1$ and $r_3$ due to their different origins. The $\mathcal{C}$-term is analytic in both the $r_1$ and $r_3$ directions, indicating that neither of the massive propagators excites CC signals, and thus, it serves purely as the background. On the other hand, the $\mathcal{A}$ and $\mathcal{B}$-terms feature oscillatory behavior in one direction, suggesting that one of the massive propagators excites CC signals, making it the signal-background mixing term. By substituting the ansatz into \eqref{ppp_1} and~\eqref{ppp_2}, we obtain the recurrence relations for the different terms which are summarized at the end of this subsection. 

With the explicit forms of the coefficients  and after performing one of the infinite summations in \eqref{ansatz_ppp}, we finally obtain the particular solution as
\begin{keyeqn}
\begin{align}
\nonumber \widehat{\mathcal{P}}_{\pm\pm\pm,\a\b}^{p_1p_2p_3} &=\frac{u^{p_1+4}}{v^{p_{12}+5}}\cdot\frac{ e^{\mp\i\frac{\pi}{2} p_{123}}}{2^{10+p_{123}}\left[\mu_\a^2+\left(p_1+5 / 2\right)^2\right] }\,\Gamma\left[\begin{array}{c}
\frac{5}{2}+p_3-\i \mu_\b, \frac{5}{2}+p_3+\i \mu_\b \\
3+p_3
\end{array}\right]\nonumber\\
&~~\times\sum_{m=0}^{\infty}\,\frac{\left(3+p_1\right)_m\,r_1^m} {\left(\frac{7}{2}+p_1-\i \mu_\a\right)_m\left(\frac{7}{2}+p_1+\i \mu_\a\right)_m}\Bigg\{ r_3^{9+p_{123}}\times{ }_3 \mathcal{F}_2\left[\begin{array}{c|c}
1,3+p_3, 9+m+p_{123}   \\
\frac{7}{2}+p_3-\i \mu_\b, \frac{7}{2}+p_3+\i \mu_\b 
\end{array}\ r_3\right]\nonumber\\
\nonumber&~~\pm\sum_{\saa=\pm}\big(\coth(\pi\saa\mu_\b){{\mp}}1\big) r_3^{\frac{13}{2}+p_{12}-\i\saa\mu_\b}\times{ }_2\mathcal{F}_1\left[\begin{array}{c|c}
\frac{1}{2}-\i\saa\mu_\b, \frac{13}{2}+m+p_{12}-\i\saa\mu_\b \\
1- 2\i \saa \mu_\b 
\end{array}\ r_3\right]\Bigg\}
\nonumber\\
&~~+\left(\begin{array}{c}
\mu_\a \leftrightarrow \mu_\b \\
p_1 \leftrightarrow p_3 \\
u\leftrightarrow v
\end{array}\right).\label{sol_P3}
\end{align}
\end{keyeqn}
\begin{tcolorbox}[toggle enlargement=none, colback=white]
The summary of recurrence relations and solutions of different coefficients.
\begin{itemize}
\item The coefficient related to the background term $\mathcal{C}_{m, n}^{\pm\pm\pm}$:
\begin{align}
&\mathcal{C}_{m+1, n}^{\pm\pm\pm}=\frac{\left(m+3+p_1\right)\left(m+n+p_{123}+9\right)}{\mu_\a^2+\left(p_1+m+7 / 2\right)^2} \mathcal{C}_{m, n}^{\pm\pm\pm}~,\\
&\mathcal{C}_{0, n}^{\pm\pm\pm}=\frac{\mp\i\,2\sqrt{\pi}}{\left[\mu_\a^2+\left(p_1+5 / 2\right)^2\right]\left[\mu_\b^2+\left(p_3+5 / 2\right)^2\right]} \frac{\left(3+p_3\right)_n \Gamma (p_{123}+9+n)}{\left(\frac{7}{2}+p_3-\i \mu_\b\right)_n\left(\frac{7}{2}+p_3+\i \mu_\b\right)_n}~,\\  
&\mathcal{C}_{m, n+1}^{\pm\pm\pm}=\frac{\left(n+3+p_3\right)\left(m+n+p_{123}+9\right)}{\mu_\b^2+\left(p_3+n+7 / 2\right)^2}\, \mathcal{C}_{m, n}^{\pm\pm\pm}~,\\
&\mathcal{C}_{m, 0}^{\pm\pm\pm}=\mathcal{C}_{0, n}^{\pm\pm\pm} \quad {\rm{with}} \quad (p_1\leftrightarrow p_3,~\a\leftrightarrow \b,n\rightarrow m)~.
\end{align}
\item The coefficient related to the background-signal mixing term $\mathcal{A}_{m, n}^{\pm\pm\pm}$:
\begin{align}
&\mathcal{A}_{m+1, n}^{\pm\pm\pm(\saa)}=\frac{\left(m+3+p_1\right)\left(m+n+p_{12}+\frac{13}{2}+\i \saa\mu_\b\right)}{\mu_\a^2+\left(p_1+m+7 / 2\right)^2} \mathcal{A}_{m, n}^{\pm\pm\pm(\saa)}~,\\
&\mathcal{A}_{m, n+1}^{\pm\pm\pm(\saa)}=\frac{\left(n+\frac{1}{2}+\i \saa\mu_\b\right)\left(m+n+p_{12}+\frac{13}{2}+\i \saa\mu_\b\right)}{(n+1)(n+1+2 \i\saa \mu_\b)} \mathcal{A}_{m, n}^{\pm\pm\pm(\saa)}~,\\
&\mathcal{A}_{0, n}^{\pm\pm\pm(\saa)}= \frac{\i\ e^{\pm \pi \saa \mu_\b} \sqrt{\pi} \operatorname{csch}(\pi\saa \mu_\b)}{\mu_\a^2+\left(p_1+5 / 2\right)^2}\nonumber\\
&\qquad\qquad~~\times\Gamma\left[\begin{array}{c}
\frac{1}{2}+n+\i \saa\mu_\b, \frac{13}{2}+n+p_{12}+\i\saa \mu_\b, \frac{5}{2}+p_3-\i \mu_\b, \frac{5}{2}+p_3+\i \mu_\b \\
1+n, 3+p_3, 1+n+2 \i\saa \mu_\b
\end{array}\right].
\end{align}
\end{itemize}
As for the other coefficient $\mathcal{B}_{m, n}^{\pm\pm\pm(\saa)}$, it satisfies the same relation as $\mathcal{A}_{m, n}^{\pm\pm\pm(\saa)}$ with the replacement $p_1\leftrightarrow p_3$ and $\a\leftrightarrow \b$.
\begin{itemize}
\item Finally, solutions of those recurrence relation are given by
\begin{align}
\nonumber \mathcal{C}_{m, n}^{\pm\pm\pm}=&\ \frac{\mp\i\, 2\sqrt{\pi}\,\Gamma(p_{123}+9+m+n)}{\left[\mu_\a^2+\left(p_1+5 / 2\right)^2\right]\left[\mu_\b^2+\left(p_3+5 / 2\right)^2\right]} \\
&\times \frac{\left(3+p_1\right)_m\left(3+p_3\right)_n }{\left(\frac{7}{2}+p_1-\i \mu_\a\right)_m\left(\frac{7}{2}+p_1+\i \mu_\a\right)_m\left(\frac{7}{2}+p_3-\i \mu_\b\right)_n\left(\frac{7}{2}+p_3+\i \mu_\b\right)_n}~,\label{C_sol}\\
\nonumber\\
\nonumber \mathcal{A}_{m, n}^{\pm\pm\pm(\saa)}=&\  \frac{\i \sqrt{\pi}e^{\pm  \pi\saa \mu_\b}\operatorname{csch} (\pi \saa\mu_\b)}{\mu_\a^2+\left(p_1+5 / 2\right)^2} \frac{\left(3+p_1\right)_m}{\left(\frac{7}{2}+p_1-\i \mu_\a\right)_m\left(\frac{7}{2}+p_1+\i \mu_\a\right)_m}\\
&\times \Gamma\left[\begin{array}{c}
\frac{5}{2}+p_3-\i \mu_\b, \frac{5}{2}+p_3+\i \mu_\b, m+n+p_{12}+\frac{13}{2}+\i \saa\mu_\b, \frac{1}{2}+n+\i\saa \mu_\b \\
1+n, 3+p_3, 1+n+2 \i\saa \mu_\b
\end{array}\right],\label{A_sol}\\
\nonumber\\
\mathcal{B}_{m, n}^{\pm\pm\pm(\saa)}=&\  \mathcal{A}_{m, n}^{\pm\pm\pm(\saa)}\quad {\rm{with}} \quad (p_1\leftrightarrow p_3,\,\a\leftrightarrow \b)~.\label{B_sol}
\end{align}
\end{itemize}
\end{tcolorbox}

\subsection{Determine the coefficients} \label{sec:coeff}
Summarizing the results so far and recalling the relation~$\mathcal{I}=u^{-4-p_1} v^{-4-p_3}\widehat{\mathcal{I}}$, the general solutions for the seed integral $\mathcal{I}$ are given by  
\begin{align}
\mathcal{I}_{\m{abc},\a\b}^{p_1p_2p_3}=\sum_{\m{d,e}=\pm}c_{\m{abc},\m{de}}u^{-\frac{5}{2}-p_1-\i\m{d} \mu_\a} v^{-\frac{5}{2}-p_3-\i\m{e} \mu_\b} \ \mathcal{F}_4\left[\begin{array}{c|c}
\frac{4+p_2-\i\left(\m{d}\mu_\a+\m{e}\mu_\b\right)}{2}, \frac{5+p_2-\i\left(\m{d}\mu_\a+\m{e}\mu_\b\right)}{2} \\
1-\i\m{d} \mu_\a, 1-\i \m{e}\mu_\b
\end{array}\  u^2, v^2\right]+\frac{\widehat{\mathcal{P}}_{\m{abc},\a\b}^{p_1p_2p_3}}{u^{p_1+4}v^{p_3+4}}~,\label{general_sol}
\end{align}
where the particular solutions are given by $\widehat{\mathcal{P}}_{\pm\mp\pm,\a\b}^{p_1p_2p_3}=0$,~\eqref{sol_P_ppm} and \eqref{sol_P3} respectively. $c_{\m{abc},\m{de}}$ are undetermined coefficients and should be fixed by boundary conditions. Previous works on single-exchange diagrams typically use the folded limit and factorised limit as boundary conditions \cite{Arkani-Hamed:2018kmz}, but obtaining the formula under such limits can be challenging in our case. Instead, we adopt an alternative method proposed in \cite{Qin:2022fbv}, where the double soft limit $u,v\rightarrow 0$ is imposed as the boundary conditions. Under this limit, the seed integral~\eqref{general_sol} becomes
\begin{align}
\lim_{u,v\ll 1}\mathcal{I}_{\m{abc},\a\b}^{p_1p_2p_3}=\sum_{\m{d,e}=\pm}c_{\m{abc},\m{de}}u^{-\frac{5}{2}-p_1-\i\m{d} \mu_\a} v^{-\frac{5}{2}-p_3-\i\m{e} \mu_\b} \ \Gamma\left[\begin{array}{c}
\frac{4+p_2-\i\left(\m{d}\mu_\a+\m{e}\mu_\b\right)}{2}, \frac{5+p_2-\i\left(\m{d}\mu_\a+\m{e}\mu_\b\right)}{2} \\
1-\i\m{d} \mu_\a, 1-\i \m{e}\mu_\b
\end{array}\right].\label{lim_I_abc}
\end{align}
Note that the particular solutions (the second term of~\eqref{general_sol}) are subdominant in this limit. On the other hand, the time integral under such limit can be readily determined by employing the partial Mellin-Barnes representation, as detailed in the Appendix~\ref{MB_app}.
By comparing the expressions, we can fix the coefficients as
\begin{align}
 c_{\pm\mp\pm,\m{de}}&=-e^{\mp \i \frac{\pi}{2}\left(p_{13}-p_2\right)}\operatorname{csch}\left(\pi \m{d} \mu_\a\right)\operatorname{csch}\left(\pi \m{e} \mu_\b\right)\times\widetilde{\Gamma}(p_1,p_2,p_3,\mu_\a,\mu_\b)~,\\
c_{\pm\pm\mp,\m{de}}&=\mp\i\, e^{\mp \i \frac{\pi}{2}\left(p_{12}-p_3\right)}\operatorname{csch}\left(\pi \m{d} \mu_\a\right)\operatorname{csch}\left(\pi \m{e} \mu_\b\right)    e^{\mp\pi\m{d}\mu_\a}\times\widetilde{\Gamma}(p_1,p_2,p_3,\mu_\a,\mu_\b)~,\\
c_{\pm\mp\mp,\m{de}}&=\ c_{\mp\mp\pm,\m{ed}} \quad {\rm{with}} \quad (p_1\leftrightarrow p_3) \quad {\rm{and}} \quad (\a\leftrightarrow \b)~,\\
c_{\pm\pm\pm,\m{de}}&= e^{\mp \i \frac{\pi}{2}p_{123}}\operatorname{csch}\left(\pi \m{d} \mu_\a\right)\operatorname{csch}\left(\pi \m{e} \mu_\b\right)  e^{\mp\pi(\m{d}\mu_\a+\m{e}\mu_\b)}\times\widetilde{\Gamma}(p_1,p_2,p_3,\mu_\a,\mu_\b)~,
\label{lim_I_ppp}
\end{align}
with the new defined function,
\begin{align}
\widetilde{\Gamma}(p_1,p_2,p_3,\mu_\a,\mu_\b)\equiv\frac{ \pi^{\frac{1}{2}}}{2^{4+p_{13}-p_2}} \Gamma\left[\begin{array}{c}
\frac{5}{2}+p_1-\i \mu_\a, \frac{5}{2}+p_1+\i \mu_\a, \frac{5}{2}+p_3-\i \mu_\b, \frac{5}{2}+p_3+\i \mu_\b \\
3+p_1, 3+p_3
\end{array}\right]~.\label{til_Gamma}
\end{align}

\subsection{Summary of solutions} \label{Summary_sol}
After lengthy calculations, we have finally arrived at the exact solution for the double massive exchange seed integral~\eqref{I_abc}.
In this section, we summarize the detailed forms of various types of seed integrals for reference. We also remind readers again of our notation. The numbers $p_i$ characterize the types of vertices in the seed~\eqref{I_abc}, and the summation is denoted by $p_{ij}=p_i+p_j$, etc. The indices $\a, \b$ label the massive fields with masses $m_\a$ and $m_\b$. The momentum ratio is defined as $u=k_1/k_{24}$ and $v=k_3/k_{24}$, where $k_{24}=k_2+k_4$, and we further introduced $r_1=2u/(1+u+v)=2k_1/k_{1234}$ and $r_3=2v/(1+u+v)=2k_3/k_{1234}$. The functions ${}_3\mathcal{F}_2$ and $\mathcal{F}_4$ represent the dressed versions of the hypergeometric function and the Appell series, defined in \eqref{dhyper} and \eqref{def_f4}, respectively. The coefficient $\widetilde{\Gamma}(p_1,p_2,p_3,\mu_\a,\mu_\b)$ is defined in Eq.~\eqref{til_Gamma}.
\newpage
\begin{itemize} 
\item The fully factorised seed integral: 
\end{itemize}
\begin{keyeqn}
\begin{align}
\nonumber \mathcal{I}_{\pm\mp\pm,\a\b}^{p_1p_2p_3} =&-e^{\mp \i \frac{\pi}{2}\left(p_{13}-p_2\right)}\,\widetilde{\Gamma}(p_1,p_2,p_3,\mu_\a,\mu_\b)\sum_{\m{a,b}=\pm}\operatorname{csch}\left(\pi \m{a} \mu_\a\right)\operatorname{csch}\left(\pi \m{b} \mu_\b\right) u^{-\frac{5}{2}-p_1-\i\m{a} \mu_\a} v^{-\frac{5}{2}-p_3-\i\m{b} \mu_\b} \\
&\times\mathcal{F}_4\left[\begin{array}{c|c}
\frac{4+p_2-\i\left(\m{a}\mu_\a+\m{b}\mu_\b\right)}{2}, \frac{5+p_2-\i\left(\m{a}\mu_\a+\m{b}\mu_\b\right)}{2} \\
1-\i\m{a} \mu_\a, 1-\i \m{b}\mu_\b
\end{array}\  u^2, v^2\right].\label{result_pmp}
\end{align}
\end{keyeqn}

\begin{itemize}
    \item The partially factorised partially nested seed integral: 
\end{itemize}

\begin{keyeqn}
\begin{align}
\nonumber
&\mathcal{I}_{\pm\pm\mp,\a\b}^{p_1p_2p_3}\\
\nonumber
&=\mp\i\,e^{\mp \i \frac{\pi}{2}\left(p_{12}-p_3\right)}\, \widetilde{\Gamma}(p_1,p_2,p_3,\mu_\a,\mu_\b)
\sum_{\m{a,b}=\pm}\operatorname{csch}\left(\pi \m{a} \mu_\a\right)\operatorname{csch}\left(\pi \m{b} \mu_\b\right)  e^{\mp\pi\m{a}\mu_\a} u^{-\frac{5}{2}-p_1-\i\m{a} \mu_\a} v^{-\frac{5}{2}-p_3-\i\m{b} \mu_\b}\\
\nonumber
&~~~\times \mathcal{F}_4\left[\begin{array}{c|c}
\frac{4+p_2-\i\left(\m{a}\mu_\a+\m{b}\mu_\b\right)}{2}, \frac{5+p_2-\i\left(\m{a}\mu_\a+\m{b}\mu_\b\right)}{2} \\
1-\i\m{a} \mu_\a, 1-\i \m{b}\mu_\b
\end{array}\  u^2, v^2\right]\\
\nonumber
&~~~+\frac{\i}{\pi^{{1}/{2}}}\frac{r_3^{p_{12}+\frac{13}{2}} }{v^{p_{123}+9}}\frac{ e^{\mp\i\frac{\pi}{2} (p_{12}-p_3)}}{2^{6+2p_{2}} }\,\widetilde{\Gamma}(p_1,p_2,p_3,\mu_\a,\mu_\b)\times\sum_{\saa=\pm}\sum_{m=0}^{\infty} 
\Gamma\left[\begin{array}{cc}
3+m+p_1 \\
\frac{7}{2}+m+p_1-\i\mu_\a, \frac{7}{2}+m+p_1+\i\mu_\a
\end{array}\right]
\\
&~~~\times \operatorname{csch}(\pi \saa\mu_\b)r_1^m r_3^{\i\saa\mu_\b}\times{ }_2 {\mathcal{F}}_1\left[\begin{array}{c|c}
\frac{1}{2}+\i\saa\mu_\b,\frac{13}{2}+m+p_{12}+\i\saa \mu_\b \\
1+2\i\saa\mu_\b
\end{array}\ r_3 \right]~.\label{result_ppm}
\end{align}
\end{keyeqn}
The results for $\mathcal{I}_{\pm\mp\mp,\a\b}^{ p_1 p_2 p_3}$ can be obtained by replacing  $u\leftrightarrow v \ (r_1\leftrightarrow r_3)$, $p_1\leftrightarrow p_3 $, and $\alpha\leftrightarrow \beta$ in  $\mathcal{I}_{\mp\mp\pm,\a\b}^{ p_1 p_2 p_3}$.
\begin{itemize}
    \item The fully nested seed integral:
\end{itemize}
\begin{keyeqn}
\begin{align}
\nonumber &\mathcal{I}_{\m{\pm\pm\pm},\a\b}^{p_1p_2p_3}\\
\nonumber 
&=e^{\mp \i \frac{\pi}{2}p_{123}}\,\widetilde{\Gamma}(p_1,p_2,p_3,\mu_\a,\mu_\b) \times\sum_{\m{a,b}=\pm}\operatorname{csch}\left(\pi \m{a} \mu_\a\right)\operatorname{csch}\left(\pi \m{b} \mu_\b\right)  e^{\mp\pi(\m{a}\mu_\a+\m{b}\mu_\b)}u^{-\frac{5}{2}-p_1-\i\m{a} \mu_\a} v^{-\frac{5}{2}-p_3-\i\m{b} \mu_\b} \\
\nonumber 
&~~~\times \mathcal{F}_4\left[\begin{array}{c|c}
\frac{4+p_2-\i\left(\m{a}\mu_\a+\m{b}\mu_\b\right)}{2}, \frac{5+p_2-\i\left(\m{a}\mu_\a+\m{b}\mu_\b\right)}{2} \\
1-\i\m{a} \mu_\a, 1-\i \m{b}\mu_\b
\end{array}\  u^2, v^2\right]\\
\nonumber 
&
~~~+\Bigg\{\frac{r_3^{p_{12}+\frac{13}{2}}}{v^{p_{123}+9}}\cdot\frac{ e^{\mp\i\frac{\pi}{2} p_{123}}}{2^{10+p_{123}}\left[\mu_\a^2+\left(p_1+5 / 2\right)^2\right] }\,\Gamma\left[\begin{array}{c}
\frac{5}{2}+p_3-\i \mu_\b, \frac{5}{2}+p_3+\i \mu_\b \\
3+p_3
\end{array}\right]\nonumber\\
&~~~\times\sum_{m=0}^{\infty}\,\frac{\left(3+p_1\right)_m\,r_1^m} {\left(\frac{7}{2}+p_1-\i \mu_\a\right)_m\left(\frac{7}{2}+p_1+\i \mu_\a\right)_m}\Bigg( r_3^{p_3+\frac{5}{2}}\,{ }_3 \mathcal{F}_2\left[\begin{array}{c|c}
1,3+p_3, 9+m+p_{123}   \\
\frac{7}{2}+p_3-\i \mu_\b, \frac{7}{2}+p_3+\i \mu_\b 
\end{array}\ r_3\right]\nonumber\\
&\left.~~~\pm\sum_{\saa=\pm}\big(\coth(\pi\saa\mu_\b){\mp}1\big) r_3^{-\i\saa\mu_\b}\times{ }_2\mathcal{F}_1\left[\begin{array}{c|c}
\frac{1}{2}-\i\saa\mu_\b, \frac{13}{2}+m+p_{12}-\i\saa\mu_\b \\
1- 2\i \saa \mu_\b 
\end{array}\ r_3\right]\Bigg)+
\left(\begin{array}{c}
\mu_\a \leftrightarrow \mu_\b \\
p_1 \leftrightarrow p_3 \\
u\leftrightarrow v
\end{array}\right)\right\}.\label{result_ppp}
\end{align}
\end{keyeqn}

\section{Continuation to All Kinematic Regions}\label{Sec_Continuation}
\subsection{Continuation and resummation}
The homogeneous solutions presented in the last section are expressed using a specific type of two-variable special function, namely the Appell function $F_4$. This function converges only when $u+v<1$. 
Meanwhile, some parts of the particular solution exhibit divergence when $u+v>1$. Therefore, the aforementioned results about the trispectrum are only applicable when the two massive propagators are relatively soft. 
Adding to this concern, the physical regions for the bispectrum are constrained to $u+v> 1$, aligning with the triangle inequality $k_1 + k_3 > k_2$.\footnote{A similar divergence 
in three-point correlators also emerges in the conformal correlation functions in momentum space, specifically in the Triple-$K$ integrals\cite{Bzowski:2013sza}.} In this section, we aim to extend the analytical results to all kinematic regions, ensuring validity even in the bispectrum limit.
Without loss of generality, we fix the power indices as $p_1 = p_2 = p_3 = -2$ in this section. The results for alternative values of $p_i$ can be derived by applying appropriate differential operators to the subsequent results \cite{Qin:2022fbv}.
Also, we will mainly focus on three representative integrals, namely $\mathcal{I}_{+-+}$, $\mathcal{I}_{++-}$ and $\mathcal{I}_{+++}$, since other components can be obtained by taking complex conjugations or permuting the variables. 

\paragraph{Seed integral $\bm{\mathcal{I_{+-+}}}$.} 
Let us begin with the most straightforward case, denoted as $\mathcal{I_{+-+}}$, which is completely factorised in the time integrals and satisfies the homogeneous bootstrap equations,~\eqref{BS_b_dS_1} and~\eqref{BS_b_dS_2}. To speed up the convergence of series, we observed that employing the variables $r_i\equiv 2 k_i/k_{1234}$ or $\tilde{r}_i\equiv 2 k_i/k_{123}$ (for the case of the bispectrum), defined as substitutes for $(u,v)$, proves to be more efficient. Note that the divergence for $u+v>1$ mentioned above appears in the region $r_1+r_3>1$ for the new variables. 

Two-variable hypergeometric series $F_4$, which are present in the homogeneous solution~\eqref{result_pmp}, can be transformed into another second-type Appell function $F_2$ with arguments $r_i$ using the formula (\ref{F4toF2}). Subsequently, $F_2$ can be expanded as a series of hypergeometric functions through (\ref{F2expansion}). This allows us to express the original expression as follows,
\begin{align}\label{Ipmp}
    \mathcal{I}_{+-+}(r_1,r_3)=&\sum_{\saa,\sb=\pm}\sum_{m=0}^{\infty} -\frac{\pi^{2}}{2^{3}}{\rm{\csch}}(2\pi \saa\mu_\a)\,{\rm{\csch}}(2\pi \sb\mu_\b)\,(r_1+r_3-2)^3 r_1^{-\frac{1}{2}+m+\i\saa\mu_{\a}}r_3^{-\frac{1}{2}+\i\sb\mu_{\b}}\nonumber\\
    &\times\Gamma
    \left[\begin{array}{c}
    \frac{1}{2}+m+\i\saa\mu_{\a} \\
    1+m,1+m+2\i\saa\mu_\a
    \end{array}\right]
    {{}_2}\mathcal{F}_1\left[\begin{array}{c|c}
    \frac{1}{2}+\i\sb\mu_\b,2+m+\i\saa\mu_\a+\i\sb\mu_\b \\
    1+2\i\sb\mu_\b
    \end{array} \ r_3\right].
\end{align}
In the obtained result, there are four series corresponding to indices $\saa,\sb=\pm$. One can check numerically that each term diverges when the sum of $r_1$ and $r_3$ exceeds one. However, intriguingly, cancellations occur when all terms are summed together. To better visualize these cancellations, we can regroup the terms in pairs and employ the hypergeometric function transformation (\ref{hyperTrans}) to rewrite the expression as
\begin{keyeqn}
    \begin{align}\label{Ipmp2}
    &\mathcal{I}_{+-+}(r_1,r_3)\nonumber\\
    &=\sum_{m=0}^{\infty} \i\,\pi^{2}{\rm{\csch}}(2\pi \mu_\a)\,{\rm{sech}}(\pi\mu_\b)\,\left(\frac{r_1}{2}+\frac{r_3}{2}-1\right)^3\Gamma
    \left[\begin{array}{c}
    \frac{1}{2}+m+\i\mu_{\a},2+m+\i\mu_\a-\i\mu_\b \\
    1+m,1+m+2\i\mu_\a,\frac{1}{2}+\i\mu_\b
    \end{array}\right]\nonumber\\
    &~~~\times r_1^{-\frac{1}{2}+m+\i\mu_{\a}}r_3^{-\frac{1}{2}+\i\mu_{\b}}
    {{}_2}\mathcal{F}_1\left[\begin{array}{c|c}
    \frac{1}{2}+\i\mu_\b,2+m+\i\mu_\a+\i\mu_\b \\
    \frac{5}{2}+m+\i\mu_\a
    \end{array} \ 1-r_3\right]+\big(\mu_\a\rightarrow-\mu_\a\big)~.
\end{align}
\end{keyeqn}

The final expression of $\mathcal{I}_{+-+}$ is now well-behaved even in regions where $r_1+r_3>1$, and thus the validity of the analytical expression is extended to all physical kinematic regions.  Regarding the bispectrum limit, one can simply replace every $r_i$ above with $\tilde{r}_i$, and the convergence is also guaranteed. Alternatively,  we can take a different perspective on the continuation procedure. In the initial steps of solving homogeneous equations, we can directly introduce an ansatz, similar to our approach to finding particular solutions. After the resummation of one variable, the ansatz for the homogeneous solution will take the form outlined above.
\paragraph{Seed integral $\bm{\mathcal{I_{++-}}}$.} The case of the partially factorised partially nested seed integral $\mathcal{I}_{++-}$ closely parallels our previous discussion. The particular solution after summing over the series of $r^n_3$ shows great convergence within the three-point region ($k_1+k_3>k_2$). Then, the only thing we need to be concerned about is the homogeneous parts $\mathcal{Y}_{++-}$, which can be addressed using the same procedure as before. After changing variables and expanding the results as the hypergeometric series, we noticed that the expressions can be systematically regrouped in pairs, which finally reads as 
\begin{align}\label{Yppm}
    &\mathcal{Y}_{++-}(r_1,r_3)\nonumber\\
    &=\sum_{m=0}^{\infty}\Bigg\{ -\frac{\pi^{2}}{2}\left[{\rm{csch}}(\pi \mu_\a)+{\rm{sech}}(\pi \mu_\a)\right]\,{\rm{sech}}(\pi\mu_\b)\,\Gamma
    \left[\begin{array}{c}
    \frac{1}{2}+m+\i\mu_{\a},2+m+\i\mu_\a+\i\mu_\b \\
    1+m,1+m+2\i\mu_\a,\frac{1}{2}-\i\mu_\b
    \end{array}\right]\nonumber\\
    &\times  r_1^{-\frac{1}{2}+m+\i\mu_{\a}}r_3^{-\frac{1}{2}-\i\mu_{\b}} \left(\frac{r_1}{2}+\frac{r_3}{2}-1\right)^3
    {{}_2}\mathcal{F}_1\left[\begin{array}{c|c}
    \frac{1}{2}-\i\mu_\b,2+m+\i\mu_\a-\i\mu_\b \\
    \frac{5}{2}+m+\i\mu_\a
    \end{array} \ 1-r_3\right]+\big(\mu_\a\rightarrow-\mu_\a\big)\Bigg\}~.
\end{align}
Adding the particular solution (\ref{sol_P_ppm}), we obtain the final results valid for all physical kinematic regions,
\begin{keyeqn}
    \begin{align}\label{Ippm}
    &\mathcal{I}_{++-}(r_1,r_3)\nonumber\\
    &=\sum_{m=0}^{\infty}\Bigg\{-\frac{\pi^{2}}{2}\left[{\rm{csch}}(\pi \mu_\a)+{\rm{sech}}(\pi \mu_\a)\right]\,{\rm{sech}}(\pi\mu_\b)\,\Gamma
    \left[\begin{array}{c}
    \frac{1}{2}+m+\i\mu_{\a},2+m+\i\mu_\a+\i\mu_\b \\
    1+m,1+m+2\i\mu_\a,\frac{1}{2}-\i\mu_\b
    \end{array}\right]\nonumber\\
    &~~~~~~~\times  r_1^{-\frac{1}{2}+m+\i\mu_{\a}}r_3^{-\frac{1}{2}-\i\mu_{\b}} \left(\frac{r_1}{2}+\frac{r_3}{2}-1\right)^3
    {{}_2}\mathcal{F}_1\left[\begin{array}{c|c}
    \frac{1}{2}-\i\mu_\b,2+m+\i\mu_\a-\i\mu_\b \\
    \frac{5}{2}+m+\i\mu_\a
    \end{array} \ 1-r_3\right]+\big(\mu_\a\rightarrow-\mu_\a\big)\Bigg\}\nonumber\\
    &+\sum_{m=0}^{\infty}\Bigg\{\i\,\pi^{2}{\rm{\csch}}(2\pi \mu_\b)\,{\rm{sech}}(\pi\mu_\a)\,\Gamma
    \left[\begin{array}{c}
    1+m \\
    \frac{3}{2}+m-\i\mu_\a,\frac{3}{2}+m+\i\mu_\a
    \end{array}\right]\nonumber\\
    &~~~~~~~\times r_1^{m}r_3^{-\frac{1}{2}+\i\mu_{\b}} \left(\frac{r_1}{2}+\frac{r_3}{2}-1\right)^3
    {{}_2}\mathcal{F}_1\left[\begin{array}{c|c}
    \frac{1}{2}+\i\mu_\b,\frac{5}{2}+m+\i\mu_\b \\
    1+2\i\mu_\b
    \end{array} \ r_3 \right]+\big(\mu_\b\rightarrow-\mu_\b\big)\Bigg\}~,
    \end{align}
\end{keyeqn}
where the first and second summations represent  homogeneous and particular solution respectively. Before moving to the discussion about the most challenging case, let us first comment on the order of summation. Initially, the particular solution involves a double summation series over $r_1^m r_3^n$, and we are able to sum up one layer of an infinite series. As we chose earlier, the initial step involves summation over $r_3^n$, resulting in a convergent series.  However, an alternative approach involves initially summing over the series related to $r_1$. In this case, the particular solution then becomes
\begin{align}\label{Pppm}
    \mathcal{P}_{++-}(r_1,r_3)=&\sum_{n=0}^{\infty}\i\,\pi^2 {\rm{\csch}}(2\pi \mu_\b)\,{\rm{sech}}(\pi\mu_\a)\left(\frac{r_1}{2}+\frac{r_3}{2}-1\right)^3\times\Gamma
    \left[\begin{array}{c}
    \frac{1}{2}+n+\i\mu_{\b} \\
    1+n,1+n+2\i\mu_\b
    \end{array}\right]\nonumber\\
    &\times r_3^{-\frac{1}{2}+n+\i\mu_\b}{{}_3}\mathcal{{F}}_2\left[\begin{array}{c|c}
    1,1,\frac{5}{2}+n+\i\mu_\b \\
    \frac{3}{2}-\i\mu_\a,\frac{3}{2}+\i\mu_\a
    \end{array} \ r_1\right]+\big(\mu_\b\rightarrow-\mu_\b\big)~,
\end{align}
but one can verify numerically that this series is divergent for regions where $r_1+r_3>1$
(more precisely, it behaves as an asymptotic series). To ensure the final contribution is convergent, instead of expanding it as a series of hypergeometric functions with the argument $r_3$ as in (\ref{Ipmp}), we should expand it with argument dependence on $r_1$, which is
\begin{align}\label{Yppm2}
    \mathcal{Y}_{++-}(r_1,r_3)&=\sum_{\saa,\sb=\pm}\sum_{m=0}^{\infty} -\i\,\pi^{2}e^{\pi\saa\mu_\a}\,{\rm{csch}}(2\pi\saa\mu_\a){\rm{csch}}(2\pi\sb\mu_\b)\,\left(\frac{r_1}{2}+\frac{r_3}{2}-1\right)^3 r_1^{-\frac{1}{2}+\i\saa\mu_{\a}}r_3^{-\frac{1}{2}+m+\i\sb\mu_{\b}}\nonumber\\
    &~~\times\Gamma
    \left[\begin{array}{c}
    \frac{1}{2}+m+\i\sb\mu_{\b}\\
    1+m,1+m+2\i\sb\mu_\b
    \end{array}\right]
    {{}_2}\mathcal{F}_1\left[\begin{array}{c|c}
    \frac{1}{2}+\i\saa\mu_\a,2+m+\i\saa\mu_\a+\i\sb\mu_\b \\
    1+2\i\saa\mu_\a
    \end{array} \ r_1\right],
\end{align}

Unfortunately, regrouping it in pairs, as in (\ref{Yppm}), to achieve convergence is not possible in this expression since the coefficient of each term does not match. This is expected, as the particular solution (\ref{Pppm}) is superficially divergent and requires another component to cancel it. We have explicitly verified that although both (\ref{Pppm}) and (\ref{Yppm2}) exhibit superficial divergences at kinematic regions relevant for three-point correlators, 
the summation of these two is actually well convergent in all physical regions, yielding the same results as (\ref{Ippm}). In other words, we should treat (\ref{Pppm}) and (\ref{Yppm2}) as a whole, which provides a complete and finite result. Let us highlight the difference in these two final expressions: for $\mathcal{I}_{++-}$, the argument inside the hypergeometric function is $r_3$, representing the summation over the $r_3$ direction being performed first. On the other hand,  for $\mathcal{P}_{++-}$ and $\mathcal{Y}_{++-}$ in equations (\ref{Pppm}) and (\ref{Yppm2}), we summed over $r_1$ first. In summary, to obtain the final convergent result, we observe that the order of summation is crucial, and both the particular and homogeneous solutions should be summed along the same direction to eliminate the superficial divergence. Otherwise, we may end up with an unphysical result.

\begin{figure}[ht]
	\centering
	\includegraphics[width=\textwidth]{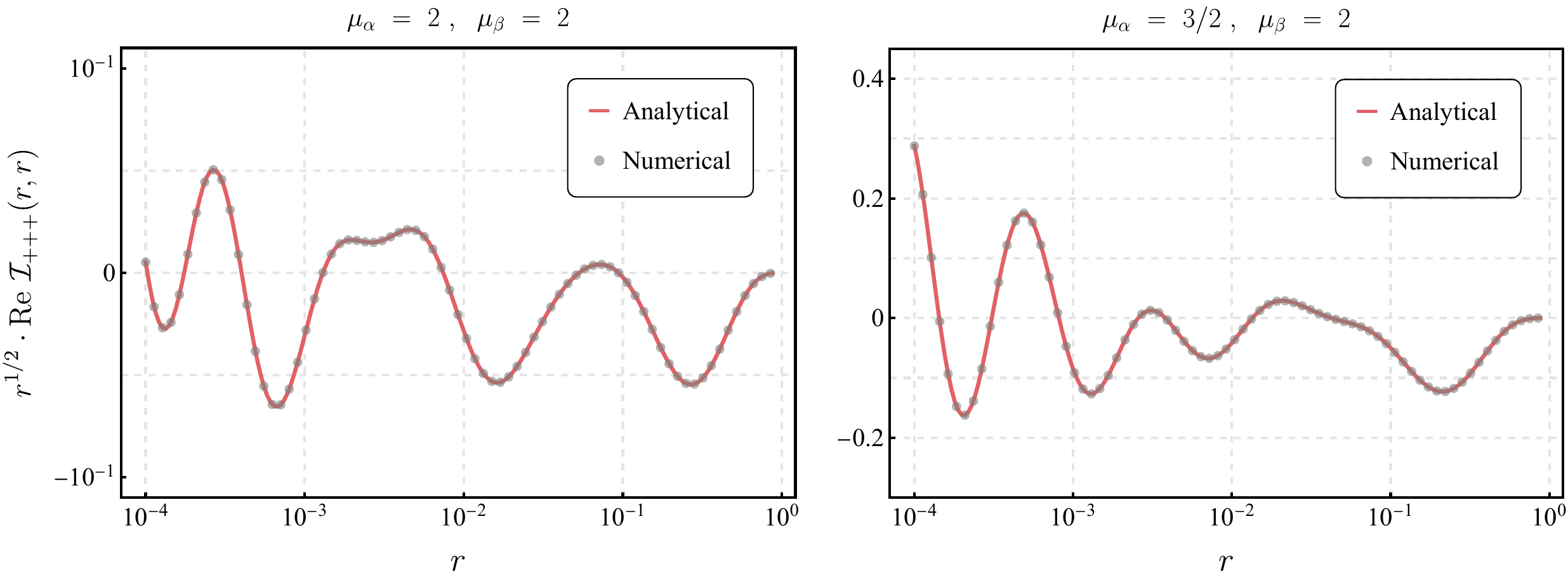}\\
	\caption{ The comparison between the numerical calculation and exact analytical result (\ref{Ipppresult}). The \textcolor{gray}{gray} dots and \textcolor{red2}{red} solid line represent numerical and analytical results, respectively. For the numerical calculation, we directly perform the time integral of the seed integral (\ref{I_abc}). In both figure we set $r_1=r_3=r$.  The masses are chosen as $\mu_\a=\mu_\b=2$ in the left panel and $\mu_\a=1.5,\mu_\b=2$ in the right panel. For better visualization, we multiplied $\mathcal{I}_{+++}$ by a factor $r^{1/2}$.
 }\label{IpppNumVSAna}
\end{figure}

\paragraph{Seed integral $\bm{\mathcal{I_{+++}}}$.} Finally, let's examine the fully nested time integral $\mathcal{I}_{+++}$, which poses the most challenging case. By employing the same procedure of expanding the original homogeneous solution, we reach the following result,
\begin{align}
    \mathcal{Y}_{+++}(r_1,r_3)&=\sum_{\saa,\sb=\pm}\sum_{m=0}^{\infty} \,\pi^{2}e^{\pi(\saa\mu_\a+\sb\mu_b)}{\rm{csch}}(2\pi\saa\mu_\a)\,{\rm{csch}}(2\pi\sb\mu_\b)\,\left(\frac{r_1}{2}+\frac{r_3}{2}-1\right)^3 r_1^{-\frac{1}{2}+m+\i\saa\mu_{\a}}r_3^{-\frac{1}{2}+\i\sb\mu_{\b}}\nonumber\\
    &~~~\times\Gamma
    \left[\begin{array}{c}
    \frac{1}{2}+m+\i\saa\mu_{\a} \\
    1+m,1+m+2\i{\saa\mu_\a}
    \end{array}\right]
    {{}_2}\mathcal{{F}}_1\left[\begin{array}{c|c}
    \frac{1}{2}+\i\sb\mu_\b,2+m+\i\saa\mu_\a+\i\sb\mu_\b \\
    1+2\i\sb\mu_\b
    \end{array} \ r_3\right].
\end{align}
This homogeneous component remains superficially divergent at kinematic regions relevant for three-point correlators and cannot be regrouped into pairs to cancel out the divergence. Similar to the preceding discussion, we therefore need to incorporate contributions from both the homogeneous and particular solutions to obtain finite and convergent results. 
By incorporating the particular solution, the final expression takes the form,
\begin{keyeqn}
    \begin{align}\label{Ipppresult}
    &\mathcal{I}_{+++}(r_1,r_3)\nonumber\\
    &=\sum_{\saa,\sb=\pm}\sum_{m=0}^{\infty} \,\pi^{2}e^{\pi(\saa\mu_\a+\sb\mu_b)}{\rm{csch}}(2\pi\saa\mu_\a)\,{\rm{csch}}(2\pi\sb\mu_\b)\,\left(\frac{r_1}{2}+\frac{r_3}{2}-1\right)^3 r_1^{-\frac{1}{2}+m+\i\saa\mu_{\a}}r_3^{-\frac{1}{2}+\i\sb\mu_{\b}}\nonumber\\
    &~~~~\times\Gamma
    \left[\begin{array}{c}
    \frac{1}{2}+m+\i\saa\mu_{\a} \\
    1+m,1+m+2\i{\saa\mu_\a}
    \end{array}\right]
    {{}_2}\mathcal{{F}}_1\left[\begin{array}{c|c}
    \frac{1}{2}+\i\sb\mu_\b,2+m+\i\saa\mu_\a+\i\sb\mu_\b \\
    1+2\i\sb\mu_\b
    \end{array} \ r_3\right]\nonumber\\
    &+\bigg\{\sum_{m=0}^{\infty}\frac{\pi^2}{2} \big[\csch(\pi\mu_\a)-{\rm{sech}}(\pi\mu_\a)\big]{\rm{sech}}(\pi\mu_\b)\left(\frac{r_1}{2}+\frac{r_3}{2}-1\right)^3r_1^{-\frac{1}{2}+m-\i\mu_\a}\nonumber\\
    &~~~~\times\Gamma
    \left[\begin{array}{c}
    \frac{1}{2}+m-\i\mu_{\a} \\
    1+m,1+m-2\i\mu_\a
    \end{array}\right] {{}_3}\mathcal{{F}}_2\left[\begin{array}{c|c}
    1,1,\frac{5}{2}+m-\i\mu_\a \\
    \frac{3}{2}-\i\mu_\b,\frac{3}{2}+\i\mu_\b
    \end{array} \ r_3\right]+\big(\mu_\a\rightarrow-\mu_\a\big)\bigg\}\nonumber\\
    &+\bigg\{\sum_{m=0}^{\infty}\frac{\pi^2}{2} \big[\csch(\pi\mu_\b)-{\rm{sech}}(\pi\mu_\b)\big]{\rm{sech}}(\pi\mu_\a)\left(\frac{r_1}{2}+\frac{r_3}{2}-1\right)^3r_1^{m} r_3^{-\frac{1}{2}-\i\mu_\b}\nonumber\\
    &~~~~\times \Gamma
    \left[\begin{array}{c}
    m+1 \\
    \frac{3}{2}+m-\i\mu_\a,\frac{3}{2}+m+\i\mu_\a
    \end{array}\right]{{}_2}\mathcal{{F}}_1\left[\begin{array}{c|c}
    \frac{1}{2}-\i\mu_\b,\frac{5}{2}+m-\i{\mu_\b} \\
    1-2\i\mu_\b
    \end{array} \ r_3\right]+\big(\mu_\b\rightarrow-\mu_\b\big)\bigg\}\nonumber\\
    &+\sum_{m=0}^{\infty}{\pi^2}{\rm{sech}}(\pi\mu_\b)\,{\rm{sech}}(\pi\mu_\a)
    \left(\frac{r_1}{2}+\frac{r_3}{2}-1\right)^3 r_1^{m}\nonumber\\
    &~~~~\times\Gamma
    \left[\begin{array}{c}
    1+m \\
    \frac{3}{2}+m-\i\mu_\a,    \frac{3}{2}+m+\i\mu_\a
    \end{array}\right]
    {{}_3}\mathcal{{F}}_2\left[\begin{array}{c|c}
    1,1,3+m \\
    \frac{3}{2}-\i\mu_\b,\frac{3}{2}+\i\mu_\b
    \end{array} \ r_3\right]~,
\end{align}
\end{keyeqn}
where the first summation comes from the homogeneous solution, while the second and third summations arise from the particular solution with a partially non-analytical structure (signal-background mixing term). The last summation is the result of the particular solution, which is analytical in both $r_1$ and $r_3$ (purely background term). As previously emphasized, the order of summation is crucial, and in the formula above, each term involves a summation along $r_3$. Alternatively, one can first sum along the $r_1$ direction, yielding similar results. In Figure \ref{IpppNumVSAna}, we compare the exact analytical results (\ref{Ipppresult}) with the corresponding numerical calculations. We have chosen $r_1=r_3=r$, while the mass parameters $\mu_\a$ and $\mu_\b$ are different in the left and right panels. We can see that the analytical expression shows excellent agreement with the numerical calculations. Interestingly, unlike the single-exchange diagram, the oscillatory pattern shown in Figure \ref{IpppNumVSAna} displays novel features, akin to the overlay of two oscillation signals. In Section \ref{Sec_pheno}, as we delve into phenomenology, we will explore these characteristics in detail.

In summary, certain series may initially exhibit superficial divergences, nevertheless, by properly organizing and treating them as a whole, the final expression yields a finite result applicable to all configurations of physical momentum. It is anticipated that certain terms in the final results can be related to each other through some mathematical formulae, potentially leading to the explicit cancellation of divergent components. For example, one potential approach involves employing the expansion series of the hypergeometric function around the arguement unity~\cite{hyper}, particularly in the regions where $r_1+r_3>1$. However, this transformation introduces significant complexity and further complicates the expression.  These transformations and simplifications are beyond the scope of this paper, and we leave a more detailed discussion to future work.
As a final note, we stress that although our final analytical results are still expressed as (single-layer) infinite summations, the series expansions are explicitly convergent.
For concrete evaluation, one may therefore incorporate any number of terms to reach an arbitrary precision.
For example, in Fig.~\ref{IpppNumVSAna}, we included all terms with $m\leq 20$, an approximation which perfectly matches the numerical calculation of the integrals, as far as the eye can tell, and the summations could always be more precise by adding more terms.
\subsection{Consistency checks at different limits}\label{sec:limits}
To further verify the correctness of the analytical expressions, in this subsection, we will perform a detailed examination of their behaviors under different limits. For example, with the standard Bunch--Davies vacuum choice, some spurious poles should cancel out, such that the total results should be smooth when $k_1=k_{234}$ or $k_3=k_{124}$, equivalently, $r_1$ or $r_3$ approaches the unity, which is similar to the folded singularity cancellation present in the single-exchange diagrams \cite{Holman:2007na,Flauger:2013hra,Arkani-Hamed:2018kmz}.
\paragraph{Double soft limits.} 
The first simple and important limit is the case when both $r_1$ and $r_3$ approach zero, which is the regime in which the non-analytic cosmological collider signals will become the dominant components.  
This double soft limit is not a consistency
check but rather an interesting limit for the phenomenology of the trispectrum which will be discussed in Section \ref{Sec_pheno}. Under such a limit, the seed integrals become,
\begin{align}
    \lim_{r_1,r_3\rightarrow 0}\mathcal{I}_{+-+}(r_1,r_3)= 
    &-\frac{\pi^3\csch(2\pi\mu_\a)\csch(2\pi\mu_\b)}{2^{2\i\mu_\a-2\i\mu_\b}}\times\Gamma
    \left[\begin{array}{c}
    2+\i\mu_\a-\i\mu_\b \\
    1+\i\mu_\a, 1-\i\mu_\b
    \end{array}\right]r_1^{-\frac{1}{2}+\i\mu_\a}r_3^{-\frac{1}{2}-\i\mu_\b}+\text{c.c.}~~~~~\nonumber\\
    \sim&~\mathcal{O}\left(e^{-\pi(\mu_\a+2\mu_\b)}\right)r_1^{-\frac{1}{2}+\i\mu_\a}r_3^{-\frac{1}{2}-\i\mu_\b}+\text{c.c.},
\end{align}
\begin{align}
    \lim_{r_1,r_3\rightarrow 0}\mathcal{I}_{++-}(r_1,r_3)= 
    &-\i\,e^{\pi\mu_\a}\frac{\pi^3\csch(2\pi\mu_\a)\csch(2\pi\mu_\b)}{2^{2\i\mu_\a-2\i\mu_\b}}\times\Gamma
    \left[\begin{array}{c}
    2+\i\mu_\a-\i\mu_\b \\
    1+\i\mu_\a, 1-\i\mu_\b
    \end{array}\right]r_1^{-\frac{1}{2}+\i\mu_\a}r_3^{-\frac{1}{2}-\i\mu_\b}~~~~~~\nonumber\\
    \sim&~\mathcal{O}\left(e^{-2\pi\mu_\b}\right)r_1^{-\frac{1}{2}+\i\mu_\a}r_3^{-\frac{1}{2}-\i\mu_\b}.
\end{align}
Without loss of generality, we have assumed $\mu_\b \ge \mu_\a$ and neglected terms with heavier Boltzmann suppression in the second lines. For the fully nested seed integral $\mathcal{I}_{+++}$, let us 
investigate separately contributions from both the homogeneous solution and the particular solution which is
\begin{align}\label{Ydoublesoft}
    \lim_{r_1,r_3\rightarrow 0}\mathcal{Y}_{+++}(r_1,r_3)= 
    &-\,e^{\pi(\mu_\a+\mu_\b)}\frac{\pi^3\csch(2\pi\mu_\a)\csch(2\pi\mu_\b)}{2^{2\i\mu_\a+2\i\mu_\b}}\times\Gamma
    \left[\begin{array}{c}
    2+\i\mu_\a+\i\mu_\b \\
    1+\i\mu_\a, 1+\i\mu_\b
    \end{array}\right]r_1^{-\frac{1}{2}+\i\mu_\a}r_3^{-\frac{1}{2}+\i\mu_\b}\nonumber\\
    \sim&~\mathcal{O}\left(e^{-\pi(\mu_\a+\mu_\b)}\right)r_1^{-\frac{1}{2}+\i\mu_\a}r_3^{-\frac{1}{2}+\i\mu_\b},
\end{align}
\begin{align}\label{Pdoublesoft}
    \lim_{r_1,r_3\rightarrow 0}\mathcal{P}_{+++}(r_1,r_3) =
    &~\frac{\pi^{3/2}}{2^{2\i\mu_\b-1}}\frac{\csch(\pi\mu_\b)+\rm{\sech(\pi\mu_\b)}}{1+4\mu_\a^2}\times\Gamma\left[\begin{array}{c}
    \frac{5}{2}+\i\mu_\b \\
    1+\i\mu_\b
    \end{array}\right] r_3^{-\frac{1}{2}+\i\mu_\b}+\left(\begin{array}{c}
    \mu_\a \leftrightarrow \mu_\b \\
    r_1\leftrightarrow r_3
    \end{array}\right)~~~~~~~
    \nonumber\\
    &-\frac{32}{\left(1+4\mu_\a^2\right)(1+4\mu_\b^2)}\nonumber\\
    \nonumber\\
    \sim&~\mathcal{O}\left(e^{-\pi\mu_\b}\right)r_3^{-\frac{1}{2}+\i\mu_\b}+\mathcal{O}\left(e^{-\pi\mu_\a}\right)r_1^{-\frac{1}{2}+\i\mu_\a}+\rm{constant}~.
\end{align}
It is clear that each signal is accompanied by a corresponding Boltzmann factor. In the homogeneous part $\mathcal{Y}_{+++}$, both massive fields contribute to the CC signals, resulting in a total amplitude of approximately $\mathcal{O}(e^{-2\pi\mu})$. Regarding the particular solutions, some components exhibit a non-analytic structure along one direction, leading to a suppression of $\mathcal{O}(e^{-\pi\mu})$, while entirely analytic terms remain unaffected by any suppression.

\paragraph{Double folded limit.} 
As we mentioned previously, an essential consistency check is to consider the limit where both $r_i$ approach the unity. With the standard Bunch--Davis initial condition, the final result should be finite under this limit. Any spurious divergence present in each term of the analytical expression should cancel out. This is similar to the single massive exchange case, where the spurious divergence under the folded limit is also eliminated \cite{Arkani-Hamed:2018kmz}. Let us start from the simplest $\mathcal{I}_{+-+}$ case. After regrouping the result as in (\ref{Ipmp2}), it is then easy to take such a limit, where
\begin{align}\label{IpmpFolded}
    &\lim_{r_3\rightarrow 1}\mathcal{I}_{+-+}(r_1,r_3)\nonumber\\
    &=\sum_{m=0}^{\infty} \i\,\frac{\pi^{2}}{2}\frac{{\rm{\csch}}(2\pi \mu_\a)\,{\rm{sech}}(\pi\mu_\b)\left(r_1-1\right)^3}{(1+2m+2\i\mu_\a)(3+2m+2\i\mu_\a)}\Gamma
    \left[\begin{array}{c}
    2+m+\i\mu_\a-\i\mu_\b,2+m+\i\mu_\a+\i\mu_\b \\
    1+m,1+m+2\i\mu_\a
    \end{array}\right]\times r_1^{-\frac{1}{2}+m+\i\mu_{\a}}\nonumber\\
    &~~~~~~~~+\big(\mu_\a\rightarrow-\mu_\a\big)\nonumber\\
    &=-\i{\pi^2}{\rm{\csch}}(2\pi \mu_\a)\,{\rm{sech}}(\pi\mu_\b)\left(\frac{r_1}{2}-\frac{1}{2}\right)^3 r_1^{-\frac{1}{2}-\i\mu_\a}\times{{}_3}\mathcal{{F}}_2\left[\begin{array}{c|c}
    \frac{1}{2}-\i\mu_\a, 2-\i\mu_\a-\i\mu_\b, 2-\i\mu_\a+\i\mu_\b \\
    \frac{5}{2}-\i\mu_\a, 1-2\i\mu_\a
    \end{array} \ r_1\right]\nonumber\\
    &~~~~~~~~+\big(\mu_\a\rightarrow-\mu_\a\big)~.
\end{align}
Clearly, there is no divergence as $r_3$ approaches unity. Regarding the limit as $r_1$  tends to one, one simply needs to sum over the $r_1$ series first in equation (\ref{Ipmp}) and then take the limit, resulting in a similar expression. The double-folded limit $r_1, r_3 \rightarrow 1$ can also be easily verified based on our results \eqref{IpmpFolded}, which is 
\begin{align}
     &\lim_{r_1, r_3\rightarrow 1}\mathcal{I}_{+-+}(r_1,r_3)=-\left(\frac{k_{24}}{k_1}\right)^3
     \pi^2\mu\, {\rm{\csch}^2(2\pi\mu)\,\rm{\sinh}(\pi\mu)}~.
\end{align}
This is $\mathcal{O}(e^{-3\pi\mu})$ and here for simplicity, we assume equal masses for both massive exchanges ($\mu_\a=\mu_\b=\mu$).
The prefactor originates from $(r_1+r_3-2)^3$ in the initial expression (\ref{Ipmp2}), and we have used the relation $r_i=2k_i/k_{1234}$. As for the seed integral $\mathcal{I_{++-}}$, through the same procedural steps, we observe the cancellation of spurious poles upon summing all contributions. Consequently, the finite terms under the double folded limit $r_1,r_3\rightarrow 1$ become
\begin{align}
     &\lim_{r_1, r_3\rightarrow 1}\mathcal{I}_{++-}(r_1,r_3)=-\left(\frac{k_{24}}{k_1}\right)^3\left[\frac{\pi^2}{32}\left(4\mu^2-3\right)\sech^2(\pi\mu)+\i\frac{\pi^2\mu}{2}\csch(2\pi\mu)\right]~,
\end{align}
which is $\mathcal{O}(e^{-2\pi\mu})$. The other partially factorised partially nested integral $\mathcal{I}_{-++}$ behaves in the same way. For the fully nested part $\mathcal{I}_{+++}$ (\ref{Ipppresult}), it can be verified that the poles under the limit $r_3\rightarrow 1$ cancel each other out order by order in the summation. However, expressing the general form for the finite terms is challenging, making it difficult to derive a closed formula under the double folded limit.  Instead, as an illustrative example and for the purpose of subsequent application, we can consider the limit relevant for three-point correlators. Then, under the squeezed limit, $\tilde{r}_3\rightarrow 1$ and $\tilde{r}_1\rightarrow 0$, therefore we only need to keep the first order in the summation (\ref{Ipppresult}) as we will demonstrate below.

\paragraph{Squeezed limit.} 
Another intriguing limit that will play significant roles in the subsequent discussion is the so-called squeezed limit where $\tilde{r}_3({r_3})\rightarrow 1$ and $\tilde{r}_1 (r_1) \rightarrow 0$. For the fully factorised seed integral, the result under ${r}_3\rightarrow 1$ is already obtained in (\ref{IpmpFolded}) and we only need to take the series expansion around $r_1=0$ which gives
\begin{align}
    \lim_{{\substack{r_3\to 1 \\ r_1\to 0}}}\mathcal{I}_{+-+}(r_1,r_3)=\frac{\pi^2}{4}\frac{\sech^2(\pi\mu)\csch(\pi\mu)}{3\i-2\mu}\times r_1^{-\frac{1}{2}+\i\mu}+\big(\mu\rightarrow-\mu\big)~,
\end{align}
with an amplitude $\mathcal{O}(e^{-3\pi\mu})$.
For two partially factorised  partially nested integrals, following the same way, the results are
  \begin{align}
    \lim_{{\substack{r_3\to 1 \\ r_1\to 0}}}\mathcal{I}_{++-}(r_1,r_3)=&~\frac{\pi^2}{2}\frac{\big[1-\tanh(\pi\mu)\big]\big[1+\coth(2\pi\mu)\big]}{3+2\i\mu}\times r_1^{-\frac{1}{2}+\i\mu}+\big(\mu\rightarrow-\mu\big)~,\\
     \lim_{{\substack{r_3\to 1 \\ r_1\to 0}}}\mathcal{I}_{-++}(r_1,r_3)=&\left(\frac{\pi^2}{4}\frac{\sech^2(\pi\mu)\big[1+\coth(\pi\mu)\big]}{2\i\mu-3}+\frac{\pi^{1/2}}{2^{4-2\i\mu}}\frac{2\mu+5\i}{2\mu+3\i}\,\Gamma\left[\frac{1}{2}-\i\mu,\i\mu\right]\sech(\pi\mu)\right)\times r_1^{-\frac{1}{2}-\i\mu}\nonumber\\
    &+\big(\mu\rightarrow-\mu\big)~,
\end{align} 
where both of them have an amplitude $\mathcal{O} (e^{-2\pi\mu
})$. For the fully nested seed integral $\mathcal{I}_{+++}$,  as mentioned previously, under $r_1 \rightarrow 0$, we will only retain the first order in the series summation. It is straightforward to show that the poles  that would diverge as $r_3 \rightarrow 1$ are canceled, yielding the final finite contributions
\begin{align}\label{singlesoft}
    \lim_{{\substack{r_3\to 1 \\ r_1\to 0}}}\mathcal{I}_{+++}(r_1,r_3)=& \Bigg(\frac{\pi^2}{2}\frac{\rm{csch}(2\pi\mu)\rm{sech}(\pi\mu)}{2 \mu-3\i}+\frac{\pi^{1/2}}{2^{4+2\i\mu}}\frac{2\i\mu+5}{2\mu-3\i}\Gamma\left[\frac{1}{2}+\i\mu,-\i\mu\right]\big(1+{\rm{tanh}} (\pi\mu)\big)\Bigg)r_1^{-\frac 1 2+\i\mu}\nonumber\\
    &+\big(\mu\rightarrow-\mu\big)~,
\end{align}
where the first term arises from the homogeneous solution and is suppressed by $\mathcal{O}(e^{-3\pi\mu})$.
In contrast, the second term comes from one of the particular solutions, wherein only one massive propagator contributes to the signal and the other massive line lies within the analytic region and contributes to the background. Hence, the overall magnitude of the second term is only suppressed by $\mathcal{O}(e^{-\pi\mu})$.
\\
For the bispectrum in the squeezed limit, the dominant signal arises from the CC signals, with the leading contribution experiencing the least Boltzmann suppression from $\mathcal{I}_{+++}$. Substituting this expression into (\ref{Bispectrum1}), we obtain an approximate result for the three-point correlator in the squeezed limit which is useful for later discussion
\begin{keyeqn}
\begin{align}
\label{squeezedlimitapprox}
    &\lim _{k_1 \rightarrow 0}\langle\varphi_{\bf{k}_1} \varphi_{\bf{k}_2}\varphi_{\bf{k}_3}\rangle '\\
    &=-\frac{\rho^2\lambda H}{ (k_1 k_2 k_3)^2}\cdot{\rm{Re}}\Bigg\{
     \left[
    \frac{\pi^{1/2}}{2^{4+2\i\mu}}\frac{2\i\mu+5}{2\mu-3\i}\Gamma\left[\frac{1}{2}+\i\mu,-\i\mu\right]\big(1+{\rm{tanh}}(\pi\mu)\big)
    +\mathcal{O}\left(e^{-2\pi\mu}\right)
    \right]
    \left(\frac{k_1}{k_3}\right)^{\frac{1}{2}+\i\mu}+\mathcal{O}\left(\frac{k_1}{k_3}\right)\Bigg\}~,\nonumber
\end{align}
\end{keyeqn}
where $\rho_\a$ and $\lambda_{\a\b}$ are reduced to $\rho$ and $\lambda$ for the equal mass case, and the $\mathcal{O}(k_1/k_3)$ term arises from the leading analytical background contribution. Since $\mathcal{I}_{+++}$ dominates in the squeezed limit, and because we will also discuss the CC signals with multiple isocurvature species in the next section, for later convenience, here we also show the expression when two propagators have different masses explicitly:
\begin{align}
    \lim_{{\substack{r_3\to 1 \\ r_1\to 0}}}\mathcal{I}_{+++}(r_1,r_3)=&-\i\frac{\pi^{3/2}}{2^{4+2\i\mu_\a}} \sech( \pi\mu_\b )\big[1+\tanh(\pi\mu_\a)\big]\times\Gamma
    \left[\begin{array}{c}
    -\i\mu_\a\\
     -1-\i\mu_\a+\i\mu_\b, -1-\i\mu_\a-\i\mu_\b 
    \end{array}\right]\nonumber\\
    &\times r_1^{-\frac{1}{2}+\i\mu_\a}{_3}\mathcal{F}_2\left[\begin{array}{c|c}
    -\frac{3}{2}-\i\mu_\a,-1-\i\mu_\a-\i\mu_\b,{-1-\i\mu_\a+\i\mu_\b}\\
    -\frac{1}{2}-\i\mu_\a,-\frac{1}{2}-\i\mu_\a 
    \end{array} \ 1\right]+\big(\mu_\a\rightarrow-\mu_\a\big)~,
\end{align}
and the three-point function in the squeezed limit becomes:
\begin{keyeqn}
\begin{align}
\label{squeezedlimitapprox2}
    &\lim _{k_1 \rightarrow 0}\langle\varphi_{\bf{k}_1} \varphi_{\bf{k}_2}\varphi_{\bf{k}_3}\rangle '\\
    &=\sum_{\a,\b}\frac{\rho_\a\rho_\b\lambda_{\a\b} H}{ (k_1 k_2 k_3)^2}\cdot{\rm{Re}}\Bigg\{
     \bigg[
    \i\frac{\pi^{3/2}}{2^{4+2\i\mu_\a}} \sech( \pi\mu_\b )\big[1+\tanh(\pi\mu_\a)\big]\times\Gamma
    \left[\begin{array}{c}
    -\i\mu_\a\\
     -1-\i\mu_\a+\i\mu_\b, -1-\i\mu_\a-\i\mu_\b 
    \end{array}\right]\nonumber\\
    &\times {_3}\mathcal{F}_2\left[\begin{array}{c|c}
    -\frac{3}{2}-\i\mu_\a,-1-\i\mu_\a-\i\mu_\b,{-1-\i\mu_\a+\i\mu_\b}\\
    -\frac{1}{2}-\i\mu_\a,-\frac{1}{2}-\i\mu_\a 
    \end{array} \ 1\right]
    +\mathcal{O}\left(e^{-2\pi {\mu_\a}},e^{-2\pi {\mu_\b}}\right)
    \bigg]
    \left(\frac{k_1}{k_3}\right)^{\frac{1}{2}+\i{\mu_\a}}+\mathcal{O}\left(\frac{k_1}{k_3}\right)\Bigg\}~.\nonumber
\end{align}
\end{keyeqn}

\section{Double Massive Exchange Phenomenology}\label{Sec_pheno}

In this section, we discuss the physical effects that can be read off our analytical solution, Eqs.~\eqref{Ipmp2},~\eqref{Ippm} and~\eqref{Ipppresult}.
We also compare it to the predictions from different techniques, namely the effective single-field description when the massive fields are integrated out and an exact numerical evolution that does not rely on a perturbative scheme in terms of quadratic mixings.
In the subsequent sections, we explore different regimes of interest.
Before that, we start by presenting the motivation for the setup and these complementary techniques.

\paragraph{Field content, interactions and relations to multifield inflation.}
For almost all concrete applications of this section, we focus on a simple field content made of the inflaton's massless fluctuations $\varphi$ and a \textit{single} extra massive scalar field $\sigma$ (see an exception in Sec.~\ref{subsec:disentangling}).
These are already mixing at the level of the quadratic interactions, as encoded by Eq.~\eqref{int_2} with $\rho_\alpha = \rho \times \delta_{\alpha 1}$ and $\sigma_1 = \sigma$.
Additional cubic interactions leading to double-exchange diagrams that we consider are described by Eq.~\eqref{int_3} with $\lambda_{\alpha\beta}=\lambda \times \delta_{\alpha 1}\delta_{\beta 1}$ and are leading to a three-point correlator of $\varphi$ given by the double-exchange seed integral in the limit $k_4 \rightarrow 0$ and with $(p_1,p_2,p_3)=(-2,-2,-2)$, see Eq.~\eqref{Bispectrum1}.
Considering these particular interactions is motivated by different though complementary approaches to multifield inflation, as well as by the potential of detectability of this channel.

First, popular multifield realizations of the inflationary paradigm, called general non-linear sigma models and encoding both potential and kinetic interactions, generically predict their existences.
In this two-field context, the quadratic mixing $\rho$ can be related to the dimensionless rate of turn of the background trajectory in field space, $\eta_\perp$, as $\rho=2 H\eta_\perp$, while the cubic interaction $\lambda$ is given as a combination of the turn rate and the scalar curvature of the field space of mass dimension $-2$, $R_\mathrm{fs}$, as $\lambda =(H^2 \eta_\perp^2 - \epsilon H^2 M_\mathrm{Pl}^2 R_\mathrm{fs})/f_\pi^2$~\cite{Garcia-Saenz:2019njm}.
In this expression, $f_\pi^2$ represents the normalisation of the curvature fluctuation with respect to the massless inflaton's fluctuations $\varphi = -(f_\pi^2/H) \times \zeta$.
Therefore, in the weak quadratic mixing regime where the power spectrum of $\varphi$ is negligibly affected by its interactions with the massive fields, one can relate the typical size of the curvature fluctuation $\Delta_\zeta$ to the one of a massless uncoupled field in de Sitter $\Delta_\varphi = H/(2\pi)$ as
\begin{equation}
\Delta_\zeta \equiv A_s^{1/2} = \frac{H}{f_\pi^2} \Delta_\varphi = \frac{H^2}{2 \pi f_\pi^2} \,,
\end{equation}
with $A_s = 2.1 \times 10^{-9}$~\cite{Planck:2018jri} at CMB scales.
This equality enables one to express $f_\pi^2$ as a large number in units of $H^2$.
Actually, for more than one additional massive field $\sigma_\alpha$, the form of the interactions depends on the choice of the basis chosen for these fluctuations, leading to a subtle interplay between flavor and mass bases eigenstates as described in Ref.~\cite{Pinol:2021aun}.
In the latter reference, the form of the interactions in the mass basis is specified, from the knowledge of the ones in the flavor basis in the context of non-linear sigma models with any number of fields, first derived in Ref.~\cite{Pinol:2020kvw}.
We will explore this more generic situation in Sec.~\ref{subsec:disentangling}.

Second, generic arguments from the effective field theory of multifield fluctuations tell us that we have to entertain the possibility of these interactions.
In this language, the adiabatic curvature fluctuation $\zeta$ is related by a gauge transformation to the Goldstone boson of broken time diffeomorphisms, $\pi$~\cite{Creminelli:2006xe,Cheung:2007st}.
At the linear level and when defining the Goldstone boson in the flat gauge, this gauge transformation reads $\zeta = - H \pi + \ldots$, where dots stand for order two and more terms in fluctuations.
One can then couple it to either massless~\cite{Senatore:2010wk} or massive~\cite{Noumi:2012vr} additional scalar fields $\sigma^\alpha$.
In the unitary gauge, the adiabatic degree of freedom is hidden in the spacetime metric fluctuation $\delta g^{00}$, that can couple to these additional fields via couplings of the form $b^{(n)}_\alpha (\delta g^{00})^n \sigma^\alpha\,,\,\, c^{(n)}_{\alpha \beta} (\delta g^{00})^n \sigma^\alpha \sigma^\beta  $, etc.
Those couplings generically lead to interactions of the forms that we considered above (amongst other ones)~\cite{Pinol:2023oux}, after the Stückelberg trick reintroducing the Goldstone boson via, e.g., $\delta g^{00} \rightarrow - 2 \dot{\pi} + \partial_\mu \pi \partial^\mu \pi$.

Third, the focus of this work is on double-exchange diagrams, where two massive field propagators are exchanged at tree level via the existence of a cubic interaction with \textit{two} powers of $\sigma$'s, motivated by its potential of detectability.
Indeed, simpler single-exchange bispectra with a single power of $\sigma$ in the cubic interaction generically lead to a small observable signal.
First, the ``Lorentz-covariant'' combination $\propto \rho \times \partial_\mu \varphi \partial^\mu \varphi \sigma$
from the unitary gauge operator $\rho \times \delta g^{00} \sigma$ has a strength fixed by symmetries to be proportional to the quadratic mixing $\rho$; the latter being treated in perturbation theory in most analytical calculations, this enforces the corresponding bispectrum to be small.
Even in the less explored regime of a large quadratic mixing, see Refs.~\cite{An:2017hlx,Iyer:2017qzw,Werth:2023pfl,Jazayeri:2023xcj}, this channel can at most lead to non-Gaussianities of order roughly unity~\cite{Pinol:2023oux}.
Second, although from the effective field theory point of view there may exist another cubic vertex leading to a single-exchange bispectrum with a size independent from the quadratic mixing (of the form $ \propto \tilde{\rho} \times (\varphi^\prime)^2 \sigma$ with $\tilde{\rho}$ potentially large), this interaction is not found in the large class of general non-linear sigma models of inflation.
On the contrary, as we have just argued in the two previous paragraphs, cubic interactions leading to double-exchange diagrams are both natural to consider from the effective field theory point of view and found in practice in concrete multifield realizations of inflation.
The size of this interaction is not fixed by the small quadratic mixing and may therefore be large while respecting the perturbative scheme used for the derivation of our analytical results.
And indeed, it was shown that this channel could lead to potentially large non-Gaussianities both in the small and large mixing regimes~\cite{Pinol:2023oux}.

\paragraph{Perturbativity bound, naturalness and size of the bispectrum.}
Although we have just argued that the signal from the double-exchange diagram may be large, it cannot be arbitrary large.
First, cubic interactions naturally define strong coupling scales, ensuring that the perturbative expansion in powers of fields' fluctuations is a meaningful series.
To be precise, the cubic vertex responsible for the double-exchange diagram is particular in the sense that $\lambda \varphi^\prime \sigma^2$ is a marginal operator (of mass dimension 4 in 4 dimensions), so that the coupling $\lambda$ appearing in the time integrals is dimensionless and cannot be used to define a strong coupling scale.
However, for consistency of the perturbative expansion in a similar fashion to the interpretation of strong coupling scales, we should still ask that this cubic interaction remains smaller than the kinetic (and mass) terms used to define the free theory~\cite{Pinol:2023oux}, from which one can derive the perturbativity bound 
\begin{equation}
    \text{Perturbativity bound:} \quad |\lambda| \lesssim 1 \,.
\end{equation}
Additionally, one may want to ask the size of the cubic interaction to be natural in order to avoid requiring fine-tuning to explain the smallness of loop corrections.
In particular for the interaction under study, it was shown in Ref.~\cite{Pinol:2023oux} that asking a mass of $\sigma$ naturally of order $H$ in the effective field theory of multifield fluctuations framework, amounts to setting
\footnote{
More precisely, this naturalness criterion corresponds to asking that loop corrections to the mass of the heavy field, from such cubic interactions, are not larger than the bare mass $m$ used in the calculation.
For $m\sim H$, this gives the upper bound~\eqref{eq:naturalness}, see Ref.~\cite{Pinol:2023oux}.
Indeed, naturalness is a statement about limiting the amount of fine-tuning required to explain the value for a quantity given by the algebraic sum of several terms.
In the reasoning above, the fine-tuning for a mass of order Hubble with a large $\lambda$ is clear: both the bare mass and the loop corrections should be large and almost cancel to give a final value $\sim H$. 
Fine-tuning for a large $\lambda$ can also be seen by specifying to concrete models of inflation.
For the large class of non-linear sigma models, the masses of isocurvature fluctuations go as~\cite{Garcia-Saenz:2019njm, Pinol:2020kvw} $m^2/H^2= -\lambda f_\pi^2 / H^2+V_{; s s} / H^2$, where $V_{; s s}$ represents consistent projections of the Hessian matrix of the potential.
Clearly, having a large $\lambda$ while maintaining $m\sim H$ requires some degree of fine-tuning.
Because the corresponding naturalness criterion, $|\lambda|\lesssim 2\pi\Delta_\zeta $ for a weak quadratic mixing, is (slightly) more model-dependent, we will not use it in the following.
}
\begin{equation}
\label{eq:naturalness}
    \text{Naturalness criterion:} \quad |\lambda|\lesssim (2\pi\Delta_\zeta)^{1/2}\,.
\end{equation}
Given the observed amplitude of curvature fluctuations at CMB scales, the bound from naturalness is $\mathcal{O}(100)$ times more constraining than the one from perturbativity.
Note however that the status of the naturalness criterion is less firm,
as asking naturalness is not a prerequisite for a theory to be well-behaved.
Indeed, (approximate) cancellations may be explained by (approximate) symmetries, accidental or not, or simply by an appropriate amount of fine-tuning.
On the contrary, perturbativity is not negotiable in our framework, and we will always enforce it.

One can read the parametric dependence of the bispectrum for the inflaton's fluctuations $\varphi$, e.g. in Eq.~\eqref{squeezedlimitapprox}, and translate it in terms of the one for the curvature fluctuation.
More precisely, we will be using the dimensionless bispectrum shape function $S$, related to the bispectrum of $\varphi$, in the weak mixing regime where $\Delta_\varphi = H/(2\pi)$, as follows:
\begin{equation}
    \langle\varphi_{\bf{k}_1} \varphi_{\bf{k}_2}\varphi_{\bf{k}_3}\rangle'  = - \frac{H^5}{f_\pi^2 (k_1 k_2 k_3)^2} S(k_1,k_2,k_3)\,.
\end{equation}
Without entering yet into the complicated mass and kinematical dependence of the entire shape function, we read its parametric dependence related to the size of the usual $f_\mathrm{NL}$'s parameters: 
\begin{equation}
\label{naivescalingofS}
    S \sim f_\mathrm{NL} \sim  (\Delta_\zeta)^{-1} \times (\rho/H)^2 \times \lambda  \,,
\end{equation}
where we used that $f_\pi^2$ is related to $\Delta_\zeta^{-1}$ in the regime of validity of our analytical calculations.
This confirms that, even in the weak mixing regime with $\rho/H \ll 1$ and respecting the perturbativity bound $|\lambda| \lesssim 1$, it is possible to have non-Gaussianities of order one or more thanks to the large factor $ (\Delta_\zeta)^{-1} \sim \mathcal{O}(10^4)$.
If one adds the requirement of naturalness, it is still possible to get order one non-Gaussianities by pushing towards intermediate values of the quadratic mixing, $\rho / H \sim \mathcal{O}(10^{-1})$, or even beyond (a regime not encompassed by our analytical approach from the bootstrap perspective but nevertheless perfectly viable theoretically).

\paragraph{Single-field effective theory.} When the mass of the additional fluctuation $\sigma$ is large enough, it may be integrated out of the theory~\cite{Tolley:2009fg,Cremonini:2010ua,Achucarro:2010da}, leading to an effectively single-field theory for the curvature fluctuation only.
This procedure including all cubic interactions for two-fields non-linear sigma models was performed in~\cite{Garcia-Saenz:2019njm} and for any number of fields in~\cite{Pinol:2023oux}.
Repeating the procedure only for the cubic interaction considered here, we find the following single-field effective theory in terms of the curvature fluctuation directly:
\begin{align}
    \mathcal{L}_\mathrm{EFT} &=  \frac{a^3 \epsilon M_{\mathrm{Pl}}^2}{c_s^2}\left[ \dot{\zeta}^2 - c_s^2 \frac{(\partial \zeta)^2}{a^2} + \left( \frac{1}{c_s^2}-1\right) A \frac{\dot{\zeta}^3}{H}\right]\,, \quad\text{with} \\
    \frac{1}{c_s^2}-1 &=\left(\frac{\rho}{H}\right)^2 \times \frac{H^2}{m^2} \quad \text{and} \quad A = -2 c_s^2 \times \frac{\lambda f_\pi^2}{H^2} \times \frac{H^2}{m^2} \,.
\end{align}
Within this single-field theory, it is straightforward to compute the bispectrum shape function in all kinematical regimes:
\begin{equation}
    S_\mathrm{EFT}(k_1,k_2,k_3) = \frac{3A}{2}\left( \frac{1}{c_s^2}-1\right)\frac{k_1k_2k_3}{k_{123}^3} \,,
\end{equation}
where $k_{123}=k_1+k_2+k_3$.
This shape not only peaks on equilateral configurations but is also very correlated with the equilateral shape template, consisting of:
\begin{equation}
    S_\mathrm{eq}(k_1,k_2,k_3)= \left(\frac{k_1}{k_2} +\text{5 perms}\right)-\left(\frac{k_3^2}{k_1 k_2}+ \text{2 perms}\right) -2 \,.
\end{equation}
Indeed, defining as usual the cosine between two shapes as their correlation\cite{Babich:2004gb,Fergusson:2008ra}
\begin{align}
\label{eq: Cosine def}
    \mathrm{Cos}\left(S_1,S_2\right) &= \frac{S_1 \cdot S_2}{\sqrt{(S_1 \cdot S_1)(S_2 \cdot S_2)}} \quad \text{with} \\ 
    S_1 \cdot S_2 &= \int_{1/2}^1 \dd x_2 \int_{1-x_2}^{x_2} \dd x_3 \, S_1(k, x_2\times k,  x_3\times k) S_2(k, x_2\times k,  x_3\times k) \,,
\end{align}
one finds $\mathrm{Cos}\left(S_\mathrm{EFT},S_\mathrm{eq}\right) \simeq 0.988 $.
We thus safely dub its amplitude at equilateral configurations the parameter $f_\mathrm{NL}^\mathrm{eq}$ that one can look for in the data with an equilateral shape template.
To put it in a nutshell, the single-field effective theory for the double-exchange diagram predicts an equilateral bispectrum shape with amplitude
\begin{equation}
\label{fNLeq}
    f_\mathrm{NL}^\mathrm{eq}  = - \frac{1}{9} \left(\frac{\rho}{H}\right)^2 \frac{1}{1+\rho^2/m^2}\times \frac{\lambda f_\pi^2}{H^2}\times \frac{H^4}{m^4} \,.
\end{equation}
Note that this single-field EFT does not rely on the assumption of a weak quadratic mixing and is therefore valid for values of $\rho/H$ larger than unity.
In the regime of a weak quadratic mixing though (that we read actually extends up to $\rho \lesssim m$), we recover the scaling with parameters naively derived in Eq.~\eqref{naivescalingofS}, together with additional information in this regime: the kinematical dependence of the shape is dominantly equilateral and the exact prefactor including mass dependence is $-H^4/(18 \pi m^4)$.
Note however that the single-field effective theory relies on the hierarchy $m^2 c_s^2 \gg H^2$~\cite{Garcia-Saenz:2019njm}, and therefore implicitly also on the absence of any other hierarchy in the theory.
A subtlety comes about when looking at soft limits of correlation functions, at which there is a kinematical hierarchy.
As far as the bispectrum is concerned, the unique soft limit is the squeezed one.
Let us fix $k_3$ the smallest momentum, in the squeezed limit the ratio $k_3/k_1$ is very small and, therefore, one may well have $(k_3/k_1) \times (m^2 c_s^2/H^2) \lesssim 1$ although the field $\sigma$ is very massive.
Said otherwise, we expect the single-field description to fail to capture not only the region of parameter space with masses of order $H$ or smaller, but also the kinematical region corresponding to soft limits.
As well known, these soft limits indeed encode striking signatures of particle production which cannot be described within the single-field framework.

\paragraph{Exact numerical evolution.} The double-exchange bispectrum diagram may also be computed exactly from first principles by computing its time evolution in the bulk of the inflationary spacetime.
It is precisely the aim of the so-called Cosmological Flow~\cite{Werth:2023pfl,Pinol:2023oux} to propose a framework for automatically solving the differential equations verified by the different two- and three-point correlators of any inflationary theory defined at the level of fluctuations.
The numerical implementation of this approach is called \textsf{CosmoFlow}, it is open source and available on \href{https://github.com/deniswerth/CosmoFlow}{GitHub} and the paper describing it is Ref.~\cite{Werth:2024aui}.
It is straightforward to implement the interactions leading to the double-exchange bispectrum diagram within it, and in the coming sections we use it as a non-trivial check of our analytical results.
Indeed, importantly, the Cosmological Flow approach does not rely on a perturbative scheme for the quadratic mixing.
Strictly speaking, it therefore does not solve for the same theory as the one described in this work, however in the common regime of validity we should expect both bootstrap analytical predictions and numerical calculations to agree.

\vspace{10pt}
In the next sections, we explore the parametric and kinematical dependencies of the double-exchange bispectrum.
We then compare them to the ones of the single-exchange channel. 

\subsection{Equilateral value and mass dependence}

We start by exploring the size of the signal on the equilateral configuration $k_1=k_2=k_3$ and its dependence on the mass of the double exchanged field.
At this particular point in kinematical space, we expect the single-field effective description to be valid as soon as $m^2 c_s^2/H^2 \gg 1$.
In practice we will focus on masses $m \gtrsim H$ and on a quadratic coupling $\rho/ H $ smaller than unity as required by consistency of the bootstrap approach, a situation for which $c_s^2 \simeq 1$.

\begin{figure}[ht!]
  \centering
  \subfloat{\includegraphics[width=0.5\textwidth]{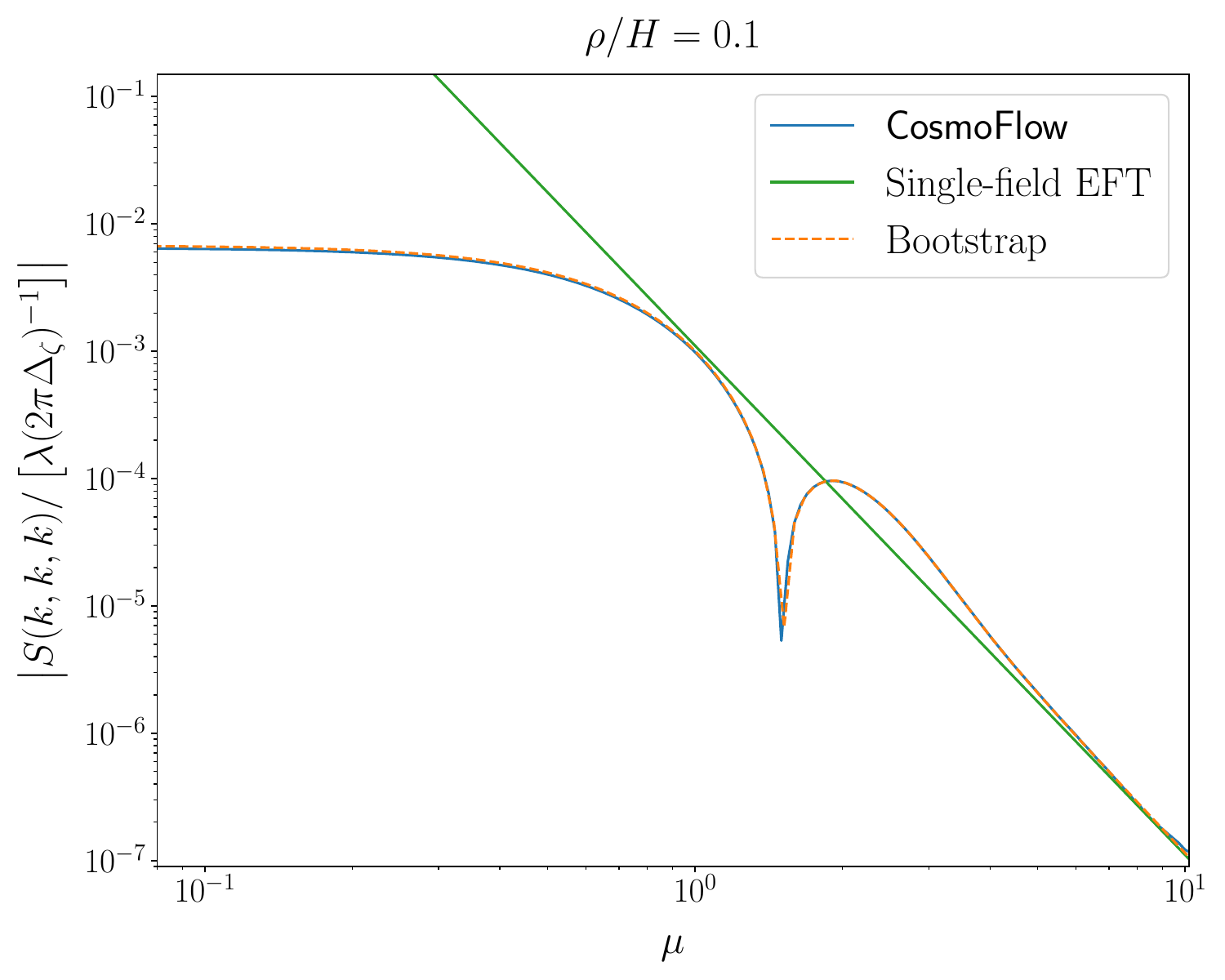}}
  \hspace*{0.2cm}
  \subfloat{\includegraphics[width=0.5\textwidth]{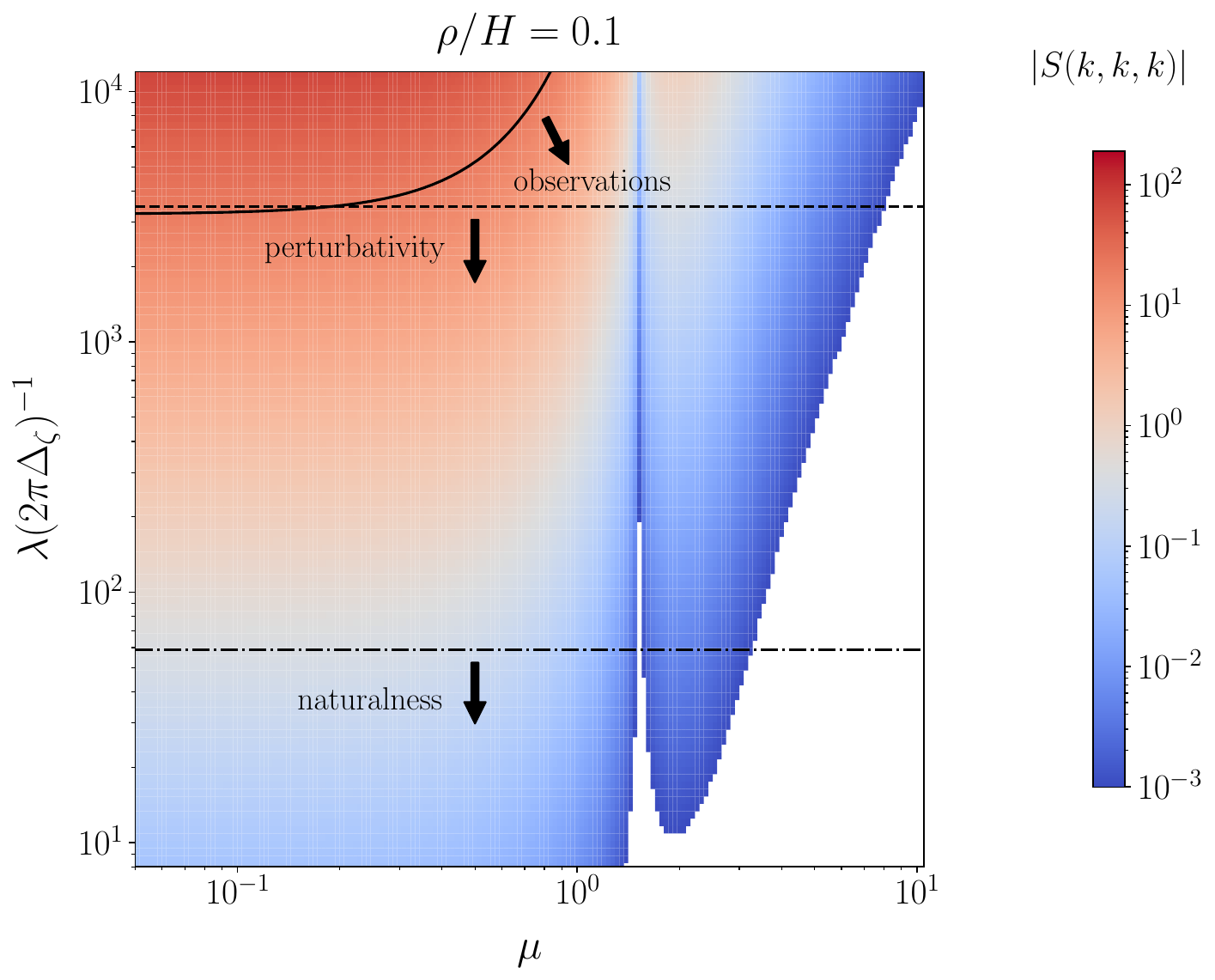}}
  \vspace*{0.2cm}
  \caption{
  \textbf{Left:} Comparison of the values of the double-exchange bispectrum at equilateral configurations from the exact numerical resolution using \textsf{CosmoFlow} (blue line), the bootstrap analytical prediction of this work (orange dashed line), as well for the corresponding single-field effective theory (green line).
   The quadratic coupling is fixed as $\rho=0.1 H$ and the mass of the double exchanged field varies along with the horizontal axis.
   The time evolution with \textsf{CosmoFlow} starts being numerically more demanding for larger values of $\mu$ so it has been cut at $\mu \sim 10$ for purely computational costs reasons.
   Note that the overall amplitude has been rescaled by the constant factor $\lambda(2\pi \Delta_\zeta)^{-1}$ that may be way larger than unity, leading to potentially large primordial non-Gaussianities even in this regime of a small quadratic mixing.
   \textbf{Right:} True size of the double-exchange bispectrum at the equilateral configuration for $\rho/H=0.1$ as a function of the mass parameter $\mu$ and the coupling constant $\lambda$.
   White regions correspond to unobservably small values $< 10^{-3}$ removed for clarity.
   The contour line ``observations'' (solid line) stands for the $1\sigma$ constraint on $f_\mathrm{NL}^\mathrm{eq}$ from Planck~\cite{Planck:2019kim}, while the other ones are theoretical constraints.
   The perturbativity one (dashed line) is necessary to enforce, while the naturalness one (dotted-dashed line) may or not be required.
   The vertical valley visible by eye corresponds to the dip in the left panel at which the bispectrum equilateral value changes from positive to negative, from left to right in the figure, when $\lambda > 0$.
  }
  \label{fig:fNLeq}
\end{figure}

In the left panel of Fig.~\ref{fig:fNLeq}, we compare the analytical prediction from the multifield bootstrap approach with the single-field effective theory description of Eq.~\eqref{fNLeq} and the complete numerical resolution not relying on a perturbative quadratic mixing with \textsf{CosmoFlow}, for $\rho/ H = 0.1$.
The agreement between the two exact methods in this regime is impressive.
As an example, both predict a change from positive to negative values with $\mu$ growing, precisely at the same threshold $\mu \simeq 1.5$.
As expected, the single-field effective theory prediction is only faithful at large values of the mass, roughly $\mu \gtrsim 5$, and always predicts a negative bispectrum at the equilateral configuration.
In this regime of a small quadratic mixing and a large mass, a simple fitting formula for our bootstrap result can be derived,
\begin{equation}
    S_\mathrm{bootstrap}(k,k,k) \underset{\mu \gg 1}{\simeq} - 0.11 \times \left(\frac{\rho}{H} \right)^2 \times \frac{\lambda f_\pi^2}{H^2} \times \frac{1}{\mu^4} \,, 
\end{equation}
which precisely matches the single-field effective theory result~\eqref{fNLeq} with the same numerical coefficient $-1/9 \simeq -0.11 $ (we remind that $\mu = \sqrt{m^2/H^2-9/4} \simeq m/H$ in the regime of a large mass).

In the right panel of Fig.~\ref{fig:fNLeq}, we show the true size of the bispectrum at equilateral configurations for $\rho/ H=0.1$ as a function of the mass parameter $\mu$ and the cubic coupling constant $\lambda$.
We compare the result to observational constraints from Planck, $f_\mathrm{NL}^\mathrm{eq}=-26\pm47$ at $1\sigma$~\cite{Planck:2019kim}, by assuming that the bispectrum can be qualitatively constrained with an equilateral shape, an assumption to which we will come back in the next section.
We find instructive to see that the necessary theoretical constraint from perturbativity is close to current observational constraints.
We also show how, if one additionally asks for naturalness, this reduces even more the possible size of the bispectrum to less than unity.

\subsection{Shape dependence}

\begin{figure}[tp!]
  \centering
  \vspace{-1.25cm}
  \subfloat{\includegraphics[width=0.49\textwidth]{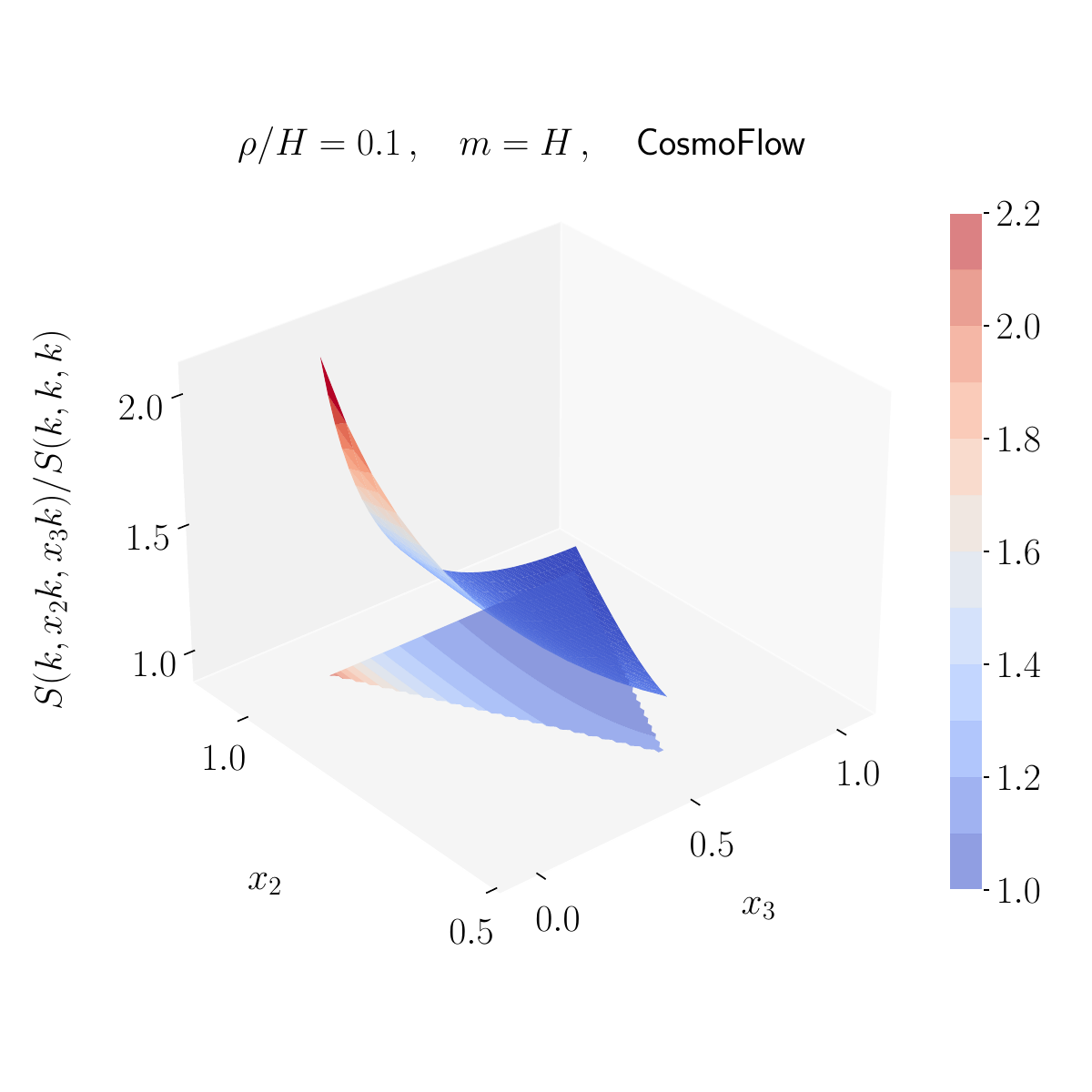}}
  \subfloat{\includegraphics[width=0.49\textwidth]{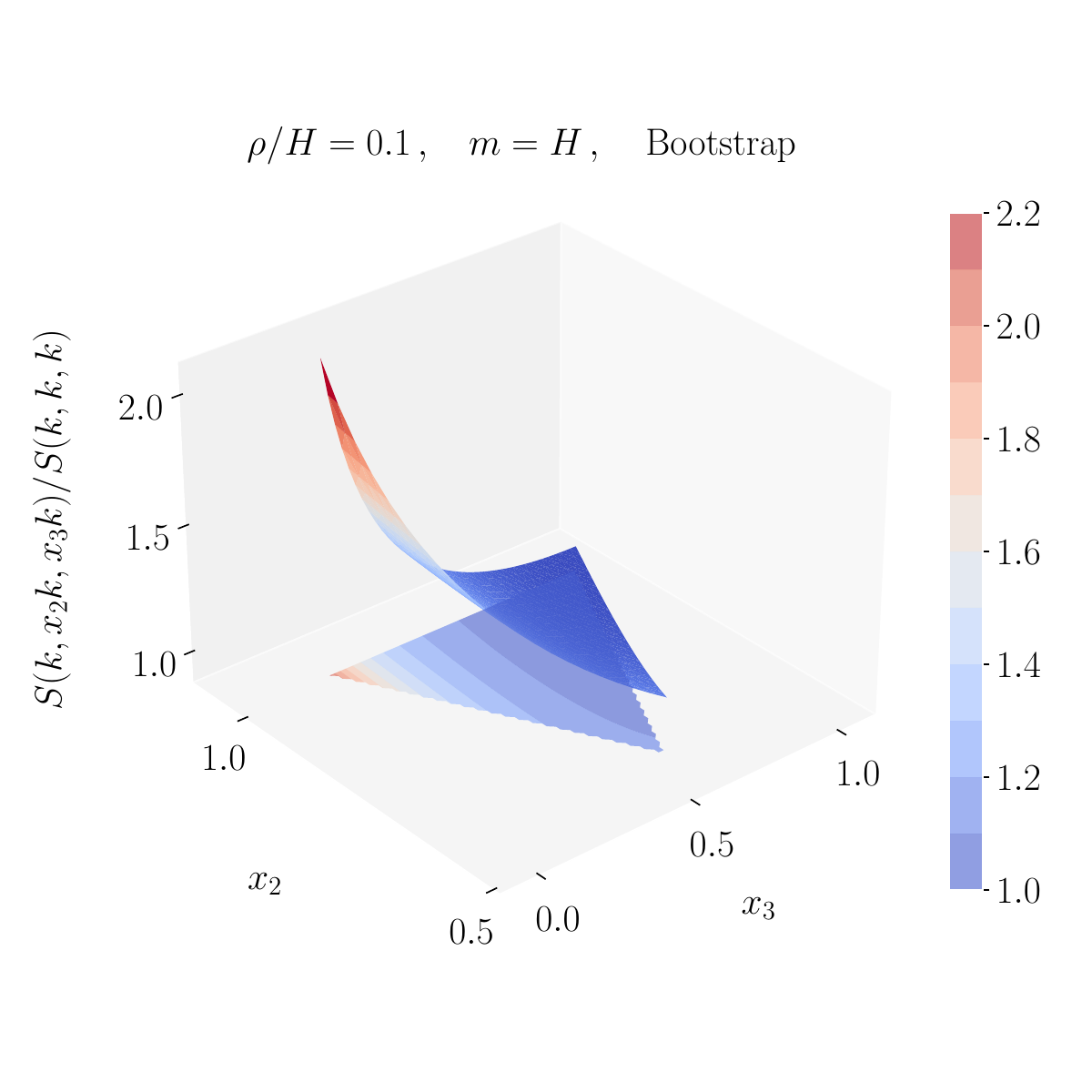}} \\
\vspace{-1.2cm}
  \subfloat{\includegraphics[width=0.49\textwidth]{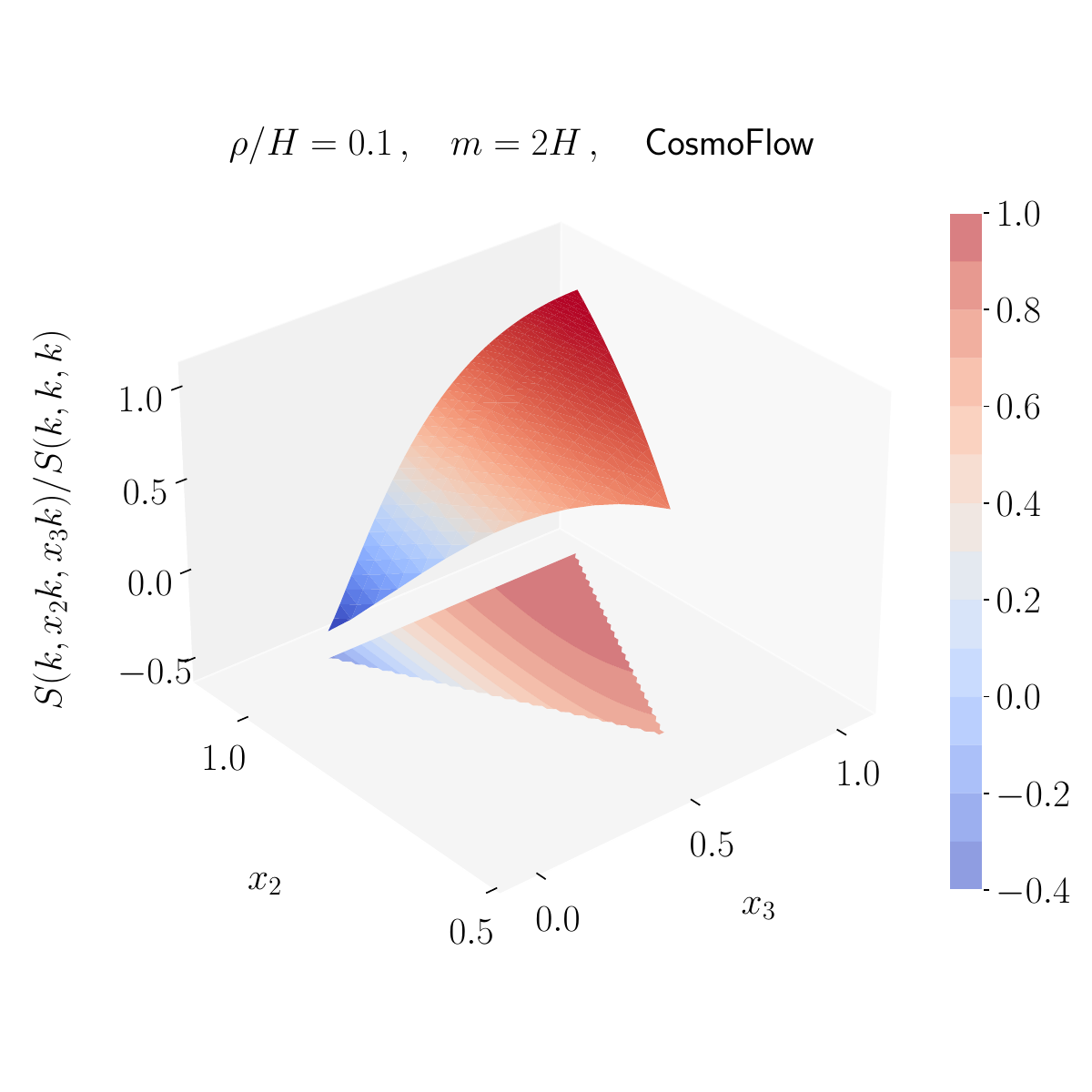}}
  \subfloat{\includegraphics[width=0.49\textwidth]{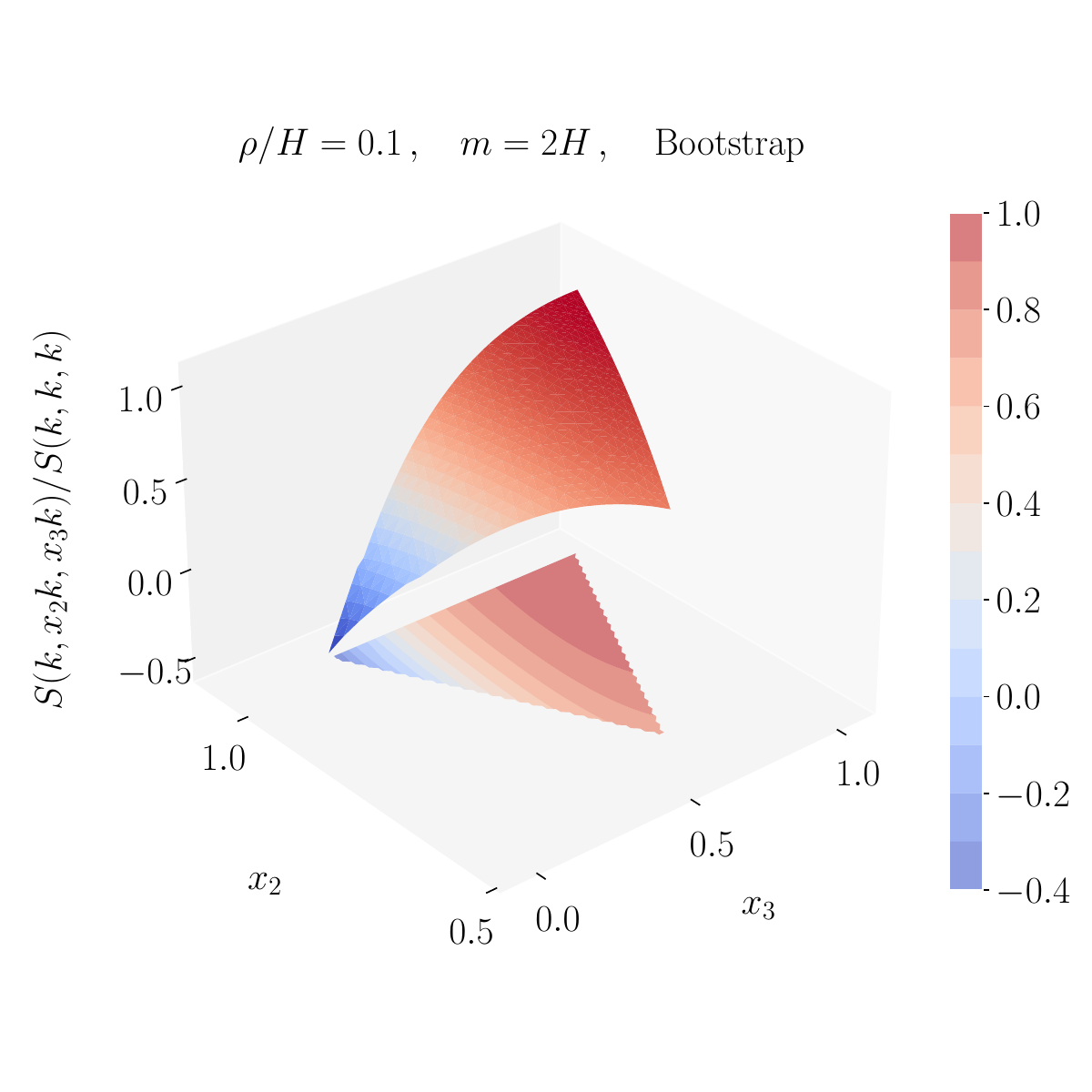}} \\
\vspace{-1.2cm}
  \subfloat{\includegraphics[width=0.49\textwidth]{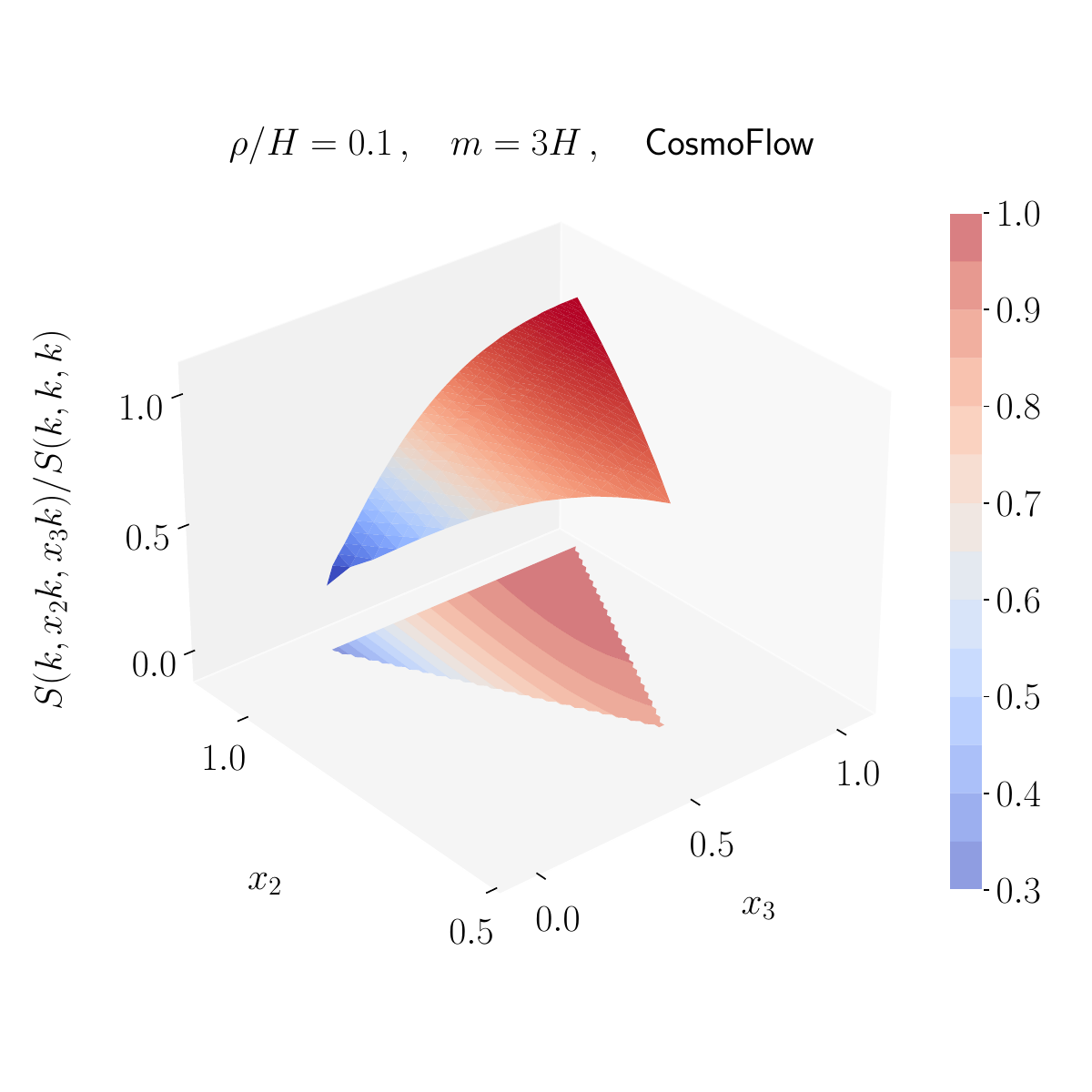}}
  \subfloat{\includegraphics[width=0.49\textwidth]{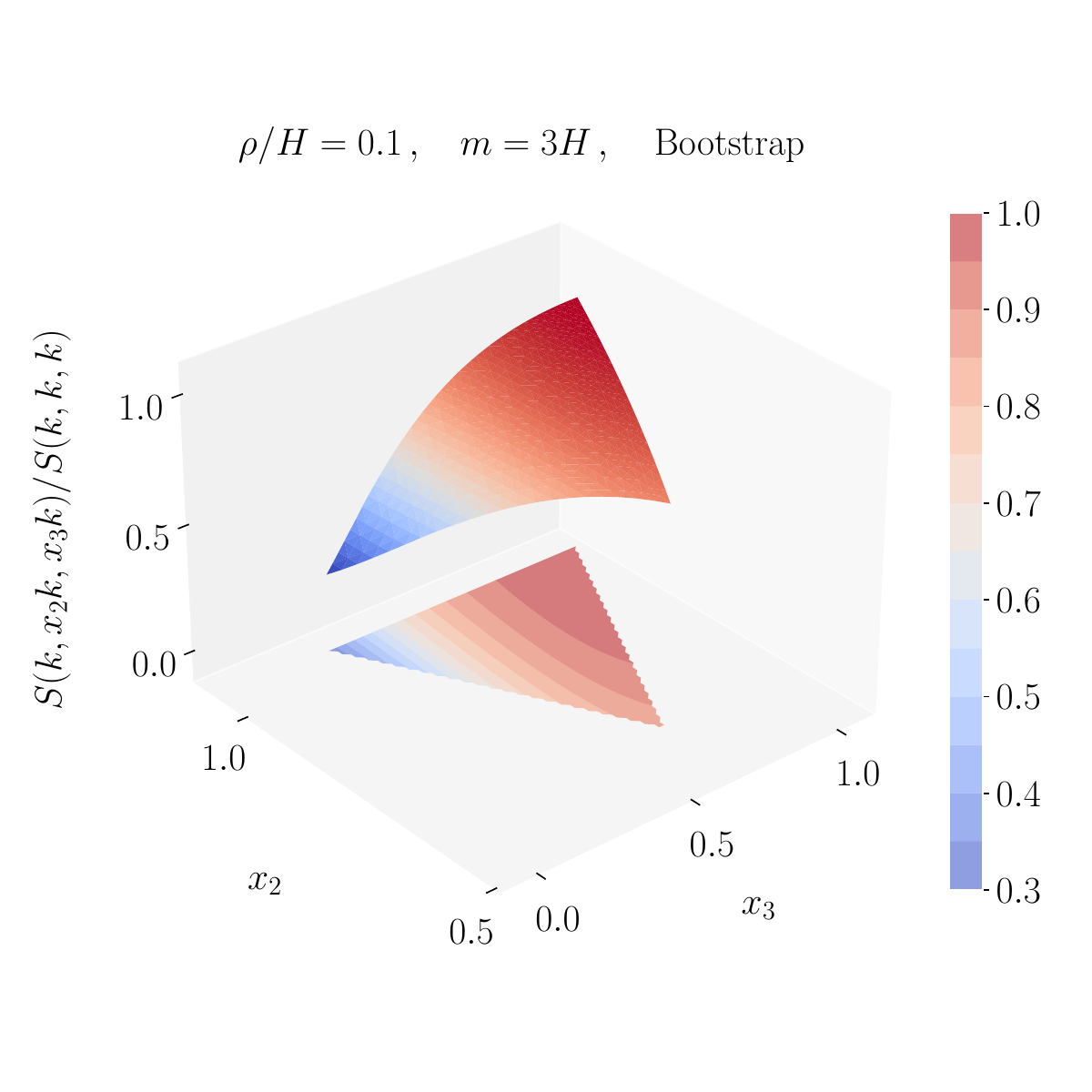}}
  \vspace{-1.2cm}
  \caption{Comparison of the shape functions for three masses of order the Hubble parameter obtained with \textsf{CosmoFlow} and the bootstrap method, for a quadratic mixing $\rho/H =0.1$. The projected surface is shown for representation purposes only and contains the same information as the 3D-shape.
  }
  \label{fig:Shape1}
\end{figure}

Our analytical expression for the result of the bootstrap equation corresponding to the double-exchange bispectrum is valid for all kinematical configurations of physical relevance.
In particular, it is valid beyond the equilateral configuration whose size dependence on the quadratic and cubic couplings, as well as the mass of the exchange field, has been studied in the previous section.
Here we instead investigate the shape dependence on  $(x_2,x_3)$ of the bispectrum $S(k,x_2 k,x_3 k)$, normalized to unity on equilateral configurations.
We fix $\rho/H =0.1$ for definiteness, although in the regime of weak mixing the shape is independent of it.
In Fig.~\ref{fig:Shape1}, we compare the exact numerical prediction using \textsf{CosmoFlow} to the analytical predictions derived from the bootstrap equation, for a few values of the mass of the exchanged field.
Unfortunately, the analytical formula converges slowly in the folded limit region, corresponding to $x_2 + x_3 \simeq 1$ (the lower ridge of the triangle).
In practice we have therefore evaluated the shape on the following kinematical region: $x_2 \in [0.58, 1]\,, x_3 \in [1.15-x_2,x_2]$, which corresponds to roughly $72\%$ of the total area.
Note that the numerical resolution does not suffer from this and can be evaluated quickly on any configuration but the very squeezed ones corresponding to $x_2 \sim 1$ and $x_3 \ll 1$ (left corner of the triangle), but for making the comparison explicit we evaluated it on the same reduced triangle region as the analytical prediction.
Agreement between the two methods is impressive.

For completeness, we also calculate the correlations of the double-exchange bispectrum shapes for a few representative values of the mass with the four shape templates that are most used in data analysis: equilateral, flattened, orthogonal and local bispectra.
For this, we use the Cosine expression of Eq.~\eqref{eq: Cosine def} and the shapes from \textsf{CosmoFlow} since we already showed agreement with the bootstrap ones and that it allows to probe a larger kinematical region; we simply cut at $x_3 = 0.1$ to avoid lengthy numerical evaluations in the squeezed limit.
The results are displayed in Fig.~\ref{fig: table} below, where we also show for reference the correlations between the shape templates for our kinematical region.
As expected, the shape resembles the local one for not-so-large masses $m \lesssim H$ and the equilateral one for masses slightly larger than Hubble, $m \gtrsim H$.
The limiting case $m=3H/2$ seems to support almost the same correlation with the local and the equilateral shape.
However, we remind that our correlations are computed in a restricted kinematical regime, $x_3 \geqslant 0.1$, and that having access to even more squeezed configurations would reveal that all shapes are less correlated with the local template.
Even the case $m=H$ would be affected, as indeed the correct template is the one of a quasi-local shape as first found in Ref.~\cite{Chen:2009we}.

\begin{figure}[ht!]
\centering
\begin{tabular}{| c | c  c  c  c  c  c |}
 \hline
 Mass value & $m=H$ & $m=3H/2$ & $m=2H$ & $m=5H/2$ & $m=3H$ & $m=7H/2$ \\
 \hline
 $\mathrm{Cos}(S , S_\mathrm{eq})$ & $0.71$ & $0.87$ & $\textbf{0.87}$ & $\textbf{0.92}$ & $\textbf{0.93}$  & $\textbf{0.93}$ \\
 $\mathrm{Cos}(S , S_\mathrm{flat})$ & $0.92$ & $0.84$ & $0.29$ & $0.72$ & $0.69$ & $0.72$ \\
 $ \mathrm{Cos}(S , S_\mathrm{orth})$ & $-0.66$ & $-0.48$ & $0.14$ & $-0.32$ & $-0.28$ & $-0.31$ \\
 $ \mathrm{Cos}(S , S_\mathrm{loc})$ & $\textbf{0.99}$ & $\textbf{0.90}$ & $0.36$ & $0.76$ & $0.75$ & $0.78$ \\
 \hline
\end{tabular}
\begin{tabular}{| c | c  c  c  c |}
 \hline
$S_i$ & $S_\mathrm{eq}$ & $S_\mathrm{flat}$ & $S_\mathrm{orth}$ & $S_\mathrm{loc}$ \\
 \hline
 $\mathrm{Cos}(S_i,S_\mathrm{eq})$ & $\textbf{1}$ & $0.46$ & $0.01$ & $0.66$ \\
 $\mathrm{Cos}(S_i,S_\mathrm{flat})$ &  & $\textbf{1}$ & $-0.88$ & $0.91$ \\
 $\mathrm{Cos}(S_i,S_\mathrm{orth})$ & &  & $\textbf{1}$ & $-0.68$  \\
$\mathrm{Cos}(S_i,S_\mathrm{loc})$ &  & & & $\textbf{1}$  \\
 \hline
\end{tabular}
\caption{Shape correlations between the double-exchange bispectrum at weak mixing, $\rho/H=0.1$, with equilateral, flattened, orthogonal and local bispectrum templates for a few representative values of the mass of the double exchanged fluctuation.
We also show for reference the correlations between these shape templates for the kinematical region that we consider, with a cut at $x_3 = 0.1$ in the squeezed limit.
The dominant component is highlighted in bold.
}
\label{fig: table}
\end{figure}

\subsection{Cosmological collider signal}

The CC signal lies in the soft limits of higher-order correlation functions, chief amongst which is the squeezed limit in the primordial bispectrum.
It consists in an imprint left by the production of heavy particles during inflation and is not encompassed by single-field effective theories.
We have already shown the explicit expression of the dominant squeezed bispectrum corresponding to the exchange of two massive fluctuations of the same field $\sigma$ in Eq.~\eqref{squeezedlimitapprox}.
Translated in terms of the primordial curvature fluctuation bispectrum shape function, it reads:
\begin{align}
\label{squeezedlimitapproxzeta}
    S^\mathrm{DE}_{\mathrm{CC},\mathrm{LO}}=\left(\frac{\rho}{H}\right)^2 \times  \lambda (2\pi \Delta_\zeta)^{-1} \times {\rm{Re}}\,\Bigg\{\left[
    \frac{\pi^{1/2}}{2^{4+2\i\mu}}\frac{2\i\mu+5}{2\mu-3\i}\Gamma\left[\frac{1}{2}+\i\mu,-\i\mu\right]\big(1+{\rm{tanh}}(\pi\mu)\big)+\mathcal{O}\left(e^{-2\pi\mu}\right)\right]\kappa^{\frac{1}{2}+\i\mu}\Bigg\}~,
\end{align}
where ``DE'' stands for Double-Exchange, ``CC'' for Cosmological Collider signal as before, and ``LO'' for Leading-Order. We introduced the more usual squeezing parameter $\kappa\equiv k_1/k_3$ and the squeezed limit corresponds to $\kappa \ll 1$.
Although this quantity~\eqref{squeezedlimitapproxzeta} indeed represents the leading-order squeezed behaviour, its mass suppression is exponential in the large $\mu$ limit: $S^\mathrm{DE}_{\mathrm{CC},\mathrm{LO}} \sim \mathcal{O} \left(\kappa^{1/2} \mu^{-1/2} e^{-\pi\mu}\right)$.
On the contrary the so-called background signal, denoted by ``BG'', that is decaying quicker in the squeezed limit, as $\kappa^1$, may be dominant in not-so-squeezed configurations since its mass suppression is only polynomial: $S^\mathrm{DE}_{\mathrm{
BG}} \sim \mathcal{O} \left(\mu^{-4}\right)$ as can be seen, for example, from its expression according to the single-field EFT and that we derived in Eq.~\eqref{fNLeq}.

\begin{figure}[!ht]
  \centering
  \subfloat{\includegraphics[width=0.5\textwidth]{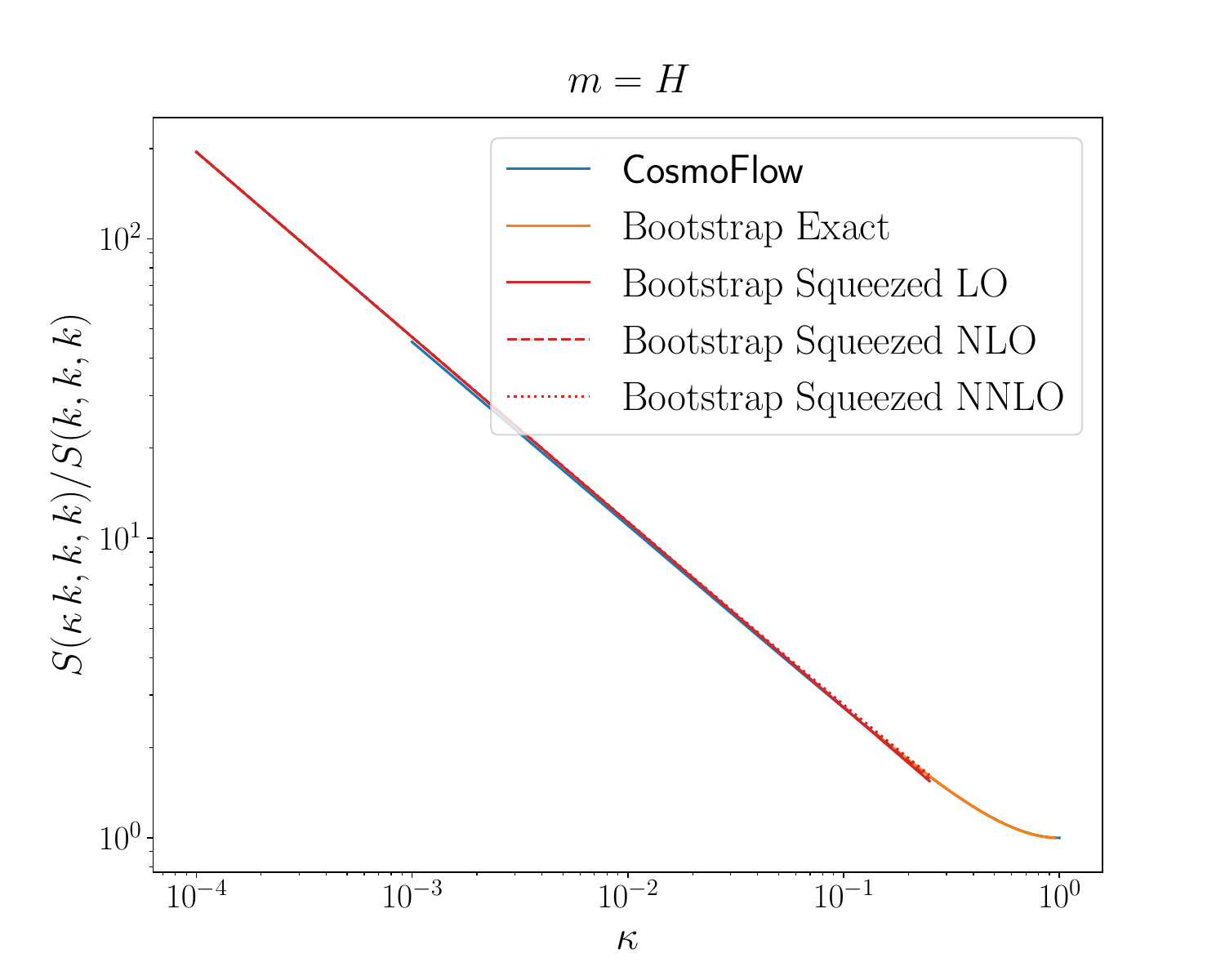}}
  \subfloat{\includegraphics[width=0.5\textwidth]{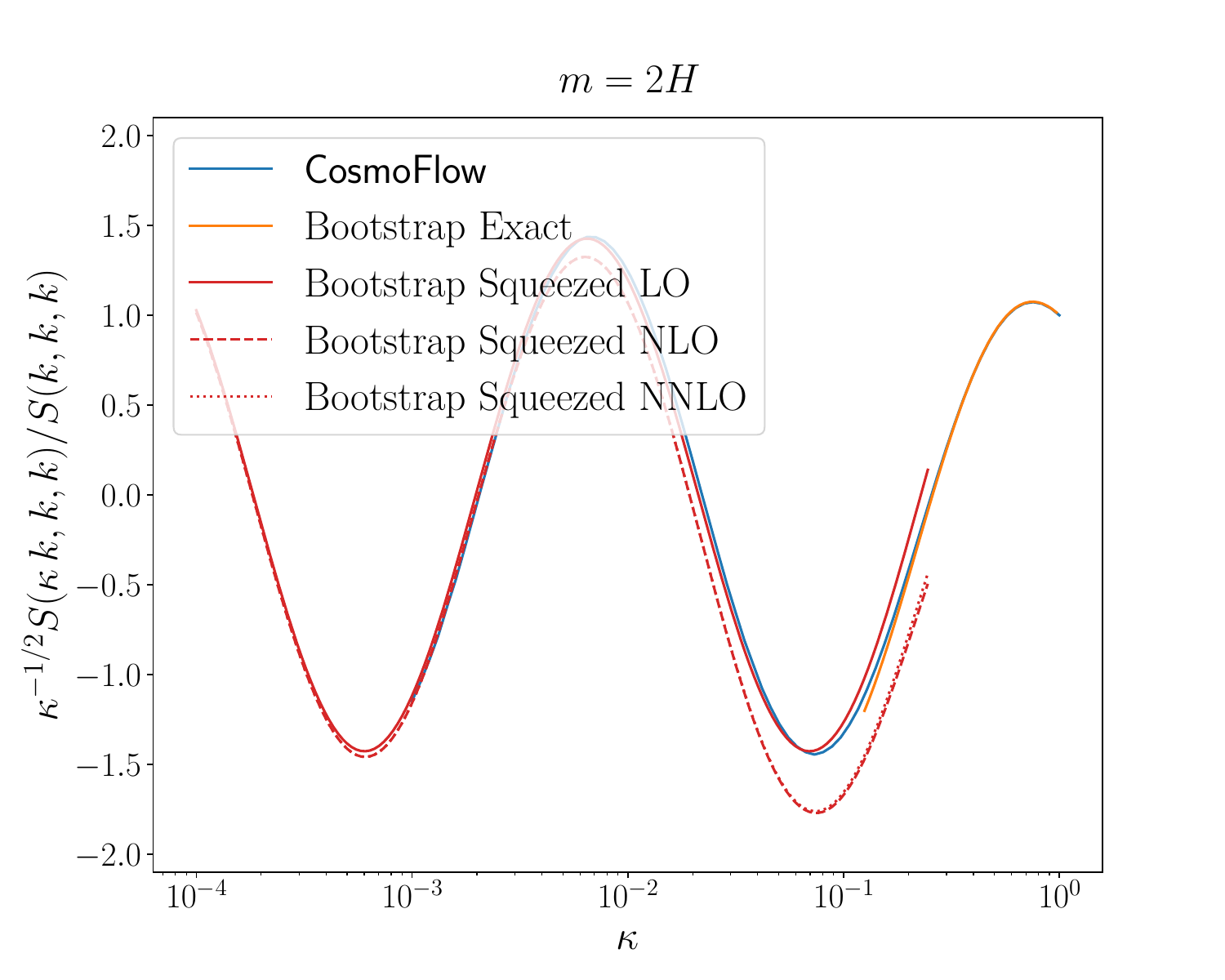}} \\
  \subfloat{\includegraphics[width=0.5\textwidth]{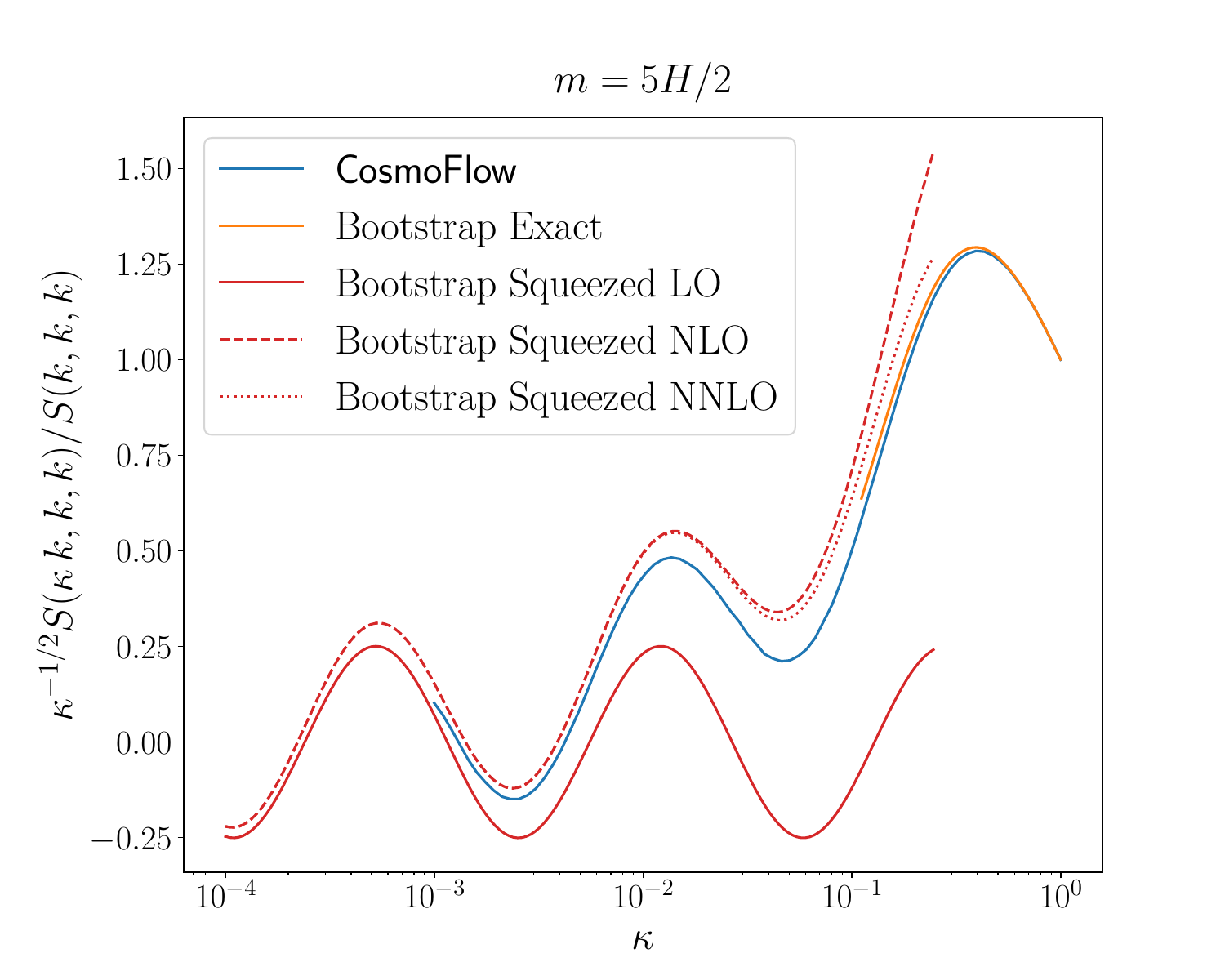}}
  \subfloat{\includegraphics[width=0.5\textwidth]{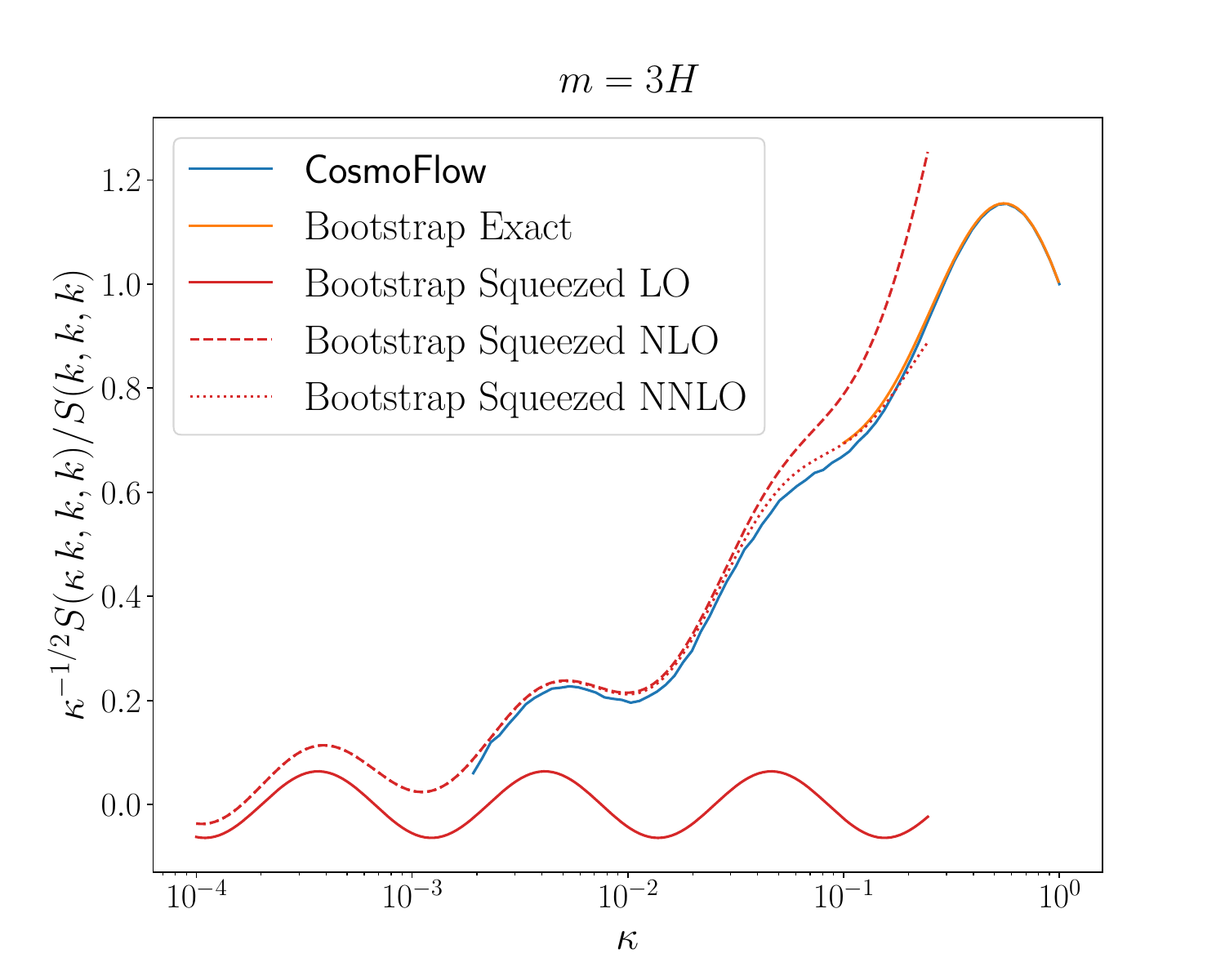}} 
  \caption{
  Cosmological collider signals in isosceles triangle configurations of the primordial bispectrum, for the double-exchange channel in the weak mixing regime with $\rho/H=0.1$, and normalized to unity in equilateral configurations, for four representative values of the mass $m$.}
  \label{fig: CC signals}
\end{figure}

For practical predictions at intermediate squeezed configurations, we find it useful to define and use next-to-leading-order (NLO) and next-to-next-to-leading-order (NNLO) corrections to the LO cosmological collider signal shown explicitly above. 
Let us clarify their meanings.
In the final bootstrap result, focusing on the squeezed limit, we are dealing with two different series expansions.
The first series is present in the analytical expressions given by Eqs.~\eqref{Ipmp2}--\eqref{Ippm}--\eqref{Ipppresult}, where we have a summation variable $m$.
Each term in the $m$ summation series is then expanded in terms of the squeezing parameter $\kappa$.
The terms ``next-'' and ``next-to-next-'' refer to the orders in $\kappa$.
For the CC signal part (containing non-analytic terms of the form $\kappa^{\i \mu}$), the series expansion of the $m$-term in the sum takes the following schematic form: $S_{\mathrm{CC},m}^\mathrm{DE}\sim\kappa^{1/2+\i\mu+m}\left(1+\mathcal{O}(\kappa) + \mathcal{O}(\kappa^2)+\ldots\right)$.
Moreover, explicit calculation shows that the $\mathcal{O}(\kappa)$ correction to the LO $m=0$ term exactly cancels with the leading-order part of the $m=1$ term.
Therefore, the total CC signal part scales as $S_{\mathrm{CC}}^\mathrm{DE}\sim\kappa^{1/2+\i\mu}\left(1+\mathcal{O}(\kappa^2)+\ldots\right)$.
Given the large suppression of the first non-zero correction to the CC signal in the squeezed limit, it is clear that we should take into account the BG part (containing only analytical terms, i.e. polynomial in $\kappa)$.
However, for the latter, each term in the $m$-summation begins at order $\kappa$ in the squeezed limit: $S_{\mathrm{BG},m}^\mathrm{DE}\sim\kappa^{1}\left(1+\mathcal{O}(\kappa) + \mathcal{O}(\kappa^2)+\ldots\right)$.
In principle, to be consistent, one should therefore perform the sum over all values of $m$.
However, the higher the order in the $m$-summation, the more they are suppressed by the mass parameter.
Therefore, in the large mass limit, $\mu \gtrsim 1$, it is consistent to keep only the first $m$-terms.
In practice, we therefore keep the first three terms corresponding to $m=\{0,1,2\}$ to approximate the BG signal, and consistently keep the dominant terms in the squeezed limit.
The NLO correction corresponds to the addition of the leading-order scaling $\mathcal{O}(\kappa)$ from this approximated BG signal.
The NNLO correction corresponds to the addition of the $\mathcal{O}(\kappa^2)$ terms from the BG signal.
Note that the NNNLO one would be the correction to the CC signal scaling as $\kappa^{5/2}$, but we do not consider it explicitly here.
Given the discussion above, it should be clear that we expect the NLO and NNLO corrections to badly encapsulate the cases where the mass is not large enough.
Indeed, this can be seen for example in the case $m=2H$ in Fig.~\ref{fig: CC signals}, for which the LO term alone performs better.
The NLO and NNLO expressions are derived from our general bootstrap result, and remain fully analytic.
They are quite long and not enough illuminating to be shown here, however they do allow for instantaneous evaluation for any value of $\kappa$ and $\mu$, in opposition with the general result for which squeezed configurations are computationally expensive to evaluate.
As mentioned previously, there exist potential methods to speed up the convergence of the exact solution around folded/squeezed regions. For instance, one approach involves employing the expansion series of the hypergeometric function around the unit~\cite{hyper} to refine the results. However, this transformation introduces considerable complexity, especially with additional layers of series. Further transformation and simplification is beyond the scope of this paper, and we leave this interesting discussion to future work.

We now discuss the relative size of the cosmological collider signal from the double-exchange channel that we have been focusing on in this work, compared to the one from the single-exchange channel.
For concrete comparison, we consider the only cubic interaction leading to a single-exchange diagram that is encountered in non-linear sigma models of inflation.
We have already mentioned that its strength is fixed by a non-linearly realized symmetry to be proportional to the quadratic mixing, $ \rho/(2 f_\pi^2) \times \partial_\mu \varphi \partial^\mu \varphi \sigma$ (see Eq.~\eqref{int_LI}).
The explicit expression of the cosmological collider signal from this single-exchange interaction, first found in Ref.~\cite{Arkani-Hamed:2015bza} and then exactly confirmed with an independent method in Ref.~\cite{Pinol:2021aun}, is shown in Eq.~\eqref{squeezedlimitapproxzeta-SE} below.
We find the ratio of the two squeezed limits, at leading-order in $\kappa$ and $\mu$, to be remarkably simple:
\begin{equation}
    \frac{S^\mathrm{DE}_{\mathrm{CC},\mathrm{LO}}}{S^\mathrm{SE}_{\mathrm{CC},\mathrm{LO}}}  = \lambda (2\pi \Delta_\zeta)^{-1} \times \frac{2}{\mu^2} + \mathcal{O}\left(\mu^{-3}\right)\,,
\end{equation}
where ``SE'' denotes this Single-Exchange channel. 
The parametric suppression in terms of the mass is exactly the same as the relative size of the equilateral background signals from both channels (indeed $S^\mathrm{SE}_{\mathrm{CC},\mathrm{LO}} \sim \mathcal{O}\left(\kappa^{1/2}\mu^{3/2}e^{-\pi\mu}\right)$ and $S^\mathrm{SE}_\mathrm{BG} \sim \mathcal{O}\left( \mu^{-2}\right)$).
In particular, the absence of further ``Boltzmann'' exponential suppression was  guessed in \cite{Tong:2021wai}, and here we have rigorously proved it.
Importantly, this implies that for values of the mass leading to a signal potentially observable, corresponding to $\mu$ of at most a few, the cosmological collider signal from the double exchange bispectrum is dominant as long as $\lambda (2\pi \Delta_\zeta)^{-1} \gtrsim 1$.
Such values are perfectly allowed by perturbativity bounds, and can even be natural, as discussed at the beginning of Sec.~\ref{Sec_pheno}, therefore making the double-exchange channel an interesting prospect for detecting the cosmological collider signal.

We now compare our exact bootstrap results (limited to close-to-equilateral configurations $\kappa \gtrsim 0.1$) with the LO, NLO and NNLO expansions in the squeezed limit, as well as the exact numerical results with \textsf{CosmoFlow} (whose only limitation is the computational cost of very squeezed configurations, in practice we restrict to $\kappa  \gtrsim 10^{-3} $).
The results are shown in Fig.~\ref{fig: CC signals}, in the weak mixing regime for $\rho/H=0.1$ where our analytical predictions hold, and for a few representative values of the mass of the massive field.
It is interesting to notice the excellent agreement between \textsf{CosmoFlow} results and i) the exact bootstrap result where it can be evaluated efficiently, close to equilateral configurations; ii) the LO expansion in the squeezed limit; iii) the NLO and NNLO expansions for not-so-squeezed configurations when the mass is sizeable ($\mu \gtrsim 1)$.
Altogether, our example showcases the utility and synergy of analytical and numerical methods, with precise and trustworthy predictions for the double-exchange bispectrum for all configuration values, ranging from equilateral to extremely squeezed ones.

\begin{figure}[ht]
  \centering
\subfloat{\includegraphics[width=0.5\textwidth]{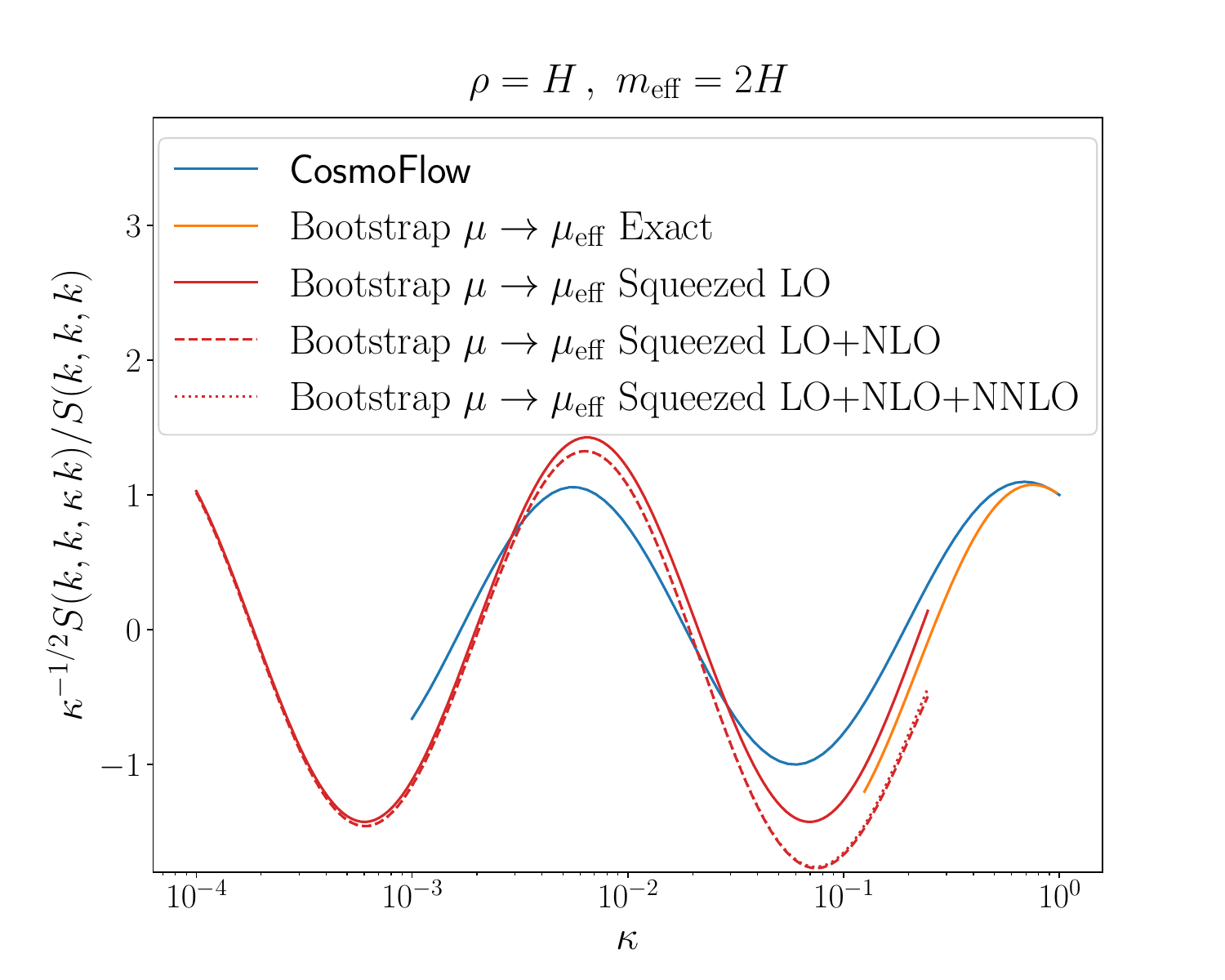}}
\subfloat{\includegraphics[width=0.5\textwidth]{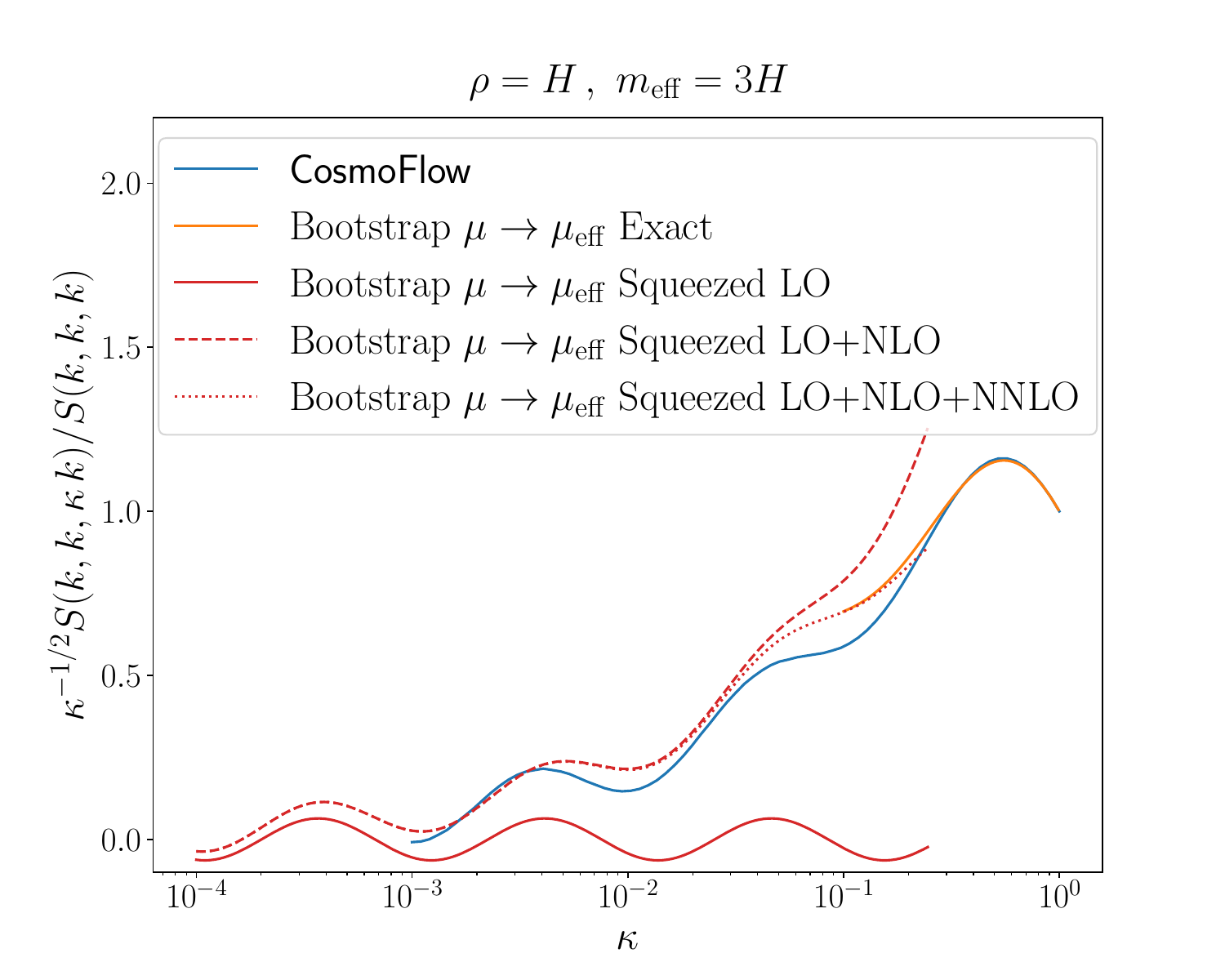}}
  \caption{
  Cosmological collider signals in isosceles triangle configurations of the primordial bispectrum, for the double-exchange channel in the strong mixing regime with $\rho/H=1$, and normalized to unity in equilateral configurations, for two representative values of the effective mass $m_\mathrm{eff}$.
  The bootstrap predictions are obtained by the naive replacement $\mu \rightarrow \mu_\mathrm{eff}$ in our analytical formulas.
  This replacement works surprisingly well, even though it misses a slight damping of the CC signal as well as a small phase shift towards larger values.
  }
  \label{fig: CC signals strong mixing}
\end{figure}

For completeness, we will also quickly explore the strong quadratic mixing regime.
The phenomenology of the cosmological collider signal in this regime has been thoroughly explored numerically in~\cite{Pinol:2023oux}.
One of the main lessons is the rescaling of the mass to an effective mass by the large quadratic mixing: $m^2 \rightarrow m_\mathrm{eff}^2 = m^2 + \rho^2$, visible in the frequency of the oscillations in the cosmological collider signal, now dictated by $\mu_\mathrm{eff}=\sqrt{\mu^2+\rho^2/H^2}$.
In Fig.~\ref{fig: CC signals strong mixing}, we explore how the analytical formula performs under such substitution compared to the exact numerical result obtained in the strong mixing regime for $\rho/H=1$ with \textsf{CosmoFlow}, though the analytical formulas are valid only for the weak quadratic mixing regime by construction. 
Note that for this value of the quadratic mixing, the true size of the bispectrum is exactly 100 times bigger than the one shown in the right panel of Fig.~\ref{fig:fNLeq}, pushing it to a level comparable to current and upcoming observational constraints, even for natural values of the cubic coupling constant $\lambda$ that are well below the perturbativity bound, thus making an impressive case for the double-exchange channel.
Interestingly, the modified analytical formulas correctly capture not only the frequency of the oscillations, but also the overall shape of the signal.
They seem to miss however a slight damping of the CC signal, as well as a small phase shift towards larger values.
It would be interesting to pursue analytical calculations tailored to the strong quadratic mixing regime and precisely predict these interesting features, but this goes beyond the scope of this article.

\subsection{Disentangling double-exchange from single-exchange channels}
\label{subsec:disentangling}

We have shown that the bispectrum signal from the double-exchange channel can easily be larger than the one from the single-exchange one.
We proved this to be correct both for the equilateral configuration value and for the cosmological collider signal lying in the squeezed limit.
However, we still lack a striking feature of the double-exchange channel that \textit{cannot} be mimicked by the single-exchange one.
In this section, we explore three directions in order to address this question: i) information in the phase of the cosmological collider signal; ii) the exchange of two massive fluctuations pertaining to two different isocurvature species with unequal masses; iii) the information in the primordial trispectrum.

\paragraph{Phase information in the CC signal.}

\begin{figure}[ht]
  \centering
  \hspace{-0.6cm}
  \includegraphics[width=0.9\textwidth]{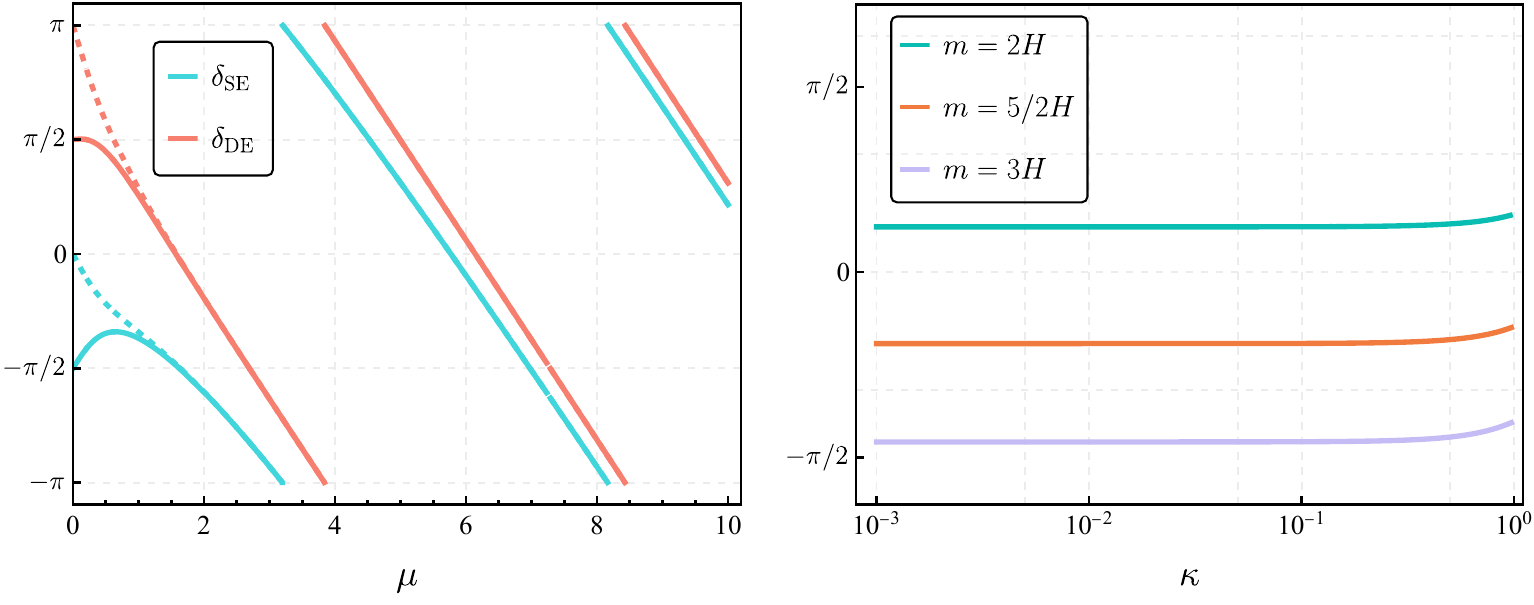}
  \caption{
  \textbf{Left:}
  Comparison of the phases of single- and double-exchange diagrams, as defined in Eqs.~\eqref{CCsignal_amp_and_phase} and \eqref{CCsignal_amp_and_phase2}.
  The figure shows the one-to-one correspondences between frequency and phase of the oscillations, unveiling the potentiality to disentangle different CC channels from the phase information.
  This is particularly true whenever $\mu$ is not too large, which is the most interesting case.
  Both dashed lines represent the leading contribution in the large $\mu$ limit, with a single power of the Boltzmann suppression, $\mathcal{O}(e^{-\pi\mu})$.
  Solid lines represent the total result including all contributions, making for a non-negligible correction in the small $\mu$ region.
  \textbf{Right:} Evolution of the phase, for three representative masses, as a function of the squeezing parameter when adding the first non-vanishing correction scaling as $\kappa^{5/2}$ in the CC signal.
  Only the region $\kappa \in [0.1,1]$ is affected by the deviation from a constant, and we expect higher-order corrections in $\kappa$ in the CC signal to also play a role when $\kappa \rightarrow 1$.
  }
  \label{fig:phase}
\end{figure}

The shape functions of single-exchange and double-exchange bispectra, on isosceles squeezed configurations, can be written as:
\begin{align}
\label{CCsignal_amp_and_phase}
    S_{\mathrm{CC},\mathrm{LO}}^\mathrm{SE} &= \left(\frac{\rho}{H}\right)^2 \times \mathrm{Re}\left[\kappa^{1/2+\i \mu} \mathcal{A}_\mathrm{SE}(\mu) e^{\i \,\delta_\mathrm{SE}(\mu)}\right] \,, \\
\label{CCsignal_amp_and_phase2}    S_{\mathrm{CC},\mathrm{LO}}^\mathrm{DE} &= \left(\frac{\rho}{H}\right)^2 \times \lambda(2\pi \Delta_\zeta)^{-1} \times \mathrm{Re}\left[\kappa^{1/2+\i \mu} \mathcal{A}_\mathrm{DE}(\mu) e^{\i \,\delta_\mathrm{DE}(\mu)}\right] \,.
\end{align}
Here $\mathcal{A}_{\mathrm{SE},\mathrm{DE}}$ denote the amplitudes of the CC signals and $\delta_{\mathrm{SE},\mathrm{DE}}$ their phases.
Importantly, in this squeezed limit and in the regime of validity (weak quadratic mixing) of the analytical calculations, they only depend on $\mu$, the mass parameter of the unique isocurvature species.
The expression for $\mathcal{A}_{\mathrm{DE}}(\mu)e^{\i \,\delta_\mathrm{DE}(\mu)}$ can be directly read off Eq.~\eqref{squeezedlimitapproxzeta}, while the one for the single-exchange channel can be found explicitly in Refs.~\cite{Arkani-Hamed:2015bza,Pinol:2021aun}:
\begin{equation}
\label{squeezedlimitapproxzeta-SE}
    \mathcal{A}_{\mathrm{SE}}(\mu)e^{\i \,\delta_\mathrm{SE}(\mu)} = - \i \frac{\pi^{3/2}}{2^{4+2i\mu}} \frac{\Gamma(7/2+\i \mu)}{(2\mu-\i)\Gamma(1+\i \mu)} \tanh(\pi \mu) e^{-\pi \mu} \left[\coth(\pi\mu)- \i \csch(\pi\mu) +1 \right]^2\,.
\end{equation}
In the regime of validity of this analysis, the frequency of the CC signal is directly given by $\mu$.
It was argued in Ref.~\cite{Qin:2022lva} that the phase of the CC signal may bring additional information to help distinguishing the channel at the origin of this signal.
In the latter reference, differences in the phases were investigated between two cubic interactions leading to single-exchange diagrams: the standard Lorentz-covariant interaction denoted as ``SE'' in our work and a more exotic Lorentz-violating one with schematic form $\tilde{\rho} \varphi^{\prime 2} \sigma$, but nevertheless allowed from an effective field theory point of view.
Here, we compare the phases between the ubiquitous single-exchange ``SE'' channel with Lorentz-covariant interaction and the double-exchange one for the first time.
As can be seen in the left panel of Fig.~\ref{fig:phase}, information about the phase, once the frequency $\mu$ is determined, indeed places interesting constraints on the possible channel to generate the CC signal.
At large values of $\mu$ though, the two phases become indistinguishable, but fortunately this is any way the regime for which the CC signal is unobservably small. 
In the right panel of Fig.~\ref{fig:phase}, we instead investigate the dependence of an effective phase $\delta_\mathrm{DE}(\mu,\kappa)$ in mildly squeezed configurations, as a function of $\kappa$ and for a few values of the mass, when adding the first non-vanishing correction to the CC signal in the squeezed limit expansion, scaling as $\kappa^{5/2}$ as already mentioned.
These not-so-squeezed configurations are of particular relevance to observations, and it is therefore important to dispose of full predictions for all kinematic regimes as we do.

\paragraph{Multiple isocurvature species: inflationary flavor oscillations.}
Before the current paragraph, all phenomenological consequences that we have shown were derived under the assumption that there exists an unique massive species $\sigma$, with mass $m$ and mass parameter $\mu$.
This choice was motivated by the long-standing tradition to explore CC signals in this restricted class of multiple fields content, and under the implicit assumption that only the lightest of massive degrees of freedom may leave an imprint on cosmological observables.
However, it was discovered in Ref.~\cite{Pinol:2021aun} that even heavier species may strongly impact the primordial bispectrum in its squeezed limit, due to an interesting interplay between the values of the masses ($m_\alpha)$ and of the mixing angles of the theory (in $\rho_\alpha$) that define mass eigenstates with respect to the flavor ones.
The most striking manifestation of multiple massive species is the appearance of modulated oscillations in place of the usual single-harmonics CC signal, signaling the presence of inflationary flavor oscillations analogous to those appearing in the Standard Model of particle physics.
By determining the frequencies and relative sizes of the different modes, it was proposed that observations could help measuring not only the masses but also the mixing angles of the inflationary theory.
Explicitly, the sum of all single-exchange diagrams with multiple isocurvature species gives the following leading-order CC signal in the squeezed limit ~\cite{Pinol:2021aun}:
\begin{align}
    S_\text{CC,LO}^\text{SE,multi}= \sum_{\alpha} \left(\frac{\rho_\alpha}{H}\right)^2 \Re\left[\kappa^{1/2+\i\mu_\alpha}\mathcal{A}_\text{SE}(\mu_\alpha)e^{\i\delta_{\text{SE}}({\mu_\alpha})}\right]\,,
    \label{eq:superposition-single}
\end{align}
where $\mathcal{A}_\text{SE}$ and $\delta_{\text{SE}}$ were already defined in Eqs.~\eqref{CCsignal_amp_and_phase} and \eqref{squeezedlimitapproxzeta-SE}.
Here, we extend the discussion of the latter reference to the double-exchange channel.
Where previously we had the exchange of two fluctuations of the same species $\sigma$, we now allow for multiple species $\sigma_\alpha$ with \textit{a priori} unequal mass parameters $\mu_\alpha$.

\begin{figure}[htp]
  \centering
  \hspace{-0.5cm}
  \includegraphics[width=\textwidth]{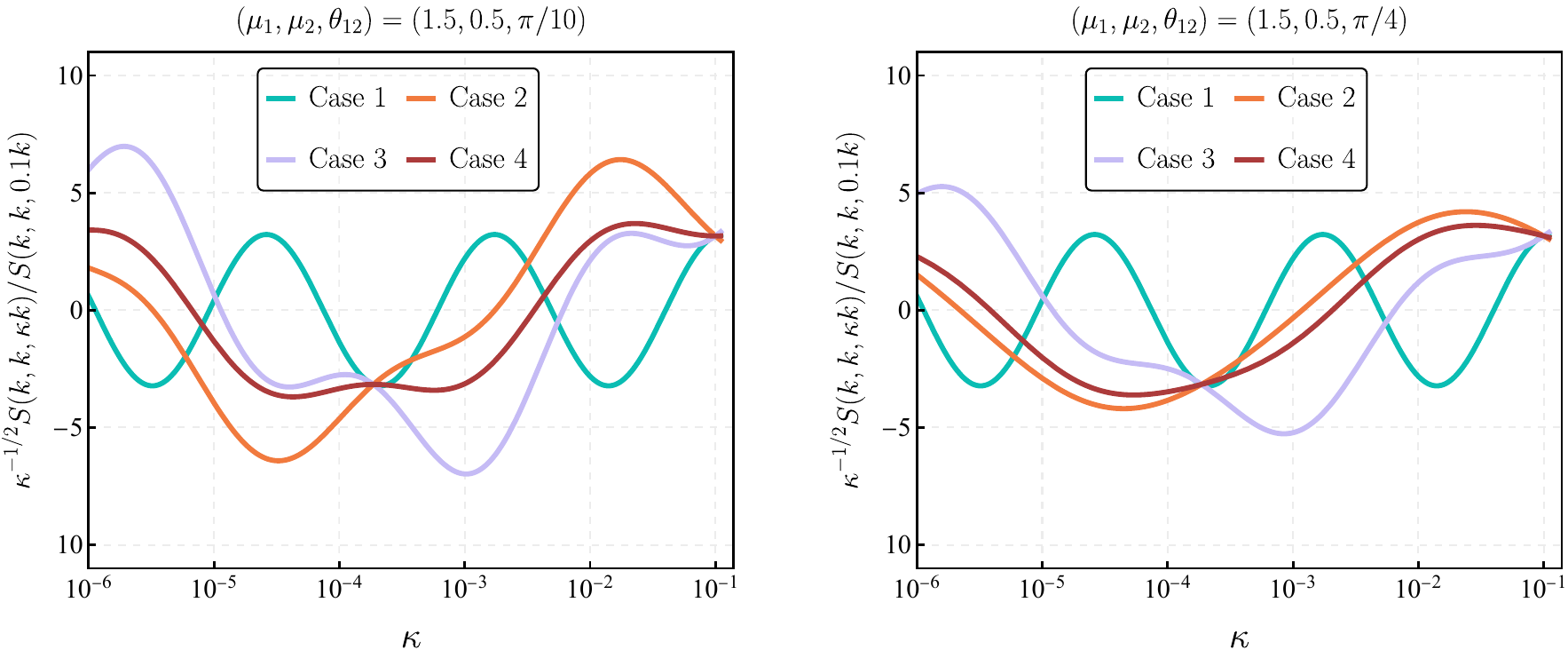}
  \caption{Inflationary flavor oscillations in the squeezed limit of the bispectrum from the double-exchange diagram. In both the left and right panels, the masses of different isocurvature species are set to $\mu_1=1.5$ and $\mu_2=0.5$, while the mixing angles $\theta_{12}$ are chosen as $\pi/10$ (left panel) and $\pi/4$ (right panel).
  The four distinct cases, corresponding to different forms of the cubic interaction matrix $\lambda_{\alpha\beta}$, are described in the main text below this figure.
  For better visualization, we have introduced a $\kappa^{-1/2}$ rescaling factor, and the shape function is arbitrarily normalized to its value at $\kappa=10^{-1}$~. 
  }
  \label{fig:flavor-oscillations}
\end{figure}

In the case of multiple isocurvature species, the double-exchange diagram results in a superposition of oscillatory signals in the squeezed limit, with the dominant contribution being given by
\begin{align}
    S_\text{CC,LO}^{\mathrm{DE},\mathrm{multi}}= \sum_{\alpha,\beta}\frac{\rho_\alpha\rho_\beta }{H^2} \times \lambda_{\alpha\beta}(2\pi \Delta_\zeta)^{-1} \times \Re\qty[\kappa^{1/2+\i\mu_\alpha}{\mathcal{A}_{\mathrm{DE}}^{\mathrm{multi}}(\mu_\alpha,\mu_\beta)}e^{\i\delta_{\mathrm{DE}}^{\mathrm{multi}}({\mu_\alpha,\mu_\beta})}],
    \label{eq:superposition}
\end{align}
where $\mathcal{A}_{\mathrm{DE}}^{\mathrm{multi}}$ and $\delta_{\mathrm{DE}}^{\mathrm{multi}}$ are uniquely determined from our analytical formulae~\eqref{squeezedlimitapprox2}.
Disentangling individual contributions $\beta$ to the same frequency signal set by $\mu_\a$ is impossible, so we also define the effective amplitude and phase of individual modes in the CC signal as:
\begin{align}
    S_\text{CC,LO}^\text{DE,multi} &= \sum_{\alpha}\left(\frac{\rho_\alpha}{H}\right)^2 \Re\left[\kappa^{1/2+\i\mu_\alpha}\tilde{\mathcal{A}}_\text{DE}^\alpha e^{\i\tilde{\delta}_{\text{DE}}^\alpha}\right]\,, \\
    \text{with} \quad  \tilde{\mathcal{A}}_\text{DE}^\alpha e^{\i\tilde{\delta}_{\text{DE}}^\alpha} &= \sum_\beta \frac{\rho_\beta }{\rho_\alpha} \times \lambda_{\alpha\beta}(2\pi \Delta_\zeta)^{-1} \times \mathcal{A}_{\mathrm{DE}}^{\mathrm{multi}}(\mu_\alpha,\mu_\beta)e^{\i\delta_{\mathrm{DE}}^{\mathrm{multi}}({\mu_\alpha,\mu_\beta})}\,, \nonumber
    \label{eq:superposition2}
\end{align}
where we have factorised out two factors of $\rho_\alpha$ for simpler comparison to the single-exchange diagram.
Although a detailed study of the phenomenology of the multiple isocurvature species double-exchange diagrams---and its comparison to the single-exchange one---is beyond the scope of this article, we already note an interesting feature.
When there is a single isocurvature species, we have seen in Fig.~\ref{fig:phase} that the phase is uniquely fixed by the frequency of the CC signal, for both kinds of diagrams.
With several isocurvature species, a similar statement still holds for the single-exchange diagram: each individual mode $\alpha$ in the CC signal has a phase uniquely fixed by its frequency.
However, this one-to-one correspondance is broken in the double-exchange case, as both the amplitude and the phase of the mode $\alpha$ depend also on all other isocurvature species's masses $\mu_\beta$ and mixing angles in $\rho_\beta$, as long as their cubic interactions as encoded in $\lambda_{\alpha\beta}$ have non-trivial non-diagonal elements.
Therefore, in the occasion that we can measure both the effective amplitude and phase of different modes in the CC signal, we could easily rule out single-exchange diagrams if the consistency condition between phase and frequency is not verified.
The opposite is harder, as a single mode in a double-exchange diagram signal can mimic the phase of a single-exchange one.
In Fig.~\ref{fig:flavor-oscillations}, we illustrate these new effects from the non-diagonal elements of the cubic interactions in $\lambda_{\alpha \beta}$. We plot the signal in the squeezed limit corresponding to inflationary flavor oscillations for a model with $\varphi$ coupled to two massive scalar fluctuations $(\sigma_1,\sigma_2)$, with mass parameters $(\mu_1,\mu_2)=(1.5,0.5)$.
The mixing angle $\theta_{12}$ between the inflationary flavor and mass eigenstates is defined by $(\rho_1,\rho_2) = \left( \rho \cos(\theta_{12})\,,\rho\sin(\theta_{12}) \right)$.
It was shown in Ref.~\cite{Pinol:2021aun} that this angle is found from the rotation matrix used to diagonalize the non-trivial mass matrix present in the flavor basis.
The left panel corresponds to $\theta_{12} = \pi/10$, which was identified in the previous work as leading to striking signatures even from the single-exchange channel.
The right panel corresponds to $\theta_{12} = \pi/4$, for which the mode mixing in the single-exchange channel can hardly be seen. 
In each of the panels, we vary the form of the cubic interaction matrix $\lambda_{\alpha\beta}\equiv \lambda \times e_{\alpha\beta}$ as follows, where we only mention the non-zero components: Case 1 has $e_{11}=1$ only and forbids the presence of mode mixing; Case 2 has  $e_{11}=e_{22}=1$ and corresponds to a situation analogous to the single-exchange diagram (mode mixing can only be seen for small mixing angles); Case 3 has $e_{11}=e_{12}=e_{21}=1$ and Case 4 has $e_{11}=e_{22}=e_{12}=e_{21}=1$, they both correspond to new features of the double-exchange channel not reproducible by the single-exchange one.
Although definitely leading to signals of different natures in the squeezed limit, which can help to disentangle single-exchange diagrams from double-exchange ones from observations in principle, it would be desirable to identify another striking signature of the exchange of two different isocurvature species.
In this perspective, we now turn to a first study of the CC signal in the trispectrum from a double-exchange diagram.

\paragraph{Towards the primordial trispectrum.} 
In the trispectrum, as evidenced by the patterns observed in Figure~\ref{IpppNumVSAna}, the oscillation features are totally different from those of the single-exchange, appearing as if two signals with different frequencies were superimposed. Therefore, at the trispectrum level, in principle, we can clearly distinguish between single-exchange and double-exchange through the oscillation patterns. 
In this subsection, we will delve into the details of these distinct features. Firstly,
let us define the dimensionless trispectrum $\mathcal{T}$ as follows
\begin{align}
\langle\varphi_{{\bf{k}}_1} \varphi_{{\bf{k}}_2}\varphi_{{\bf{k}}_3}\varphi_{{\bf{k}}_4}\rangle '= \frac{H^8}{f_\pi^4}\frac{(k_{1234}/4)^3}{(k_1k_2k_3k_4)^3}\mathcal{T}({\bf{k}}_1,{\bf{k}}_2,{\bf{k}}_3,{\bf{k}}_4)~.
\end{align}
The interactions chosen here are specified in (\ref{int_2}) and (\ref{int_4}), then the four-point functions can be related to the seed integral through the simple relation (\ref{Trispectrum1}). With these coupling choices, the $\tau_{\text{NL}}$ typically denoting the size of the trispectrum, has the following parametric dependence:
\begin{align}
 \mathcal{T}\sim\tau_{\text{NL}}\sim \left({\rho}/{H}\right)^2\times \left(H^2\tilde{\lambda}\right) \times(2\pi\Delta_\zeta)^{-2}~.
\end{align}
The kinematic dependence of the trispectrum from this diagram is relatively simple, with no dependence on the diagonals of the quadrangle formed in Fourier space by the four wavevectors $\bf{k}_i$, so we will simply write $\mathcal{T}(k_1,k_2,k_3,k_4)$.
Similar to what we did in the bispectrum case, in the regime of perturbative quadratic mixing and the large mass, we can easily find a simple fitting formula from the analytical result in regular configurations:
\begin{equation}
    \mathcal{T}(k,k,k,k) \underset{\mu \gg 1}{\simeq} 0.07 \times \rho^2\tilde{\lambda}\times\frac{f_\pi^4}{H^4}\times\frac{1}{\mu^4}~.
\end{equation}
To investigate the phenomenon indicated in Figure~\ref{IpppNumVSAna}, we will focus on one particular channel of the trispectrum, setting $k_2=k_4=k$ and $k_1=k_3=rk$, with $r \rightarrow 0$ representing the double soft limit.
To gain a clear understanding of the origin of this phenomenon, instead of dealing with the complicated exact solution, we will rely on some approximate formulae discussed in Section~\ref{sec:limits}.
Specifically, the approximation here arises from a series expansion around $r\sim0$ of the exact solutions. 
Under this limit, we divide the approximate expression into three distinct parts, each with their own interpretations,
\begin{align}
\mathcal{T}_{\text{approx}}=\mathcal{T}_{\text{ss}}+\mathcal{T}_{\text{sb}}+\mathcal{T}_{\text{bb}}~.\label{T_app}
\end{align}
$\mathcal{T}_{\text{ss}}$ arises when both momenta associated with massive propagators are soft, leading to a CC signal from each. Given that the time-ordered part $\mathcal{I}_{+++}$ dominates due to lighter Boltzmann suppression, let us simplify and focus solely on this term for clarity. 
In this context, $\mathcal{T}_{\text{ss}}$ originates from the homogeneous solution $\mathcal{Y}^{\text{-2}~0~\text{-2}}_{+++}$, under the double soft limit, analogous to Eq.~(\ref{Ydoublesoft}) discussed earlier. 
For $\mathcal{T}_\text{sb}$, the CC signal is generated only from one propagator, while the other contribution arises from the particular solution, similar to Eq.~(\ref{Pdoublesoft}) under the double-soft limit. 
Additionally, due to permutations, there is another contribution where only one massive propagator becomes soft, approximated by Eq.~(\ref{singlesoft}). The last term $\mathcal{T}_{\text{bb}}$, represents the leading-order analytical background.
\begin{figure}[htp]
	\centering
        \hspace{-1.8cm}
	\includegraphics[width=0.6\textwidth]{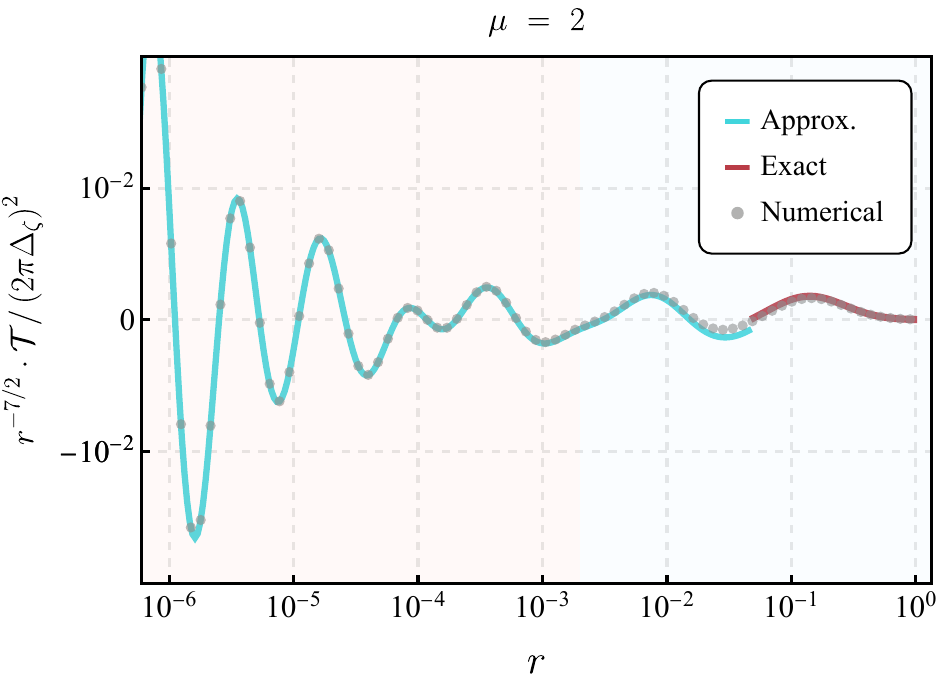}\\
	\caption{ The comparison between the numerical calculation and  analytical result. The \textcolor{gray}{gray} dots,  \textcolor{blue2}{blue} solid line and \textcolor{red4}{red} line represents numerical results,  approximation expression and exact result from (\ref{Ipppresult}) respectively. In the figure we set $k_2=k_4=k$ amd $k_1=k_3=r k$,  the masses are chosen as $\mu_\a=\mu_\b=2$
    and the coupling $\tilde{\lambda}\cdot\rho^2$ is set to $10^{-2}$. The left region (light red) is dominated by the term $r^{2\i\mu}$ while the right region (light blue) is dominated by the oscillation $r^{\i\mu}$. For better visualization, we have multiplied the dimentionless trispectrum $\mathcal{T}(rk,k,rk,k)$ by a factor $r^{-7/2}$. 
    }\label{Trispectrum_figure}
\end{figure}

\begin{table}[ht]
\renewcommand\arraystretch{1.3}
\tabcolsep=0.4cm
\centering
\begin{tabular}{ |c|c|c|  }
\hline
different terms& mass dependence & $r$ dependence \\
\hline
$\mathcal{T}_{\text{ss}}$ & $\mu^{3/2}e^{-2\pi\mu}$ & $r^{3+2\i\mu}$ \\
\hline
$\mathcal{T}_{\text{sb}}$ & $\mu^{3/2}e^{-\pi\mu}$ & $r^{7/2+\i\mu}$ \\
\hline
$\mathcal{T}_{\text{bb}}$ &$\mu^{-4}$ & $r^4$ \\
\hline
\end{tabular}
 \caption{The leading order mass and $r$ dependence of each term in Eq.~\eqref{T_app}.}
    \label{tab:massrTable}
\end{table}

Because the formula is lengthy even for the leading-order series expansion, we do not intend to present details of in this section, but it can be directly obtained using the methods introduced in Section~\ref{sec:limits}. 
Here, we summarize the key information about the magnitude and dependence on $r$ of each term in Table~\ref{tab:massrTable}, where we have only kept the leading order in the series expansion around $r=0$. 
Now, it is clear to see the origins of the unique oscillation pattern. $\mathcal{T}_{\text{ss}}$ excites CC signals in both massive propagators, thus the signal is suppressed by $\mathcal{O}(e^{-2\pi\mu})$, and the frequency is doubled to $2\mu$. As a compensation, it features a dominant scaling with respect to the soft variable $r$ when it approaches zero, and the overall $r$ dependence is $r^{3+2\i\mu}$. In contrast, $\mathcal{T}_{\text{sb}}$ is suppressed by $\mathcal{O}(e^{-\pi\mu})$ and oscillates as $r^{7/2+\i\mu}$. Consequently, there is a competition between mass and $r$ dependence. In very soft regions, the term with the lower power in $r$ will dominate, whereas in intermediate regions, the term with less Boltzmann suppression will prevail.
To illustrate this idea, in Figure \ref{Trispectrum_figure}, we compare the analytical expression with the numerical results. 
The approximation depicted by the \textcolor{blue2}{blue} line closely matches the numerical calculations represented by the \textcolor{gray}{gray} dots, except in the region close to the equilateral configuration. In this region, we employed the exact solution shown as the \textcolor{red4}{red} line, which actually matches perfectly the numerical results. In this figure, we have assumed equal masses with $\mu=2$
and we have multiplied the dimensionless trispectrum $\mathcal{T}(rk,k,rk,k)$ by a factor of $r^{-7/2}$ for better visualization. 
Now, based on our discussion above, we can understand the superimposed oscillation shape  in Figure~\ref{Trispectrum_figure}. This shape arises from the competition between the oscillation signals $\mathcal{O}(e^{-2\pi\mu})\times r^{2\i\mu+3}$ and $\mathcal{O}(e^{-\pi\mu})\times r^{\i\mu+7/2}$. In the very soft region (with a light red background in Figure \ref{Trispectrum_figure}), the term from the homogeneous solution dominates due to the additional $r^{-1/2}$ factor, and the frequency is $2\mu$. In the blue regions, due to lighter Boltzmann suppression, $r^{\i\mu+7/2}$ dominates with a frequency of $\mu$. This explains why the frequency on the left side of Figure \ref{Trispectrum_figure} is double that of the right region, making for a definitely striking feature of the double-exchange channel.

\section{Conclusions and Outlooks }
\label{sec:conclusion}
Cosmological correlators inherit information about all primordial fields, with any masses, spins, mixing angles, and including all kinds of interactions with the curvature perturbation and the two polarisations of the gravitational waves.
In order to have at hand full predictions in any given inflationary scenario, as well as for not biasing the interpretation of (upcoming) cosmological data, channels involving multiple massive exchanges \textit{must} be taken into account. However, obtaining analytical solutions for these channels is highly challenging. 
In this work, we present the \textit{first} example wherein we apply recently developed bootstrap equations to find the exact and explicit solutions of four-point and three-point correlation functions with the exchange of two massive fields.
The approach employed in this paper involves applying differential operators associated with each massive propagator individually, thereby expressing derivatives of the double-exchange seed integral as the corresponding single-exchange one without this propagator. By utilizing known results from prior research as input, we ultimately derive exact analytical solutions for these bootstrap equations. 
Through proper selection of variables, our final results exhibit a simplified form, comprising only \textit{one} layer of series summation.
This simplification facilitates faster convergence and makes it easier to investigate the analytic structure. 
The primary results of the four-point function exhibit divergences in certain terms for specific kinematic regions, making the task of taking the three-point function limit challenging.
However, we observed that by employing the continuation of certain special functions and regrouping the divergences in pairs, treating all contributions as a whole, the final expression is explicitly convergent. 
We also conducted a detailed analysis of the phenomenology related to double-exchange correlation functions, together with a careful comparison with an independent numerical method using \textsf{CosmoFlow}.
We concluded that the primordial bispectrum generated from this channel may be large, both at equilateral configurations and in the squeezed limit where the cosmological collider dominates, thereby opening the thrilling possibility to probe the existence of massive primordial fields via cosmological observations.
We explored diverse phenomenological aspects that make the double-exchange channels particularly intriguing, including specific phase information in the CC signal, new features in the inflationary flavor oscillations pattern and a unique transition to double frequency in the double soft limit of the primordial trispectrum.
More generally, our work showcases the utility of using diverse methods for the calculations of cosmological correlators, and taking full advantage of their synergies.

There are many aspects that deserve further investigation in the future. Here we outline a few of them.
First, as previously mentioned in the main text, it would be desirable to find simplified analytical expressions when the special functions are evaluated on arguments with values close to unity.
Second, as demonstrated by the example of the triple-exchange channel provided in Appendix~\ref{triple} and for which we derived preliminary results, our technique can be used to obtain fully analytical results for diagrams involving more-than-two exchanges of massive fields, definitely providing fascinating directions for future work. 
Third, it would be intriguing to investigate whether the well-known Suyama--Yamaguchi inequality~\cite{Suyama:2007bg}, relating the collapsed limit of the trispectrum to the squeezed one of the bispectrum, still holds in the context of heavy field exchanges, depending on the number of such exchanges.
Fourth, as already acknowledged in the main text and following the now standard tradition in computing cosmological correlators, we focused here on the exchange of massive fields in pure dS, but future works on multiple exchange diagrams should also include symmetry-breaking cases, such as those with non-unit sound speed, chemical potential, IR effects, and time-dependent masses and coupling constants. 
Fifth, although we did mention inflationary flavor oscillations in the context of double-exchange channels and uncovered first interesting unique features, it remains yet to conduct a thorough investigation of how to tell different channels apart in practice from (to-be) observations. 
Sixth, with the analytical results of tree-level double-exchange here, and the known result of the 1-loop bubble diagram \cite{Xianyu:2022jwk} that also contains two massive propagators, it would be interesting to compare the signals originating from these different channels.
Seventh, we explained in the phenomenology Section~\ref{Sec_pheno} that the cubic interaction $\lambda$, relevant for double-exchange channels, can be related to the field-space curvature in general non-linear sigma models of inflation; it would be thrilling to explore double-exchange correlators in the context of concrete models of moduli field spaces in string compactifications or the coset field space of (pseudo-)Nambu-Goldstones from spontaneous symmetry breaking patterns.
Lastly, only very recently was the CC signal arising from varied single-exchange diagrams searched for in the cosmological data~\cite{Cabass:2024wob,Sohn:2024xzd}, but given the relatively higher prospects of detectability of the double-exchange channels as we explained at length, it seems important to extend the search to templates proposed in this work.
We plan to address several of these directions in future works.

\paragraph{Acknowledgements.}
We would like to thank Kohei Fujikura, Toshifumi Noumi, Zhehan Qin, Sébastien Renaux-Petel, David Stefanyszyn, Xi Tong, Denis Werth and Zhong-Zhi Xianyu for useful discussion. L.P. would like to acknowledge
Denis Werth for helpful discussions regarding the \textsf{CosmoFlow} code used in this work.
S.A. is supported by IBS under the project code, IBS-R018-D1.
L.P. acknowledges funding support from the Initiative Physique des Infinis (IPI), a research training program of the Idex SUPER at Sorbonne Université.
F.S. acknowledges financial aid from the Institute for Basic Science, and JSPS Grant-in-Aid for Scientific Research No. 23KJ0938. M.Y. is supported by IBS under the project code, IBS-R018-D3, and by JSPS Grant-in-Aid for Scientific Research Number JP21H01080. Y.Z. is supported
by the IBS under the project code, IBS-R018-D3.

\begin{appendix}
\section{Mathematical Formulae}
\label{sec: formula}
In this Appendix, we provide a summary of definitions and formulae frequently utilized in this work. Most of these formulae can also be found in Mathematical functions handbook \cite{NIST:DLMF}.

Following Refs.~\cite{Qin:2022fbv, Qin:2023ejc, Xianyu:2023ytd}, we use shorthand notations for products of Gamma function,
\begin{align}
&\Gamma\left[z_1, \cdots, z_m\right] \equiv \Gamma\left(z_1\right) \cdots \Gamma\left(z_m\right),\\
&\Gamma\left[\begin{array}{c}
z_1, \cdots, z_m \\
w_1, \cdots, w_n
\end{array}\right] \equiv \frac{\Gamma\left(z_1\right) \cdots \Gamma\left(z_m\right)}{\Gamma\left(w_1\right) \cdots \Gamma\left(w_n\right)}.
\end{align}
Several types of hypergeometric functions are used in the main text, which we collect their definitions here. 
First, the (generalized) hypergeometric function ${}_p\rm{F}_q$ is defined by 
\begin{align}
{ }_p {\rm{F}}_q\left[\begin{array}{c|c}
a_1, \cdots, a_p \\
b_1, \cdots, b_q
\end{array} \ x\right]=\sum_{n=0}^{\infty} \frac{\left(a_1\right)_n \cdots\left(a_p\right)_n}{\left(b_1\right)_n \cdots\left(b_q\right)_n} \frac{x^n}{n !},  \label{def_hyper}  
\end{align}
where $(a)_n\equiv \Gamma[a+n]/\Gamma[a]$ is the Pochhammer symbol. 

We also use its ``dressed" version, which is defined by
\begin{align}
\nonumber { }_p\mathcal{F}_q\left[\begin{array}{c|c}
a_1, \cdots, a_p \\
b_1, \cdots, b_q
\end{array} \ x\right]&\equiv \Gamma\left[\begin{array}{c}
a_1, \cdots, a_p \\
b_1, \cdots, b_q
\end{array}\right]{ }_p \mathrm{F}_q\left[\begin{array}{c|c}
a_1, \cdots, a_p \\
b_1, \cdots, b_q
\end{array} \ x\right] \\
&=\sum_{n=0}^{\infty} \Gamma\left[\begin{array}{c}
a_1+n, \cdots, a_p+n \\
b_1+n, \cdots, b_q+n
\end{array}\right] \frac{x^n}{n !} .\label{dhyper}
\end{align}
There are many formulae changing the variables of hypergeometric functions. The one frequently used in the main text is
\begin{align}
    &\frac{\sin{(\pi(c-a-b))}}{\pi}{ }_2\mathcal{F}_1\left[\begin{array}{c|c}
a,b\\
c
\end{array} \ x\right]\nonumber\\
&=\frac{1}{\Gamma\left[c-a,c-b\right]}\,{ }_2\mathcal{F}_1\left[\begin{array}{c|c}
a,b\\
a+b-c+1
\end{array}\ 1-x\right]-\frac{(1-x)^{c-a-b}}{\Gamma\left[c-a,c-b\right]}{ }_2\mathcal{F}_1\left[\begin{array}{c|c}
c-a,c-b\\
c-a-b+1
\end{array} \ 1-x\right]~. \label{hyperTrans}
\end{align}

Next, let us move to the hypergeometric series of two variables. In this paper, we used two types of Appell series, $F_2$ and $F_4$, whose definitions are given by
\begin{align}
F_2\left[a\left|\begin{array}{l}
b_1, b_2 \\
c_1, c_2
\end{array}\right| x, y\right]&=\sum_{m, n=0}^{\infty} \frac{(a)_{m+n}\left(b_1\right)_m\left(b_2\right)_n}{\left(c_1\right)_m\left(c_2\right)_n m ! n !} x^m y^n,\\
F_4\left[\begin{array}{c|c}
a, b \\
c_1, c_2
\end{array}\ x, y\right]&=\sum_{m, n=0}^{\infty} \frac{(a)_{m+n}(b)_{m+n}}{\left(c_1\right)_m\left(c_2\right)_n m ! n !} x^m y^n ,  
\end{align}
and their dressed versions are expressed as
\begin{align}
\mathcal{F}_2\left[a\left|\begin{array}{l}
b_1, b_2 \\
c_1, c_2
\end{array}\right| x, y\right]&=\sum_{m, n=0}^{\infty} \Gamma\left[\begin{array}{c}
a+m+n, b_1+m, b_2+n \\
c_1+m, c_2+n
\end{array}\right] \frac{x^m y^n}{m ! n !}, \label{def_f2}\\   
\mathcal{F}_4\left[\begin{array}{c|c}
a, b \\
c_1, c_2
\end{array}\ x, y\right]&=\sum_{m, n=0}^{\infty} \Gamma\left[\begin{array}{c}
a+m+n, b+m+n \\
c_1+m, c_2+n
\end{array}\right] \frac{x^m y^n}{m ! n !} .    \label{def_f4}
\end{align}
Note that the function $F_2$ is convergent for $|x|+|y|< 1$, while the function  $ F_4$ for $\sqrt{|x|} +\sqrt{|y|} < 1$.
Here are some useful conversion formulae. The Appell $F_4$ function is related to Appell $F_2$ through 
\begin{align}\label{F4toF2}
    F_4\left[\begin{array}{c|c}
\frac{a}{2},\frac{a+1}{2} \\
b_1+\frac{1}{2}, b_2+\frac{1}{2}
\end{array}\ x^2, y^2\right]=(1+x+y)^{-a}F_2\left[a\left|\begin{array}{l}
b_1, b_2 \\
2b_1, 2b_2
\end{array}\right| \frac{2x}{x+y+1}, \frac{2y}{x+y+1}\right]~,
\end{align}
and the Appell $F_2$ can be expanded as summation of Gauss hypergeometric function~${}_2{\rm{F}}_1$ as 
\begin{align}\label{F2expansion}
	F_2\left[a\left|\begin{array}{l}
b_1, b_2 \\
c_1, c_2
\end{array}\right| x, y\right]=\sum_{m=0}^{\infty}\frac{(a)_m(b_1)_m}{(c_1)_m }\frac{x^m}{m!}{ }_2 {\rm{F}}_1\left[\begin{array}{c|c}
a+m,b_2 \\
c_2
\end{array} \ y\right]~.
\end{align}

Finally, the $F_C$-type of Lauricella’s hypergeometric functions of $m$ variables $x_1,\cdots,x_m$ appears in Appendix~\ref{triple}, which is defined by~\cite{matsumoto_2020},
\begin{align}
F_C\left[\begin{array}{c|c}
a,b \\
c_1, \cdots,c_m
\end{array}\ x_1,\cdots,x_m\right]=\sum_{n_1, \ldots, n_m=0}^{\infty} \frac{(a)_{n_1+\cdots+n_m}(b)_{n_1+\cdots+n_m}}{\left(c_1\right)_{n_1} \cdots\left(c_m\right)_{n_m} n_{1} ! \cdots n_{m} !} x_1^{n_1} \cdots x_m^{n_m},    \label{L_C}
\end{align}
with convergent radius
\begin{align}
\sqrt{\left|x_1\right|}+\cdots+\sqrt{\left|x_m\right|}<1.
\end{align}
Note that Lauricella’s $F_C$ function is reduced to Appell's and Gauss hypergeometric functions $F_4$ and ${}_2{\rm{F}}_1$ for $m=2$ and $m=1$, respectively.

\section{Seed Integral with Single Massive Exchange}\label{diagram_a}

Here we show analytic expressions of the seed integral with {\it{single}} exchange, which appear as source terms of bootstrap equations for double-exchange: the right-hand side of \eqref{BS_b_dS_ppm1}, \eqref{BS_b_dS_ppp1}, and \eqref{BS_b_dS_ppp2}. The analytical solution of the single-exchange tree-level diagram is well-established in the literature, as we mentioned in the introduction. Further, in cases where one vertex involves linear mixing, the solution provides an exact closed form. This solution as the source terms is utilized extensively in our main text discussion. The comprehensive derivation can be found in detail in \cite{Qin:2023ejc}. Here, we just collect the results in a convenient format for our purpose.  

To do so, let us transform the seed integral with single exchange~\eqref{I_dS_a} by changing the variables 
\begin{align}
z_1\equiv -k_{124}\tau_1,\quad z_2=-k_3\tau_2,    
\end{align}
and introducing
\begin{align}
D_{\m{a b}}^\a\left(k_3 ; \tau_1, \tau_2\right)=k_3^{-3} \widehat{D}^\a_{\m{a b}}\left(\frac{R}{2-R} z_1, z_2\right), \label{hat_r}  
\end{align}
where 
\begin{align}
R\equiv \frac{2k_3}{k_{1234}}.
\end{align}
Under these change, Eq.~\eqref{I_dS_a} can be written as
\begin{align}
\mathcal{I}_{\m{ab},\a}^{p_1p_2}(R) &=H^{-2} \left(\frac{R}{2-R}\right)^{1+p_1}(-\m{ab})   \int_0^{\infty} \mathrm{d} z_1 \mathrm{d} z_2\  z_1^{p_1}z_2^{p_2}e^{-\i \m{a} z_1-\i \m{b} z_2} \widehat{D}^\a_{\m{ab}}\left(  \frac{R}{2-R} z_1, z_2\right). \label{I_dS_a_2}
\end{align}
Note that the seed integral depends only on a specific momentum ratio $R$.

A summarized version of the explicit expression for Eq.~\eqref{I_dS_a_2} as a function of $R$ is given by
\begin{keyeqn}
\begin{align}
\nonumber \mathcal{I}_{ \pm\mp,\a}^{p_1 p_2}(R)=&\ \frac{ e^{\mp\i\frac{\pi}{2} \bar{p}_{12}}}{2^{6+p_{12}-2\i  \mu_\a} \pi^{\frac{1}{2}}} \Gamma\left[\begin{array}{c}
\frac{5}{2}+p_2+\i \mu_\a, \frac{5}{2}+p_2-\i \mu_\a, \frac{5}{2}+p_1-\i \mu_\a, \i  \mu_\a \\
3+p_2
\end{array}\right]\\
&\times R^{\frac{5}{2}+p_1-\i   \mu_\a} { }_2 {\rm{F}}_1\left[\begin{array}{c|c}
\frac{1}{2}-\i  \mu_\a, \frac{5}{2}+p_1-\i  \mu_\a \\
1-2\i \mu_\a
\end{array}\ R\right]+(\mu_\a\rightarrow -\mu_\a),\label{result_Ipm}\\
\nonumber \mathcal{I}_{ \pm\pm,\a}^{p_1 p_2}(R)=&\left\{\frac{\pm\i\  e^{\mp\i\frac{\pi}{2} p_{12}}e^{\mp \pi \mu_\a}}{2^{6+p_{12}-2\i \mu_\a} \pi^{\frac{1}{2}}} \Gamma\left[\begin{array}{c}
\frac{5}{2}+p_2+\i \mu_\a, \frac{5}{2}+p_2-\i \mu_\a, \frac{5}{2}+p_1-\i  \mu_\a, \i  \mu_\a \\
3+p_2
\end{array}\right]\right.\\
\nonumber &\left.\times R^{\frac{5}{2}+p_1-\i   \mu_\a} { }_2 {\rm{F}}_1\left[\begin{array}{c|c}
\frac{1}{2}-\i  \mu_\a, \frac{5}{2}+p_1-\i  \mu_\a \\
1-2\i \mu_\a
\end{array}\ R\right]+(\mu_\a\rightarrow -\mu_\a)\right\}\\
&+\frac{e^{\mp \i\frac{\pi}{2}  p_{12}} \Gamma\left(p_{12}+5\right)}{\mu_\a^2+\left(\frac{5}{2}+p_2\right)^2 }  \left(\frac{R}{2}\right)^{5+p_{12}}{}_3{\rm{F}}_2\left[\begin{array}{c|c}
1,5+p_{12}, 3+p_2 \\
\frac{7}{2}+p_2-\i \mu_\a, \frac{7}{2}+p_2+\i \mu_\a
\end{array}\  R\right],\label{result_Ipp}
\end{align}
\end{keyeqn}
where $p_{12}\equiv p_1+p_2$ and $\bar{p}_{12}\equiv p_1-p_2$.

\section{Boundary Conditions of Seed Integrals}\label{MB_app}
In the main text, in Section~\ref{sec:coeff}, we imposed the soft limit ($u,v\rightarrow0$) as the boundary conditions to determine the coefficients of bootstrap equations. In this Appendix, we present the calculation of the time integrals under these limits. In principle, one can employ the late-time expansion of the Hankel function to directly evaluate the time integral under these limits. Instead, as suggested by previous work \cite{Qin:2022fbv}, the partial Mellin-Barnes (MB) representation proves to be an efficient approach for handling such time integrals. So in this Appendix we will adopt the MB representation to find the boundary condition. We will not extensively discuss the derivation details, but instead, we will present the final results in a format suitable for our purposes. Further details on the use of the partial MB method to evaluate double-exchange correlators can be found in \cite{Xianyu:2023ytd}~.
\\
The key trick of this method is to use the following representation of Hankel function,
\begin{align}
\mathrm{H}_{\i\mu}^{(j)}(-k\tau)=\int_{-\mathrm{i} \infty}^{\mathrm{i} \infty} \frac{\mathrm{d} s}{2 \pi \mathrm{i}} \frac{( -k\tau/ 2)^{-2 s}}{\pi} e^{(-1)^{j+1}(2 s-\i\mu-1)  \mathrm{i} \frac{\pi}{2}} \Gamma\left[s-\frac{\i\mu}{2}, s+\frac{\i\mu}{2}\right], \quad(j=1,2) \label{MB}
\end{align}
by which the time-integral in the seed integral can be trivially performed. Then, the remaining complex integral with variable $s$ can be done by the residue theorem. With \eqref{MB}, the factorised propagators 
in the seed integral~\eqref{I_abc} can be expressed as
\begin{align}
\nonumber D^\a_{\pm \mp}\left(k_1 ; \tau_1, \tau_2\right) =&\ \frac{H^2\pi e^{-\pi \mu_\a}}{4}\left(\tau_1 \tau_2\right)^{3 / 2} H_{\mp \i \mu_\a}^{(2,1)}\left(-k_1 \tau_1\right) H_{\pm\i \mu_\a}^{(1,2)}\left(-k_1 \tau_2\right)\\
\nonumber =&\ \frac{H^2}{4 \pi} \int_{-\mathrm{i} \infty}^{\mathrm{i} \infty} \frac{\mathrm{d} s_1}{2 \pi \mathrm{i}} \frac{\mathrm{d} s_2}{2 \pi \mathrm{i}} e^{\mp \mathrm{i} \pi\left(s_1-s_2\right)}\left(\frac{k_1}{2}\right)^{-2 s_{12}}\left(-\tau_1\right)^{-2 s_1+3 / 2}\left(-\tau_2\right)^{-2 s_2+3 / 2}\\
&\times \Gamma\left[s_1-\frac{\mathrm{i} \mu_\a}{2}, s_1+\frac{\mathrm{i} \mu_\a}{2}, s_2-\frac{\mathrm{i} \mu_\a}{2}, s_2+\frac{\mathrm{i} \mu_\a}{2}\right],\label{D_pm_MB}
\end{align}
and similarly for $D^\b_{\pm\mp}\left(k_3;\tau_2,\tau_3\right)$. For time-ordered propagators, it has been observed that a more effective way involves dividing them into factorised and time-ordered components through
\begin{align}
 &D^\a_{\pm \pm}\left(k_1 ; \tau_1, \tau_2\right)=D^\a_{\pm\mp}\left(k_1 ; \tau_1, \tau_2\right)+\left[D^\a_{\mp\pm}\left(k_1 ; \tau_1, \tau_2\right)-D^\a_{\pm\mp}\left(k_1 ; \tau_1, \tau_2\right)\right] \theta\left(\tau_1-\tau_2\right) , \\
 & D^\b_{\pm \pm}\left(k_3 ; \tau_2, \tau_3\right)=D^\b_{\mp\pm}\left(k_3 ; \tau_2, \tau_3\right)+\left[D^\b_{\pm\mp}\left(k_3 ; \tau_2, \tau_3\right)-D^\b_{\mp\pm}\left(k ; \tau_2, \tau_3\right)\right] \theta\left(\tau_3-\tau_2\right).
\end{align}
Then accordingly, we define
factorised and time-ordered parts of the seed integral by \cite{Xianyu:2023ytd}
\begin{align}
&\mathcal{I}_{\pm \mp \pm}~,\\
&\mathcal{I}_{\pm \pm\mp}=\mathcal{I}_{\pm \pm\mp, \mathrm{F}}+\mathcal{I}_{\pm \pm\mp, \mathrm{T}}~,\\ 
&\mathcal{I}_{\pm \mp\mp}=\mathcal{I}_{\pm \mp\mp, \mathrm{F}}+\mathcal{I}_{\pm \mp\mp, \mathrm{T}}~,\\ 
&\mathcal{I}_{\pm \pm\pm}=\mathcal{I}_{\pm \pm\pm, \mathrm{FF}}+\mathcal{I}_{\pm \pm\pm, \mathrm{TF}}+\mathcal{I}_{\pm \pm\pm, \mathrm{FT}}+\mathcal{I}_{\pm \pm\pm, \mathrm{TT}}~,
\end{align}
where $\mathcal{I}_{\pm \mp \pm}$ is already factorised from the beginning. The subscript ${}{{}_\mathrm{F}}$(${}_{\mathrm{T}}$) indicates that it originates from the factorised (time-ordered) part of the time-ordered propagators. In the soft limit, the (fully) factorised part will dominate \cite{Xianyu:2023ytd}.
To clarify further, the dominant parts are
\begin{align}
\nonumber \mathcal{I}_{\pm\mp\pm}=&\pm\i H^{-4}k_{24}^{9+p_{123}} \int^0_{-\infty}  \mathrm{d} \tau_1\mathrm{d} \tau_2\mathrm{d} \tau_3(-\tau_1)^{p_1}(-\tau_2)^{p_2}(-\tau_3)^{p_3}e^{\pm\mathrm{i} k_{1} \tau_1\mp\i k_{24} \tau_2\pm\i k_{3} \tau_3}\\
&\times  D^{\a}_{\pm\mp}\left(k_1 ; \tau_1, \tau_2\right)D^{\b}_{\mp\pm}\left(k_3 ; \tau_2, \tau_3\right),    
\end{align}
\begin{align}
\nonumber \mathcal{I}_{\pm\pm\mp,\mathrm{F}}=&\pm\i H^{-4} k_{24}^{9+p_{123}}\int^0_{-\infty}  \mathrm{d} \tau_1\mathrm{d} \tau_2\mathrm{d} \tau_3(-\tau_1)^{p_1}(-\tau_2)^{p_2}(-\tau_3)^{p_3}e^{\pm\mathrm{i} k_{1} \tau_1\pm\i k_{24} \tau_2\mp\i k_{3} \tau_3}\\
&\times  D^{\a}_{\pm\mp}\left(k_1 ; \tau_1, \tau_2\right)D^{\b}_{\pm\mp}\left(k_3 ; \tau_2, \tau_3\right),    
\end{align}
\begin{align}
\nonumber \mathcal{I}_{\pm\mp\mp,\mathrm{F}}=&\mp\i H^{-4} k_{24}^{9+p_{123}}\int^0_{-\infty}  \mathrm{d} \tau_1\mathrm{d} \tau_2\mathrm{d} \tau_3(-\tau_1)^{p_1}(-\tau_2)^{p_2}(-\tau_3)^{p_3}e^{\pm\mathrm{i} k_{1} \tau_1\mp\i k_{24} \tau_2\mp\i k_{3} \tau_3} \\
&\times D^{\a}_{\pm\mp}\left(k_1 ; \tau_1, \tau_2\right)D^{\b}_{\pm\mp}\left(k_3 ; \tau_2, \tau_3\right),    
\end{align}
\begin{align}
\nonumber \mathcal{I}_{\pm\pm\pm,\mathrm{FF}}=&\mp\i H^{-4} k_{24}^{9+p_{123}}\int^0_{-\infty}  \mathrm{d} \tau_1\mathrm{d} \tau_2\mathrm{d} \tau_3(-\tau_1)^{p_1}(-\tau_2)^{p_2}(-\tau_3)^{p_3}e^{\pm\mathrm{i} k_{1} \tau_1\pm\i k_{24} \tau_2\pm\i k_{3} \tau_3} \\
&\times D^{\a}_{\pm\mp}\left(k_1 ; \tau_1, \tau_2\right)D^{\b}_{\mp\pm}\left(k_3 ; \tau_2, \tau_3\right).    
\end{align}
As mentioned above, the things to do are to insert the MB representations of the SK propagators~\eqref{D_pm_MB} into the seed integrals above, perform $\tau_i\ (i=1,2,3)$-integral, and evaluate the complex $s_j\ (j=1,2,3,4)$-integral by residue theorem. After straightforward calculation, we gets the results summarized below. 
\\
\\
As for the fully factorised seed integral $\mathcal{I}_{\pm \mp\pm,\a\b}^{ p_1 p_2 p_3}$, the final results are 
\begin{align}
\nonumber \mathcal{I}_{\pm \mp\pm,\a\b}^{ p_1 p_2 p_3} =&\ \frac{e^{\mp \i\frac{ \pi}{2} (p_{13}-p_2)}}{2^{7+p_{13}}\pi} u^{-p_1-\frac{5}{2}} v^{-p_3-\frac{5}{2}}\Gamma\left[\begin{array}{c}
\frac{5}{2}+p_1-\i \mu_\a, \frac{5}{2}+p_1+\i \mu_\a, \frac{5}{2}+p_3-\i \mu_\b, \frac{5}{2}+p_3+\i \mu_\b \\
3+p_1, 3+p_3
\end{array}\right] \\
&\nonumber\times \sum_{\m{a},\m{b}=\pm}\ \sum_{n_{2}, n_3=0}^{\infty}\left(\frac{u}{2}\right)^{2 n_2-\i \m{a}\mu_\a}\left(\frac{v}{2}\right)^{2 n_3-\i \m{b}\mu_\b} \frac{(-1)^{n_{23}}}{n_{2} ! n_{3} !} \\
&\times \Gamma\left[-n_2+\i \m{a}\mu_\a,-n_3+\i \m{b}\mu_\b, p_2+4+2 n_{23}- \i (\m{a}\mu_\a+\m{b}\mu_\b)\right],
\end{align}
where $n_{23}=n_2+n_3$.
Under the limit $u,v\ll 1$ for boundary conditions of bootstrap equations, the dominant contribution comes from $n_{2,3}=0$ in summations. Then, utilizing the following formulae for the Gamma function,
\begin{align}
\Gamma[z, 1-z]=\frac{\pi}{\sin \pi z},\quad\Gamma\left[\begin{array}{c} 2 z \\ z, z+\frac{1}{2} \end{array}\right]=\frac{2^{2 z-1}}{\pi^{1 / 2}},    
\end{align}
one may summarize 
\begin{align}
\nonumber \lim _{u,v \ll 1}\mathcal{I}_{\pm \mp\pm,\a\b}^{ p_1 p_2 p_3}
=&-\sum_{\m{a,b}=\pm}e^{\mp \i \frac{\pi}{2}\left(p_{13}-p_2\right)}\operatorname{csch}\left(\pi \m{a} \mu_\a\right)\operatorname{csch}\left(\pi \m{b} \mu_\b\right) \widetilde{\Gamma}(p_1,p_2,p_3,\mu_\a,\mu_\b)\\
&\times u^{-\frac{5}{2}-p_1-\i\m{a} \mu_\a} v^{-\frac{5}{2}-p_3-\i\m{b} \mu_\b} \ \Gamma\left[\begin{array}{c}
\frac{4+p_2-\i\left(\m{a}\mu_\a+\m{b}\mu_\b\right)}{2}, \frac{5+p_2-\i\left(\m{a}\mu_\a+\m{b}\mu_\b\right)}{2} \\
1-\i\m{a} \mu_\a, 1-\i \m{b}\mu_\b
\end{array}\right],\label{lim_+-+}
\end{align}
with
 \begin{align}
\widetilde{\Gamma}(p_1,p_2,p_3,\mu_\a,\mu_\b)\equiv\frac{ \pi^{\frac{1}{2}}}{2^{4+p_{13}-p_2}} \Gamma\left[\begin{array}{c}
\frac{5}{2}+p_1-\i \mu_\a, \frac{5}{2}+p_1+\i \mu_\a, \frac{5}{2}+p_3-\i \mu_\b, \frac{5}{2}+p_3+\i \mu_\b \\
3+p_1, 3+p_3
\end{array}\right]~.
\end{align}
A similar procedure can be applied to both the partially factorised (partially nested) and fully nested seed integrals. Under double soft limits, they yield:
\begin{align}
\nonumber \lim _{u,v \ll 1}\mathcal{I}_{\pm \pm\mp,\a\b,}^{ p_1 p_2 p_3} 
=&\mp\i \sum_{\m{a,b}=\pm}e^{\mp \i \frac{\pi}{2}\left(p_{12}-p_3\right)} \operatorname{csch}\left(\pi \m{a} \mu_\a\right)\operatorname{csch}\left(\pi \m{b} \mu_\b\right)  e^{\mp\pi\m{a}\mu_\a}\widetilde{\Gamma}(p_1,p_2,p_3,\mu_\a,\mu_\b)\\
&\times u^{-\frac{5}{2}-p_1-\i\m{a} \mu_\a} v^{-\frac{5}{2}-p_3-\i\m{b} \mu_\b} \ \Gamma\left[\begin{array}{c}
\frac{4+p_2-\i\left(\m{a}\mu_\a+\m{b}\mu_\b\right)}{2}, \frac{5+p_2-\i\left(\m{a}\mu_\a+\m{b}\mu_\b\right)}{2} \\
1-\i\m{a} \mu_\a, 1-\i \m{b}\mu_\b
\end{array}\right],\label{lim_++-}
\end{align}
\begin{align}
\nonumber \lim _{u,v \ll 1}\mathcal{I}_{\pm \pm\pm,\a\b}^{ p_1 p_2 p_3} =&\sum_{\m{a,b}=\pm}e^{\mp \i \frac{\pi}{2}p_{123}}\operatorname{csch}\left(\pi \m{a} \mu_\a\right)\operatorname{csch}\left(\pi \m{b} \mu_\b\right)  e^{\mp\pi(\m{a}\mu_\a+\m{b}\mu_\b)}\widetilde{\Gamma}(p_1,p_2,p_3,\mu_\a,\mu_\b)\\
&\times u^{-\frac{5}{2}-p_1-\i\m{a} \mu_\a} v^{-\frac{5}{2}-p_3-\i\m{b} \mu_\b} \ \Gamma\left[\begin{array}{c}
\frac{4+p_2-\i\left(\m{a}\mu_\a+\m{b}\mu_\b\right)}{2}, \frac{5+p_2-\i\left(\m{a}\mu_\a+\m{b}\mu_\b\right)}{2} \\
1-\i\m{a} \mu_\a, 1-\i \m{b}\mu_\b
\end{array}\right].\label{lim_+++}
\end{align}
The results for $\mathcal{I}_{\pm\mp\mp,\a\b}^{ p_1 p_2 p_3}$ can be directly obtained by replacing  $u\leftrightarrow v$, $p_1\leftrightarrow p_3 $, and $\alpha\leftrightarrow \beta$ in the expression of $\mathcal{I}_{\mp\mp\pm,\a\b}^{ p_1 p_2 p_3}$.
\section{Toward Triple Exchange} \label{triple}
In principle, the bootstrap equation approach we applied to the double-exchange diagram in this work can be easily generalized to any tree-level diagrams.
In this section, let us see how it works for the triple-exchange diagram, keeping in mind that obtaining the complete analytic form remains a non-trivial task for future work. In particular, the bispectrum with triple massive exchange would be phenomenologically important because it can give a sizable signal known, as first emphasized in the context of quasi-single-field inflation~\cite{Chen:2009zp}.

\begin{figure}[ht]
\centering
\includegraphics[width=0.7\textwidth]{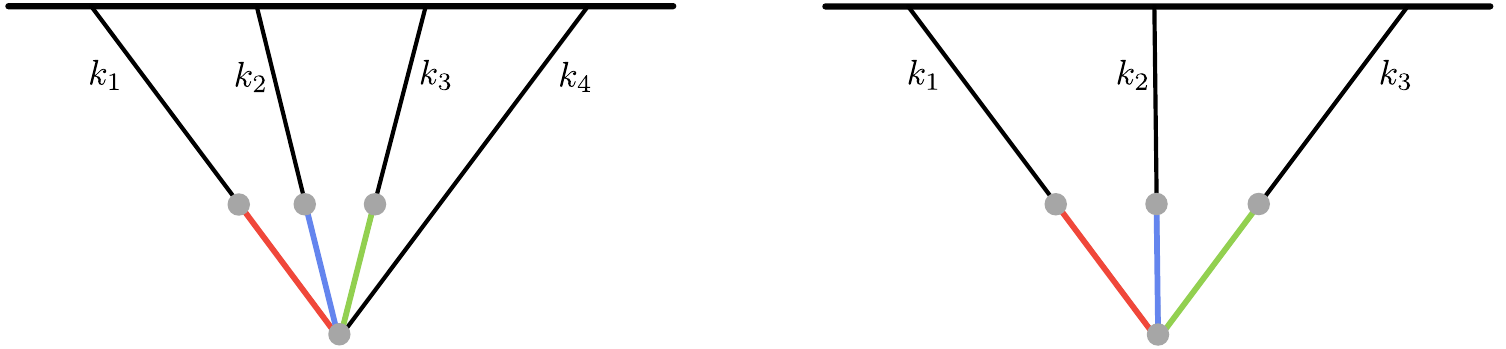}\\
	\caption{Four-point and three-point diagrams with triple massive exchange. For generality,  distinct colors here denote massive scalar fields \textcolor{red}{$\sigma^\a$}, \textcolor[RGB]{100,133,238}{$\sigma^\b$} and \textcolor[RGB]{146,209,79}{$\sigma^\c$} with different mass.  \label{TripleDiagram} }
\end{figure}

Here again, we start from the seed integral for the four-point inflaton correlator, but from a triple exchange, which can be defined by 
\begin{keyeqn}
\begin{align}
\nonumber \mathcal{I}_{\m{abcd},\a\b\c}^{p_1p_2p_3p_4}=&\  H^{-6}k_{4}^{13+p_{1234}}( \m{abcd})  \int^0_{-\infty}\ \prod_{i=1}^4\  \mathrm{d} \tau_i\ (-\tau_i)^{p_i}e^{\i \m{a} k_{1} \tau_1+\i\m{b} k_{2} \tau_2+\i\m{c} k_{3} \tau_3+\i\m{d} k_{4} \tau_4}\\
&\times  D^{\a}_{\m{ad}}\left(k_1 ; \tau_1, \tau_4\right)D^{\b}_{\m{bd}}\left(k_2 ; \tau_2, \tau_4\right)D^{\c}_{\m{cd}}\left(k_3 ; \tau_3, \tau_4\right).\label{I_abcd}
\end{align}
\end{keyeqn}
It can be diagrammatically expressed as shown in Figure~\ref{TripleDiagram}. The three-point correlator with triple exchange can be obtained by setting the external momentum soft, i.e., $k_4\rightarrow 0$. In the following, we consider a situation where three massive fields contribute to the seed from the mixing vertex $\sigma^\a\sigma^\b\sigma^\c$. Without ambiguity and for the sake of brevity, we will omit mass indices ($\a,\b,\c$) in the following text.

Similarly to the procedure shown in the main text, we transform the seed \eqref{I_abcd} to a more convenient form through changing the integration variables by
\begin{align}
-k_i \tau_i=z_i, \quad (i=1, 2,3, 4),    
\end{align}
and define the hat propagator in the same way as Eqs.~\eqref{hat_mp}-\eqref{hat_pp}. Then, the seed integral~\eqref{I_abcd} can be written as
\begin{align}
\nonumber \mathcal{I}_{\m{abcd}}^{p_1p_2p_3p_4}=&\ \frac{H^{-6}( \m{abcd})}{w_1^{4+p_1} w_2^{4+p_2} w_3^{4+p_3}} \int_0^{\infty} \prod_{i=1}^4\ \mathrm{d} z_i\ z_i^{p_i}e^{-\i\m{a} z_{1} -\i\m{b} z_2-\i\m{c} z_3-\i\m{d} z_4}  \widehat{D}_{\m{ad}}\left(z_1, w_1z_4\right)\widehat{D}_{\m{bd}}\left(z_2, w_2z_4\right)\widehat{D}_{\m{cd}}\left(z_3, w_3z_4\right),\\
\equiv&\ \frac{1}{w_1^{4+p_1} w_2^{4+p_2} w_3^{4+p_3}}\widehat{\mathcal{I}}_{\m{abcd}}^{p_1p_2p_3p_4}(w_1,w_2,w_3)~, \label{hatI_abcd}
\end{align}
where  
\begin{align}
w_i \equiv \frac{k_i}{k_4}~, \quad (i=1,2,3).    
\end{align}
\\
Now let us derive the bootstrap equations for $\widehat{\mathcal{I}}$. 
For example,  repeating the same procedure as in the main text, we find $\widehat{\mathcal{I}}_{\pm\pm\pm\mp}$ to satisfy the following homogeneous differential equations
\begin{align}
\mathcal{D}^{(3)}_{w_i}\,\widehat{\mathcal{I}}_{\pm\pm\pm\mp}^{p_1p_2p_3p_4}=0, ~~\qquad (i=1,2,3)~, \label{BS_pppm}
\end{align}
where the differential operators $\mathcal{D}^{(3)}_{w_i}\ (i=1,2,3)$ are defined by
\begin{keyeqn}
\begin{align}
\nonumber \mathcal{D}^{(3)}_{w_i} &\equiv   w_i^2\partial_{w_i}^2-2w_i\partial_{w_i}+\mu_{i}^2+\frac{9}{4}-w_i^2\left( \sum_{j=1}^3 w_j\partial_{w_j}+p_4+2\right)\left(\sum_{j=1}^3 w_j\partial_{w_j}+p_4+1\right)\\
\nonumber &=w_i^2\left(1-w_i^2\right) \partial_{w_i}^2-2 w_i^3 \partial_{w_i} \sum_{j\neq i}w_j\partial_{w_j}-w_i^2 \sum_{j, k \neq i} w_j w_k \partial_{w_j} \partial_{w_k}-2 w_i\left(1+\left( p_4+2\right)w_i^2\right) \partial_{w_i}\\
&\ \ \ \ -2\left( p_4+2\right)w_i^2 \sum_{j\neq i} w_j \partial_{w_j}+\mu_{i}^2+\frac{9}{4}-(p_4+2)(p_4+1)w_i^2\,,\label{D_triple}
\end{align}
\end{keyeqn}
where $\mu_i$ is the mass parameter of the massive field associated with the momentum $k_i$.
Remarkably, this can be analytically solved by
\begin{align} 
\nonumber \widehat{\mathcal{I}}_{\pm\pm\pm\mp}^{p_1p_2p_3p_4}=&\ \sum_{\m{abc}=\pm}c_{\pm\pm\pm\mp, \m{abc}}\ w_1^{\frac{3}{2}-\i\m{a}\mu_1}w_2^{\frac{3}{2}-\i\m{b}\mu_2}w_3^{\frac{3}{2}-\i\m{c}\mu_3}\\
&\times F_C\left[\begin{array}{c|c}
\frac{2p_4+11-2\i (\m{a}\mu_1+\m{b}\mu_2+\m{c}\mu_3)}{4}, \frac{2p_4+13-2\i (\m{a}\mu_1+\m{b}\mu_2+\m{c}\mu_3)}{4} \\
1-\i \m{a} \mu_1, 1-\i \m{b}\mu_2,1-\i\m{c} \mu_3
\end{array}\ w_1^2, w_2^2, w_3^2\right], \label{sol_pppm}
\end{align}
where the function~$F_C[\cdots]$ is Lauricella’s hypergeometric function with three variables defined in~\eqref{L_C}. The eight coefficients~$c_{\pm\pm\pm\mp, \m{abc}}$ could be determined by proper boundary conditions.

\begin{figure}[htp]
\centering
\includegraphics[width=0.75\textwidth]{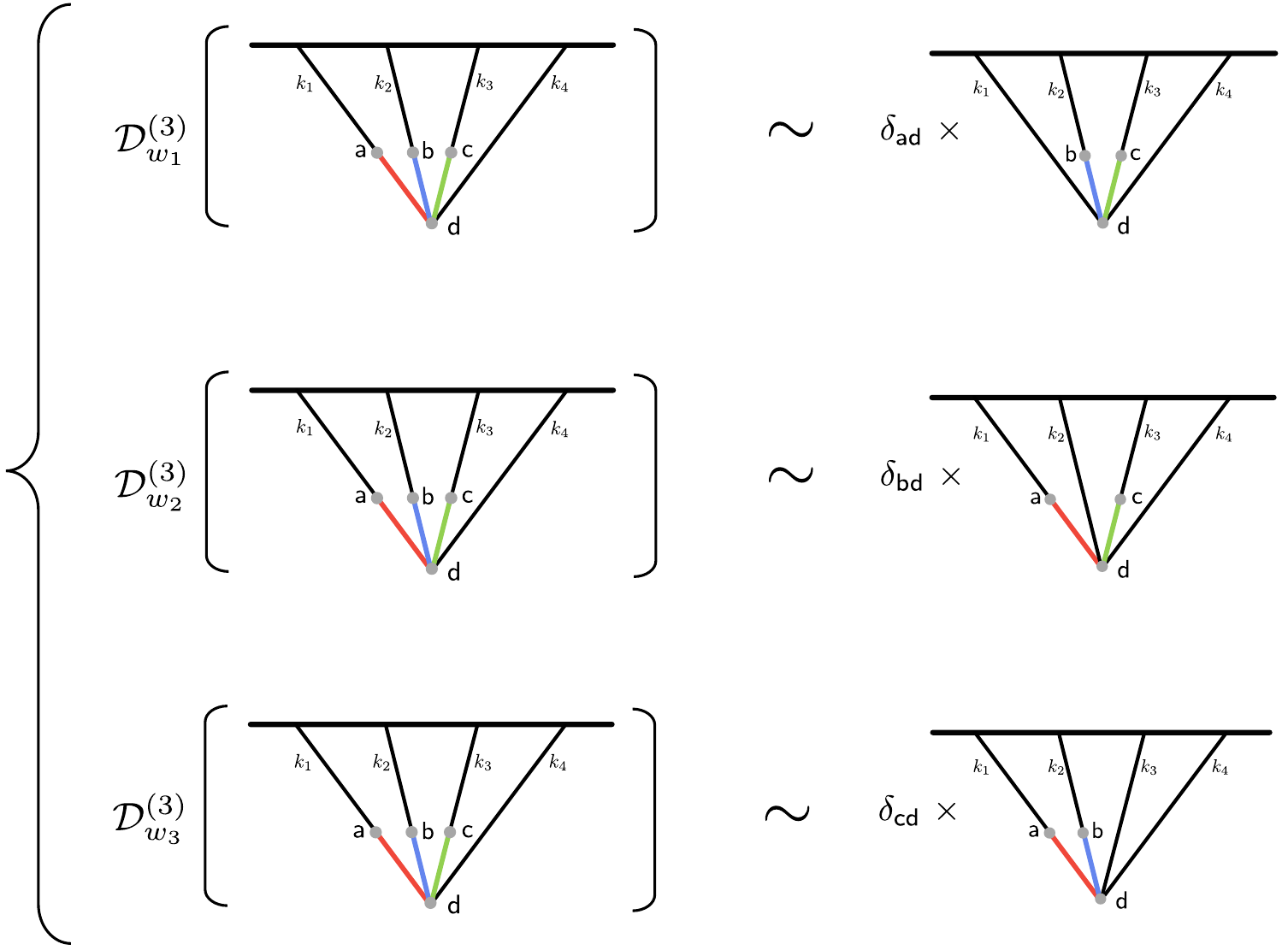}\\
    \caption{The schematic diagram illustrates the triple-exchange bootstrap equations. $\m{a},\m{b},\m{c}$ and $\m{d}$ are the SK indices. When the two SK indices of one massive propagator are opposite, applying the corresponding differential operator results in homogeneous equations. Conversely, if the two indices are the same, an additional $\delta$-term introduces a source term, which can be expressed using the seed integral of the double-exchange diagram.
    }\label{BSeqTriple}
\end{figure}

As a next example, let us see $\widehat{\mathcal{I}}_{\mp\pm\pm\mp}$ including a single time-ordered propagator for~$\sigma^\a $, $\widehat{D}_{\mp\mp}(z_1,r_1z_4)$. In this case, we obtain an inhomogeneous term in the bootstrap equation after acting with the differential operator~$\mathcal{D}^{(3)}_{w_1}$, 
\begin{align}
&\mathcal{D}^{(3)}_{w_1}\,\widehat{\mathcal{I}}_{\mp\pm\pm\mp,\a\b\c}^{p_1p_2p_3p_4}=\frac{w_1^{4+p_1} w_2^{4+p_2} w_3^{4+p_3}}{\left(1+w_1\right)^{p_{1234}+13}}\mathcal{I}_{ \pm \mp \pm}^{p_2, p_{14}+4, p_3} \left(u=\frac{w_2}{1+w_1}, v=\frac{w_3}{1+w_1}\right)~,\label{BS_mppm}
\end{align}
while the other equations with $\mathcal{D}^{(3)}_{w_2}$ and $\mathcal{D}^{(3)}_{w_3}$ still satisfy the homogeneous ones,
\begin{align}
\mathcal{D}^{(3)}_{w_2}\,\widehat{\mathcal{I}}_{\mp\pm\pm\mp,\a\b\c}^{p_1p_2p_3p_4}=\mathcal{D}^{(3)}_{w_3}\,\widehat{\mathcal{I}}_{\mp\pm\pm\mp,\a\b\c}^{p_1p_2p_3p_4}=0~.
\end{align}
Note that the right-hand side of the first equation~\eqref{BS_mppm} is written as the seed integral of double-exchange diagram, which we obtained analytic forms in the main text. 
This structure is generic: the seed integral with $n$-massive field exchange, denoted as $\mathcal{I}_n$, emerges as a source term in the bootstrap equations of $\mathcal{I}_{n+1}$.
Consequently, we can systematically solve more complex tree-level diagrams, although the detailed calculations remain non-trivial tasks: as shown in \eqref{sol_pppm}, the homogeneous solutions can be written down as known special functions, which is true even for more general diagram with $n$-massive exchanges. 
However, determining particular solutions would be a non-trivial task since the source term (the right-hand side of \eqref{BS_mppm} for triple exchange) becomes more and more complicated as $n$ increases, and we leave the detailed derivation for future work.
\\
Similarly, deriving the remaining bootstrap equations for other seed integrals follows a straightforward procedure. However, we do not intend to delve into all the tedious details within this work. Instead, we illustrate the structure of the bootstrap equations in Figure \ref{BSeqTriple}.
\end{appendix}

	\bibliographystyle{JHEP}
	\bibliography{ref}

	\end{document}